\documentclass[fleqn,usenatbib]{mnras}

\usepackage{newtxtext,newtxmath}

\usepackage[T1]{fontenc}

\DeclareRobustCommand{\VAN}[3]{#2}
\let\VANthebibliography\thebibliography
\def\thebibliography{\DeclareRobustCommand{\VAN}[3]{##3}\VANthebibliography}

\newcommand{\HI}{H{\sc i}}

\usepackage{graphicx}
\usepackage{amsmath}
\usepackage{epstopdf}
\usepackage{subcaption}
\usepackage{caption}
\usepackage{rotating}
\usepackage{pdflscape}
\usepackage{flafter}
\usepackage{afterpage}
\usepackage{lipsum}
\newcommand{\myedit}{}

\let\ACMmaketitle=\maketitle
\renewcommand{\maketitle}{\begingroup\let\footnote=\thanks \ACMmaketitle\endgroup}

\title[A serendipitous discovery of HI-rich galaxy groups]{A serendipitous discovery of HI-rich galaxy groups with MeerKAT}

\author[M. Glowacki]{
M. Glowacki$^{1}$\thanks{E-mail: marcin.glowacki@curtin.edu.au}, L. Albrow$^{2}$, T. Reynolds$^{3}$, E. Elson$^{4}$, E. K. Mahony$^{5}$, J. R. Allison$^{6}$
\\
$^{1}$International Centre for Radio Astronomy Research (ICRAR), Curtin University, Bentley, WA 6102, Australia\\
$^{2}$University of Canterbury, Department of Physics and Astronomy, Private Bag 4800, Christchurch 8020, New Zealand\\
$^{3}$International Centre for Radio Astronomy Research (ICRAR), The University of Western Australia, 35 Stirling Hwy, Crawley, WA 6009, Australia\\
$^{4}$Department of Physics \& Astronomy, University of the Western Cape, Robert Sobukwe Rd, Bellville, 7535, South Africa\\
$^{5}$ATNF, CSIRO Space and Astronomy, PO Box 76, Epping, NSW 1710, Australia \\
$^{6}$First Light Fusion Ltd., Unit 9/10 Oxford Pioneer Park, Mead Road, Yarnton, Kidlington OX5 1QU, UK \\
}

\date{Accepted 2024 March 1. Received 2024 March 1; in original form 2023 September 30.}

\pubyear{2024}

\begin{document}
\label{firstpage}
\pagerange{\pageref{firstpage}--\pageref{lastpage}}
\maketitle

\begin{abstract}
We report on the serendipitous discovery of 49 \HI-rich galaxies in a 2.3 hour Open Time observation with MeerKAT. We present their properties including their \HI\ masses, intensity and velocity maps, and spectra. We determine that at least three \HI-rich galaxy groups have been detected, potentially as part of a supergroup. Some members of these galaxy groups show clear interaction with each other in their \HI\ emission. We cross-match the detections with PanSTARRS, WISE and GALEX, and obtain stellar masses and star formation rates. One source is found to be a potential OH megamaser, but further follow-up is required to confidently determine this. For 6 sources with sufficient spatial resolution in \HI\ we produce rotation curves with BBarolo, generate mass models, and derive a dark matter halo mass. While the number of galaxies detected in this relatively short pointing appears to be at the high end of expectations compared to other MeerKAT observations and group HIMF studies, this finding highlights the capability of MeerKAT for other serendipitous discoveries, and the potential for many more \HI-rich galaxies to be revealed within both existing and upcoming Open Time datasets.
\end{abstract}




\section{Introduction}

Galaxy groups occupy up to $\sim$50\% of the local Universe \cite[$z \sim 0$;][]{Eke2004, Robotham2011} and are important parts of the hierarchical structure of the Universe \citep{Springel2018} where they trace filamentary large-scale structure, as they lie between the low-density environments of the field and high-density clusters. Galaxy systems form through episodic mergers, with groups thought to act as the building blocks for the more massive clusters. Groups number between 3--100 members within a dark matter halo of mass between 10$^{12}$--10$^{14}$\,M$_{\odot}$ \cite[e.g.][]{Catinella2013}. However, the distinction between galaxy groups and clusters is inconsistently defined in literature \cite[see review by][]{Lovisari2021}. 

Galaxy groups provide a window into the baryon cycle and the circumgalactic medium \citep{Tumlinson2017,Nielsen2020}. External and internal feedback mechanisms such as active galactic nuclei (AGN) feedback and tidal interactions dominate \cite[e.g.][]{Ponman2003,McCarthy2010}, meaning galaxy groups are not merely `scaled-down' versions of galaxy clusters. The distinction between galaxy groups and clusters has not been clearly made in previous literature, but galaxy clusters (often numbering 100+ galaxy members) are larger than groups ($\sim$3--50 galaxy members). The properties of both are dependent on the density of the environment, although galaxy clusters are generally denser. 
Interactions between group members and their surrounding environment can also shape the evolution of galaxies as they transition from star-forming spirals to massive, quenched galaxies dominating in clusters \cite[e.g.][]{Baldry2004,Driver2011,Davies2019D}. For example, there is evidence that galaxies have undergone `pre-processing' in group environments before they fall into galaxy clusters \citep{Mahajan2013,Bianconi2018,Kleiner2021,Loubser2024}, which can potentially quench star formation in group galaxies. 

Neutral hydrogen (\HI) emission is a useful tool for studying group environments because it can trace the tidal interactions between group members, and extends well beyond the stellar component of the galaxy \cite[e.g.][]{Broeils1997, Leroy2008}. \HI\ is the fuel for star formation, and its distribution provides key insights into the quenching or triggering of star formation in galaxies. \HI\ is a sensitive dynamical tracer and can be used to observe environmental processes such as tidal interactions or ram-pressure stripping, the latter seen for larger groups \citep{Oosterloo2005, Serra2013, Saponara2018}. By observing \HI\ in group galaxies, much work has been done to understand the evolution of \HI\ within this environment \citep{Hess2013,Jones2019,Kleiner2021}. For example, \cite{Brown2017} conducted a statistic study on the \HI\ content as a function of environment density for gas-poor to gas-rich regimes, and found gas starvation alone cannot account for an observed decrease of gas content in group and cluster environments. \cite{Stevens2023} examined the TNG50 simulation suite and reproduced the effects of gas truncation from a Virgo-like cluster environment on the gas content of satellite galaxies.

New radio telescopes serving as pathfinders for the upcoming Square Kilometer Array (SKA) have already demonstrated their ability in detecting new \HI-rich galaxy groups, as well as enhancing previous studies of galaxy groups, thanks to improved sensitivity and survey speeds. The APERture Tile In Focus \cite[Apertif;][]{Cappellen2022} system uses a phased array feed (PAF) upgrade of the Westerbork Synthesis Radio Telescope. The first year of survey observations with Apertif has covered approximately one thousand square degrees of sky \citep{Adams2022}. The Australian Square Kilometre Array Pathfinder telescope \cite[ASKAP;][]{Deboer2009,Hotan2021} also employs PAFs which enables Widefield ASKAP L-band Legacy All-sky Blind surveY \cite[WALLABY;][]{Koribalski2020}. WALLABY has publicly released its phase 1 dataset \citep{Westmeier2022} and already conducted studies of individual galaxy groups \cite[e.g.][]{Lee-Waddell2019,Reynolds2019}. \cite{For2021} also used WALLABY data to study the Eridanus Supergroup, where a supergroup is defined as a group of groups that may eventually merge to form a cluster.

The third of the SKA pathfinder telescopes is MeerKAT \citep{Jonas2016}. MeerKAT has a lower survey speed than Apertif or ASKAP, but has a greater sensitivity, and hence is able to detect lower \HI-mass galaxies than other SKA pathfinders. The Fornax Survey with MeerKAT \citep{Serra2023} is a targeted survey of the Fornax cluster, with \cite{Kleiner2021} presenting results on the Fornax A group. The MeerKAT International GigaHertz Tiered Extragalactic Exploration survey \cite[MIGHTEE;][]{Jarvis2017} survey includes a \HI\ emission component \cite[MIGHTEE-HI;][]{Maddox2021}. \cite{Ranchod2021} reported on the discovery of a galaxy group with 20 members from the MIGHTEE-HI data. In addition, the Looking At the Distant Universe with the MeerKAT Array \citep[LADUMA;][]{Blyth2016} survey is to be the deepest \HI\ emission study by targeting a single field for over 3,000 hours.  

These results from major science surveys are not the only avenue for detecting and studying \HI-rich galaxies. For example, \cite{Healy2021} presented over 200 galaxies detected in a 15 hour observation of a galaxy cluster with MeerKAT Open Time. However, \textit{untargeted} studies of known clusters and galaxy groups through MeerKAT Open Time or ASKAP Guest Science Time can provide yet another avenue for the detection and analysis of \HI-rich galaxies, provided such observations are in spectral-line modes. For example, \HI\ absorption searches towards radio-bright quasars with these telescopes can also contain \HI\ emission associated with other unrelated galaxies within the field of view.

We present a serendipitous discovery of 49 detections of \HI\ emission in a single pointing with MeerKAT. In Section~\ref{sec:obs} we describe the MeerKAT observations and ancillary datasets. In Section~\ref{sec:results} we present the MeerKAT detections and properties of the galaxies, and determine the detections reside in multiple \HI-rich groups. We briefly discuss the serendipity of the detection in Section~\ref{sec:serendipity}, and summarise our conclusions in Section~\ref{sec:conclusions}. Throughout, we adopt optical velocities (c$z$) in the heliocentric reference frame, the AB magnitude convention, and we assume a flat $\Lambda$CDM cosmology with $H_{\rm 0}$~=~67.7~km\,s$^{-1}$\,Mpc$^{-1}$ \citep{Planck2016}.

\section{Observation and data analysis}\label{sec:obs}

\subsection{Observation with MeerKAT} 

Observation 1617470178 of the target J0944-1000 (NVSS\,J084452-100103), at 8:44:52.67, --10:00:59.5, was carried out on 2021 April 3rd with MeerKAT as part of proposal ID SCI-20210212-MG-01 (see Table~\ref{tab:obsdetails}). The observation was taken in L-band (856~MHz bandwidth centred at 1284~MHz) at 32K spectral resolution (26.123~kHz wide channels). The intention of the observation was to confirm a tentative \HI\ absorption feature seen with the Australia Telescope Compact Array (ATCA) near the optical spectroscopic redshift of $z$~=~0.04287 \citep{Glowacki2017}. The SARAO SDP continuum image quality report gave an root mean square (RMS) noise of 41~$\mu$Jy. 

The raw data were transferred to the ilifu supercomputing cloud system and reduced there. Bandpass, flux, and phase calibration, along with self-calibrated continuum imaging, was performed using the {\sc processMeerKAT} pipeline\footnote{\url{https://idia-pipelines.github.io/docs/processMeerKAT}}, which is written in Python, uses a purpose-built CASA \citep{McMullin2007} Singularity container, and employs MPICASA (a parallelized form of CASA). Data was at this stage rebinned by a factor of 4 (i.e. to `8K' mode, 104.49~kHz-wide channels), and the 1304--1420~MHz segment of the L-band was extracted for the results presented in this paper. Model continuum visibility data were subtracted from the corrected visibility data using the CASA task \textit{uvsub}. A second-order polynomial fit to the continuum was then calculated and subtracted using the CASA task \textit{uvcontsub} for all channels  to remove residual continuum emission from the spectral line data. Finally, spectral line cubes were created using \textit{tclean} with robust~=~0.5 and no cleaning. The RMS per 104.49~kHz channel was consistent between 1304--1420~MHz with the per-channel noise of 0.16~mJy\,beam$^{-1}$ at the centre (1362~MHz). All channels were convolved to a common synthesized beam of $14.0^{\prime\prime} \times 9.8^{\prime\prime}$ at a position angle of $-$20.1$^{\circ}$. 

\begin{table}
\small
\caption{Details of the MeerKAT observation for proposal ID SCI-20210212-MG-01.}
\centering
\begin{tabular}{ll}
\hline\hline
SBID & 1617470178\\
Phase centre & 8:44:52.67, --10:00:59.5\\
Bandpass calibrator & J0408-6545\\
Gain calibrator & J0730-1141\\
Channel width & 26.123~kHz (rebinned by factor of 4)\\
Pixel size & 2 arcsec\\
On-source integration time & 8340 sec (2.32 hr)\\
RMS & 0.16~mJy per channel ($\sim$1362~MHz)\\ 
Beam & $14.0^{\prime\prime} \times 9.8^{\prime\prime}$; PA of $-$20.1$^{\circ}$\\
\hline
\end{tabular}
\label{tab:obsdetails}
\end{table}

\subsection{Source finding}

While the putative HI absorption feature was not confirmed, upon initial visual inspection of the spectral line cube with CARTA \citep{Comrie2021}, \HI\ emission was noticed in multiple instances, both spatially and spectrally away from the centre of the cube.  A more careful investigation followed, with each channel inspected for \HI\ emission visually in CARTA. In addition to manual visual source finding, we employed two other methods. The first was a matched-filter approach which was separately developed to find galaxy candidates in \HI\ data cubes. The MeerKAT cube was box-smoothed spatially with a kernel of 20~arcsec by 20~arcsec, and spectrally with a kernel of width equal to two channels (approximately 45~km\,s$^{-1}$).  Thereafter, the average flux within the full 3D smoothing kernel (20~arcsec x 20~arcsec x 45~km\,s$^{-1}$) was calculated at every (x, y, z) location (i.e., voxel) in the smoothed cube. The distribution of the resulting means was well modelled by a Gaussian of mean zero and standard deviation $\sigma$. Any voxels in the smoothed cube that had a filter-averaged flux greater than 6$\sigma$ were taken to contain potential galaxy emission. 

\begin{landscape}
\begin{figure}
\includegraphics[width=1.14\textwidth]{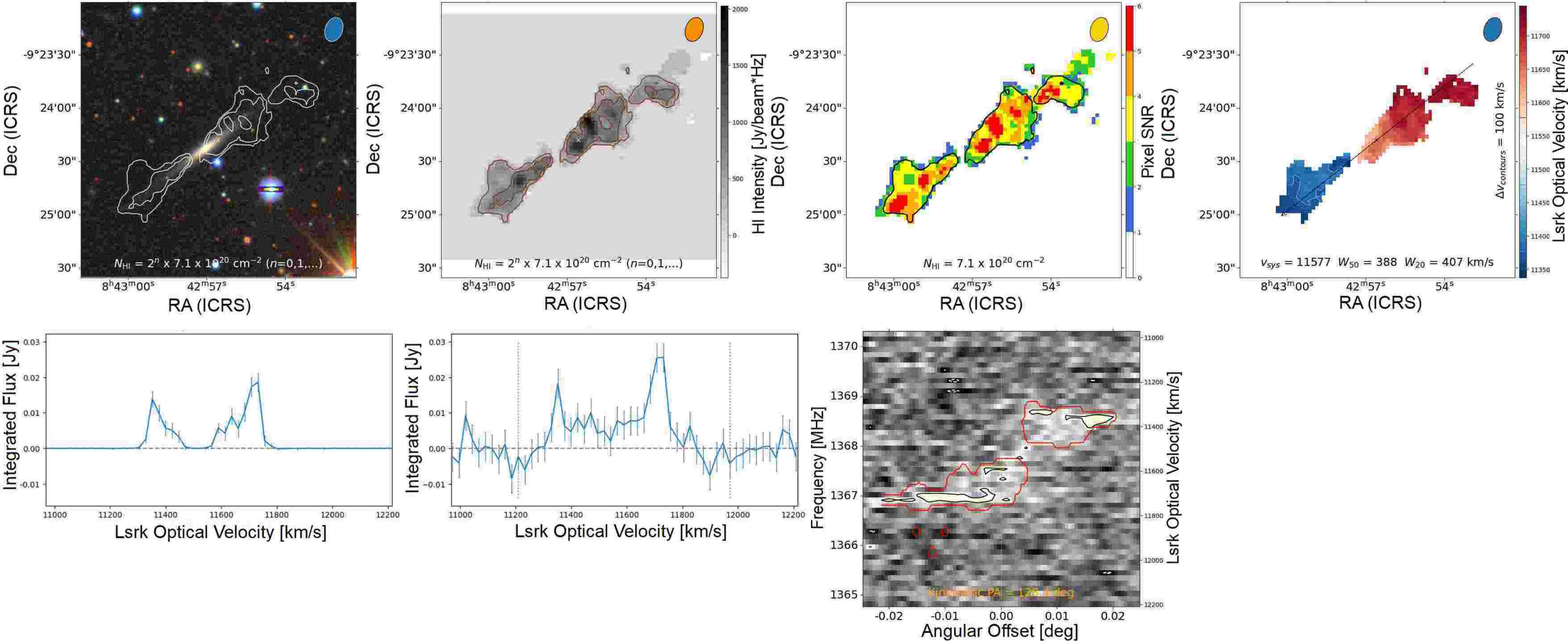}
\includegraphics[width=1.14\textwidth]{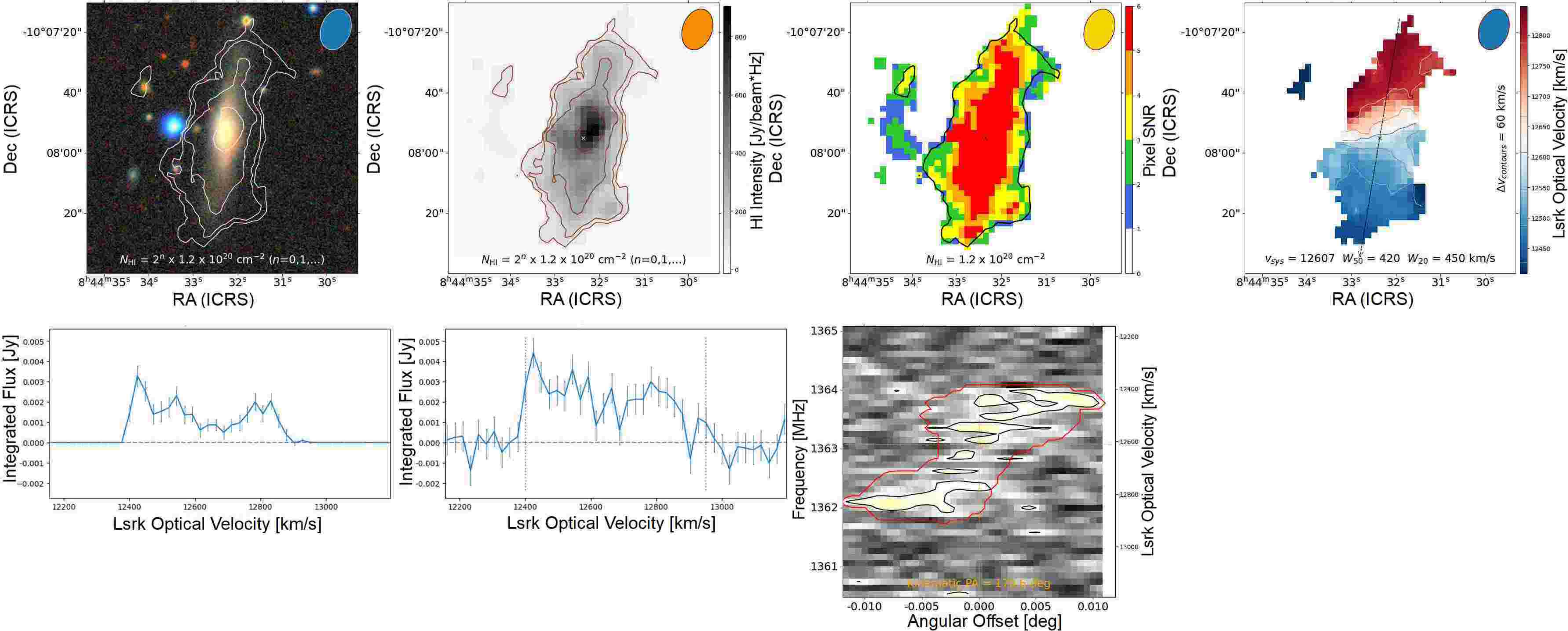}
\caption{SIP outputs for IDs 23 and 34. Left to right, top to bottom, is the \HI\ moment 0 map overlaid on a DECaLS DR10 image, the moment 0 map isolated, the pixel SNR map, the moment 1 (velocity) map, the SoFiA masked and unmasked spectrum, and the p-v slice diagram. SIP outputs for the remaining galaxies are given in the Appendix 
available online. Contour levels are noted on the individual subplots.}
\label{fig:combo}
\end{figure}
\end{landscape}


In general, we found the matched-filter source-finding approach to reliably extract all emission of real galaxies, often including extended features such as \HI\ tails.  Additionally, the Source Finding Application 2 \cite[SoFiA 2;][]{Serra2015,Westmeier2021} was employed to verify detections and create individual subcubes. 

The results of the three methods were then combined and individually verified by eye. As part of verification, we considered data outputs from the SoFiA Image Pipeline \cite[SIP;][]{Hess2022}, where optical imaging from either the Dark Energy Camera Legacy Survey \cite[DECaLS;][]{Dey2019} DR10 release, or the PanSTARRS \citep{Chambers2016} survey was used for overlaying \HI\ contours, using default settings (where the lowest contour is at minimum 2$\sigma$ significance). Two examples of SIP outputs are given in Figure~\ref{fig:combo}. Tentative detections were removed e.g. single-channel detections or ones difficult to distinguish from noise, and additionally without an optical counterpart. SoFiA identified a couple additional \HI\ sources not originally identified by the other methods, although it had initially failed to also detect a couple other sources identified through visual source finding and matched filter approaches when run on the entire cube.

\HI\ masses are calculated from SoFiA data products using equation 48 from \cite{Meyer2017}.

\subsection{Ancillary datasets}

To complement our \HI\ datasets, we looked at available ancillary data for stellar mass and star formation rate (SFR) estimates. We used a mixture of optical, mid-infrared, and near ultraviolet (NUV) datasets. In each case we follow the methodology described by \cite{Reynolds2022}, where we adapted scripts provided on \url{https://github.com/tflowers15/wallaby-analysis-scripts}. We summarise the steps involved below. 

\subsubsection{Optical}\label{sec:optical}

For each \HI\ detection we derive stellar masses from the PanSTARRS $g-$ and $r-$band images. In summary, we obtain image cutouts at the position of each \HI\ detection through the PanSTARRS cutout server \footnote{\url{https://ps1images.stsci.edu/cgi-bin/ps1cutouts}}. Photometry was derived through the {\sc Python} package {\sc Photutils} on the $r$-band image, segmentation maps via {\sc Segmentation} for masking other sources, and isophotes fitted via {\sc Isophote}. The empirical relation from \cite{Taylor2011} was then used for stellar mass calculation:

\begin{equation}
\begin{split}
    {\rm log(}M_{*}{\rm /M_{\odot)}} & = - 0.840 + 1.654(g - r) + 0.5(D_{\rm mod} + M_{\rm sol} - m) \\ 
    & - {\rm log(1 +}z) - 2{\rm log(}h/0.7),
\end{split}
\label{eqn:stellarmass}
\end{equation}

where 0.840 and 1.654 are empirically determined constants \citep{Zibetti2009}, the $g - r$ colour is in the SDSS photometric system, $m$ is the $r$-band apparent magnitude in the SDSS photometric system, $D_{\rm mod}$ is the distance modulus (used to convert from apparent to absolute magnitude), $h$ is the Hubble Constant, and $M_{\rm sol}$ = 4.64 is the absolute magnitude of the Sun in the $r$-band \citep{Willmer2018}. Calculated stellar masses have uncertainties of $\sim$0.16 dex as by \cite{Reynolds2022}. 

\subsubsection{NUV and mid-infrared}

We estimate the star formation rate from two surveys. The first is the Galaxy Evolution Explorer \cite[\textit{GALEX}][]{Martin2005}, where the NUV luminosity traces emission from young stars \citep{Kennicutt1998, Kennicutt2012}. Next is the Wide-field Infrared Survey Explorer \cite[WISE;][]{Wright2010}, where the fractions of polycyclic aromatic hydrocarbons (PAH) traced by the W3 (12~$\mu$m) band are high in regions of active star formation, thought to be produced in molecular clouds and growing on dust grains. Alternatively the warm dust continuum at the W4 (22~$\mu$m) band, sensitive to reprocessed radiation from star formation, is another avenue for SFR measurements. 

As with PanSTARRS, the GALEX and WISE images are masked as described by \cite{Reynolds2022}. Image units for photometry are converted to magnitudes, with both NUV and WISE magnitudes with SNR $<$ 5 considered to be upper limits. After converting to luminosity $L_{\rm NUV}$, the SFR from the GALEX measurements is defined from \cite{Schiminovich2007} as

\begin{equation}
    {\rm SFR_{NUV}/M_{\odot}yr^{-1} = 10^{-28.165}} L_{\rm NUV}/{\rm erg s^{-1} Hz^{-1}},
\end{equation}

For the mid-infrared SFR, we consider W4, or W3 if the less sensitive W4 band is not available.  After converting to the corresponding luminosity ($L_{\rm W4}$ or $L_{\rm W3}$), the SFR is calculated as by \cite{Jarrett2013} via:

\begin{equation}
    {\rm SFR_{W3}/M_{\odot}yr^{-1} = 4.91 \times 10^{-10}}(L_{\rm W3} - 0.201L_{\rm W1}/{\rm L_{\odot}}),
\end{equation}

and

\begin{equation}
    {\rm SFR_{W4}/M_{\odot}yr^{-1} = 7.50 \times 10^{-10}}(L_{\rm W4} - 0.044L_{\rm W1}/{\rm L_{\odot}}),
\end{equation}

where a subtraction happens due to contamination from old stellar populations done via subtracting a fraction of the W1 (3.4~$\mu$m) luminosity $L_{\rm W1}$.

The total SFR is then the sum of SFR$_{\rm NUV}$ and SFR$_{\rm W4(3)}$, as done by \cite{Reynolds2022}. The derived SFRs have uncertainties of $\leq$0.1~dex.

\begin{figure*}
\centering
\includegraphics[width=0.99\textwidth]{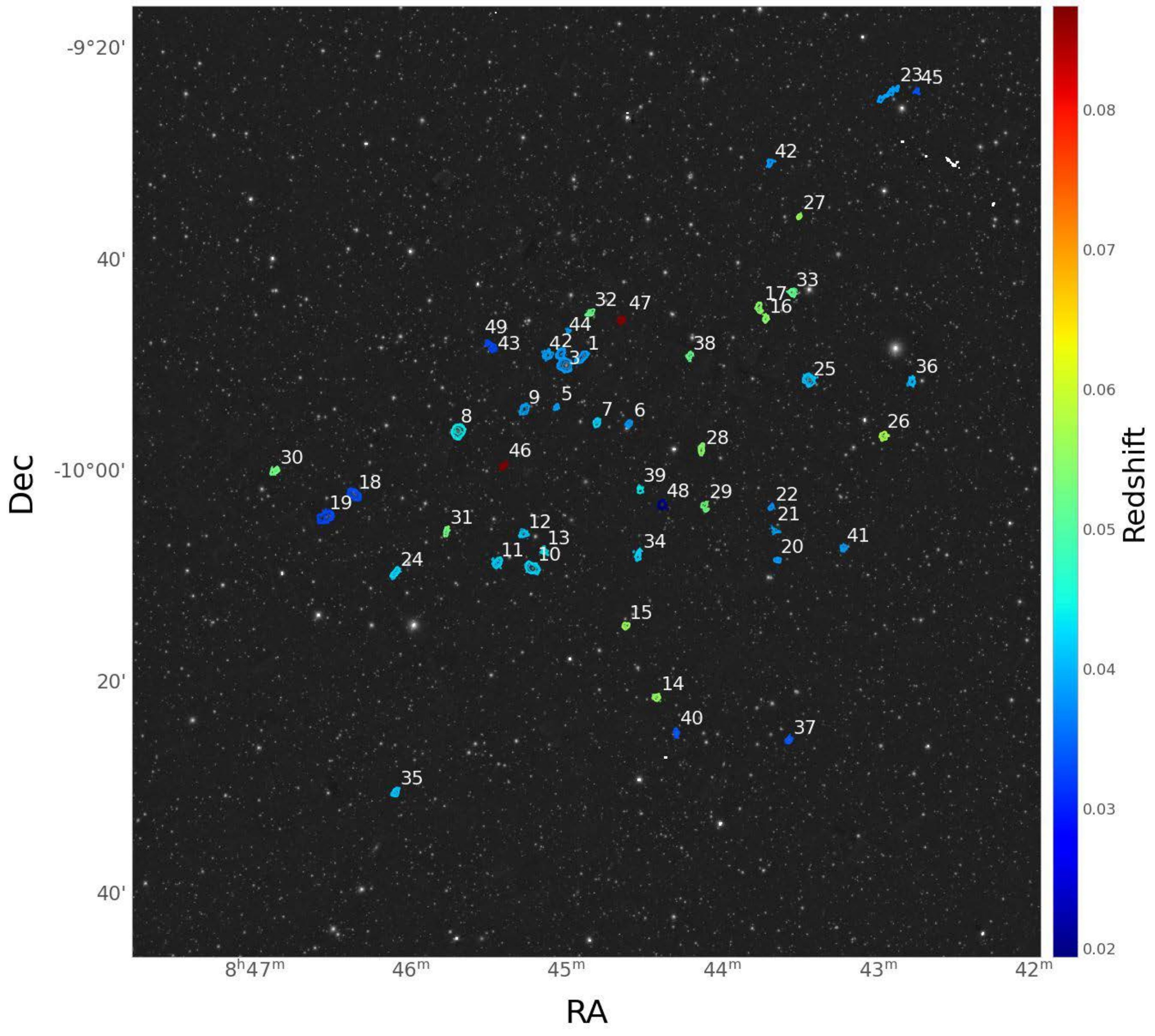}
\caption{The 49 \HI-rich galaxies detected by MeerKAT in a single 2.3~hr pointing. Their intensity (moment 0) maps are overlaid as contours coloured by \HI\ redshift on a PanSTARRS $r$-band image. Numbers correspond to the galaxy's ID number, as defined in Table~\ref{tab:properties}. Contour levels are of column density multiples of (2, 4, 8, 16, 32)~$\times$~10$^{20}$\,cm$^{-2}$.}
\label{fig:overlayall}
\end{figure*}

\section{Results}\label{sec:results}

In total, we obtain 49 detections of 21-cm emission within the MeerKAT data. Henceforth we refer to individual galaxies by their assigned ID number. We present the combined outputs for SIP for two galaxies, IDs 23 and 34, in Fig.~\ref{fig:combo}. In the Appendix 
(available online) we give the remaining SIP images (Figure~A1), where the optical image is either a three-colour image from DECaLS~DR10 or PanSTARRS, and notes on individual sources. A combined intensity (moment 0) map for the whole field of view overlaid on a PanSTARRS $r$-band image is presented in Fig.~\ref{fig:overlayall}, where contours are coloured by redshift assuming a \HI\ emission line (rather than e.g. hydroxyl; see Section~\ref{sec:ohm}). 

In Table~\ref{tab:properties} we present the \HI\ redshift and masses, stellar masses, SFRs from the WISE mid-infrared and GALEX photometries where available, and \HI\ gas fractions and specific SFR (sSFR) for each host detected in MeerKAT. Three galaxies (ID3, ID20, ID25) had optical spectroscopic redshifts available. The \HI\ redshift, as determined using SoFiA, agrees within 3$\sigma$ with the optical redshift and associated uncertainty (given as 0.00015). For a further 7 galaxies photometric redshifts estimates were available using WISE magnitudes from the literature. All 7 photometric redshifts from the literature here agree with \HI\ emission rather than 
other radio spectral transitions (e.g. the hydroxyl 1665-1667 doublet), with slightly higher (by $<$0.025) photometric redshifts seen for ID8, ID19, ID23 and ID24, and a lower photometric redshift for ID30. However, for the majority of sources, no redshift information is available in the literature. In Table~\ref{tab:radiocont} we give the measured total radio continuum flux density for our sources from the MeerKAT observation, generated with the {\sc processMeerKAT} pipeline at 1362~MHz with a 9.1 x 6.8" radio beam. We detect radio continuum for 15 of our emission-line detections (13 unresolved). The remaining emission sources are not detected at a sensitivity of 0.02~mJy.

No \HI\ data for these detections had been found in the literature, with sources typically at too high redshift to be within the HIPASS survey footprint \citep{Barnes2001}, or otherwise had too low a \HI\ mass to be detected (HIPASS' 3$\sigma$ $M_{\rm HI}$ sensitivity is stated to be 10$^{6}$\,$d_{\rm Mpc}^{2}$\,M$_{\odot}$; table 1 of \cite{Barnes2001}). Therefore, all \HI\ detections presented here are new. 

\begin{table*}[t]
\centering
\caption{Properties of the 49 \HI\ galaxies detected in the 2.3\,hr pointing centered on J0844--1000. We give the ID of the galaxy, 
the right ascension and declination based on the \HI\ emission as measured by SoFiA, the redshift (assuming a \HI\ emission line), the optical spectroscopic or photometric redshift where available in the literature, \HI\ mass, stellar mass, \HI\ gas fraction, and where available the star formation rates (SFR) based on either the WISE W4 or W3 mid-infrared magnitude, SFR from GALEX (near-UV data), the total SFR (mid-infrared and NUV values added, using W4 if it and W3 are both available), and the specific SFR (sSFR). }
\begin{tabular}{rllcccccccccc}
 \hline
 \hline
ID & RA & Dec & $z_{\rm HI}$ & $z_{\rm lit}$ & log$_{\rm 10}$($M_{\rm HI}$) & log$_{\rm 10}$($M_{*}$) & log$_{\rm 10}$($f_{\rm HI}$) & SFR$_{\rm W3}$ & SFR$_{\rm W4}$ & SFR$_{\rm NUV}$ & SFR$_{\rm tot}$ & log$_{\rm 10}$(sSFR) \\
 & & & & & M$_{\odot}$ & M$_{\odot}$ & & M$_{\odot}$\,yr$^{-1}$ & M$_{\odot}$\,yr$^{-1}$ & M$_{\odot}$\,yr$^{-1}$ & M$_{\odot}$\,yr$^{-1}$ & yr$^{-1}$\\
\hline
       1 & 08:44:53.72 & -09:49:10.16 &  0.0387 &                &          9.96 &            8.88 &           1.08 &             0.01 &             0.08 &    0.21 &              0.28 &  -9.43 \\
       2 & 08:45:02.01 & -09:48:54.00 &  0.0383 &                &         10.12 &            9.49 &           0.63 &             0.11 &             0.30 &    0.16 &              0.46 &  -9.83 \\
       3 & 08:45:00.28 & -09:49:56.91 &  0.0385 &       0.038557$^{a}$ &         10.64 &           10.46 &           0.18 &             3.24 &             4.79 &    2.32 &              7.11 &  -9.61 \\
       4 & 08:45:07.14 & -09:48:58.61 &  0.0389 &                &          9.69 &            8.10 &           1.59 &             0.00 &                  &    0.11 &            0.11       &    -9.06    \\
       5 & 08:45:03.82 & -09:53:55.94 &  0.0392 &                &          9.25 &            8.10 &           1.14 &                  &                  &         &                   &        \\
       6 & 08:44:35.89 & -09:55:30.16 &  0.0387 &                &          9.36 &            7.32 &           2.03 &                  &             0.00 &         &              0.00 &  -9.72 \\
       7 & 08:44:48.32 & -09:55:25.42 &  0.0424 &                &          9.76 &            9.52 &           0.25 &             0.10 &             0.39 &    0.19 &              0.58 &  -9.76 \\
       8 & 08:45:41.55 & -09:56:12.56 &  0.0439 & $\sim$0.054$^{b}$ &         10.40 &           10.44 &          -0.04 &             0.38 &             0.72 &    0.72 &              1.44 & -10.28 \\
       9 & 08:45:16.10 & -09:54:08.20 &  0.0391 &                &          9.77 &            8.98 &           0.79 &                  &                  &         &                   &        \\
      10 & 08:45:13.10 & -10:09:09.35 &  0.0423 &                &         10.37 &            9.92 &           0.45 &             0.01 &             0.09 &    0.31 &              0.40 & -10.32 \\
      11 & 08:45:26.42 & -10:08:38.50 &  0.0422 &                &          9.97 &            9.20 &           0.77 &             0.03 &             0.15 &    0.35 &              0.50 &  -9.50 \\
      12 & 08:45:16.36 & -10:05:55.03 &  0.0416 &   &          9.67 &            9.81 &          -0.14 &             0.14 &             0.28 &    0.46 &              0.74 &  -9.95 \\ 
      13 & 08:45:08.56 & -10:07:32.75 &  0.0430 &                &          9.42 &            9.52 &          -0.11 &             0.03 &                  &         &            0.03       &   -11.04     \\
      14 & 08:44:25.25 & -10:21:24.08 &  0.0557 &                &          9.97 &            8.99 &           0.98 &                  &                  &         &                   &        \\
      15 & 08:44:37.08 & -10:14:38.52 &  0.0558 &                &          9.87 &            9.10 &           0.77 &             0.01 &             0.15 &         &              0.15 &  -9.94 \\
      16 & 08:43:43.43 & -09:45:33.57 &  0.0548 &                &          9.92 &            9.34 &           0.58 &             0.03 &             0.05 &    0.49 &              0.54 &  -9.61 \\
      17 & 08:43:45.98 & -09:44:29.32 &  0.0547 &                &          9.95 &            9.21 &           0.74 &                  &             0.21 &    0.11 &              0.32 &  -9.70 \\
      18 & 08:46:21.42 & -10:02:12.45 &  0.0331 &                &         10.22 &            9.97 &           0.26 &             0.29 &             0.36 &    0.23 &              0.60 & -10.19 \\
      19 & 08:46:32.23 & -10:04:18.73 &  0.0329 & $\sim$0.055$^{b}$ &         10.36 &           10.02 &           0.33 &             0.03 &             0.42 &    0.17 &              0.59 & -10.25 \\
      20 & 08:43:38.83 & -10:08:23.96 &  0.0385 &        0.03886$^{a}$ &          9.48 &           10.63 &          -1.15 &             0.08 &                  &         &               0.08    &   -11.73     \\
      21 & 08:43:39.64 & -10:05:39.58 &  0.0390 &                &          9.10 &            7.59 &           1.51 &             0.01 &                  &         &              0.01     &     -9.59   \\
      22 & 08:43:41.26 & -10:03:23.66 &  0.0378 &                &          9.26 &            8.83 &           0.43 &                  &             0.04 &         &              0.04 & -10.19 \\
      23 & 08:42:56.57 & -09:24:17.91 &  0.0386 &        $\sim$0.048$^{b}$ &         10.27 &           10.33 &          -0.06 &             0.07 &             0.06 &         &              0.06 & -11.57 \\
      24 & 08:46:05.74 & -10:09:35.30 &  0.0423 & $\sim$0.058$^{b}$ &         10.02 &           10.17 &          -0.15 &             0.40 &             1.12 &    0.18 &              1.30 & -10.06 \\
      25 & 08:43:26.70 & -09:51:22.16 &  0.0412 &       0.040802$^{a}$ &         10.21 &           10.19 &           0.02 &             0.59 &             1.35 &         &              1.35 & -10.06 \\
      26 & 08:42:58.09 & -09:56:37.80 &  0.0572 &                &         10.01 &            8.94 &           1.07 &             0.00 &                  &         &                   &        \\
      27 & 08:43:30.57 & -09:35:52.50 &  0.0562 &                &          9.45 &            8.64 &           0.81 &             0.01 &                  &         &           0.01        &    -10.64    \\
      28 & 08:44:08.10 & -09:57:56.62 &  0.0546 &                &         10.08 &            9.46 &           0.62 &             0.04 &             0.29 &         &              0.29 &  -9.99 \\
      29 & 08:44:06.75 & -10:03:19.94 &  0.0528 &                &          9.70 &            8.26 &           1.43 &             0.00 &                  &         &                   &        \\
      30 & 08:46:51.85 & -09:59:56.26 &  0.0525 &  $\sim$0.041$^{b}$ &         10.17 &           10.62 &          -0.44 &             0.80 &             0.60 &    1.35 &              1.95 & -10.33 \\
      31 & 08:45:46.14 & -10:05:40.65 &  0.0529 &                &          9.76 &            8.95 &           0.81 &             0.14 &             0.16 &    0.17 &              0.33 &  -9.43 \\
      32 & 08:44:50.83 & -09:45:00.91 &  0.0519 &                &          9.79 &            9.62 &           0.17 &             0.06 &             0.27 &    0.16 &              0.43 &  -9.99 \\
      33 & 08:43:33.29 & -09:43:02.85 &  0.0518 &                &          9.97 &            9.57 &           0.40 &             0.20 &             0.30 &    1.04 &              1.34 &  -9.44 \\
      34 & 08:44:32.36 & -10:07:55.08 &  0.0421 &  $\sim$0.042$^{b}$ &          9.76 &           10.80 &          -1.04 &             0.39 &             0.87 &         &              0.87 & -10.86 \\
      35 & 08:46:05.67 & -10:30:20.14 &  0.0412 &                &          9.99 &            9.43 &           0.56 &             0.04 &             0.10 &    0.45 &              0.55 &  -9.70 \\
      36 & 08:42:47.31 & -09:51:26.29 &  0.0406 &                &          9.71 &            9.19 &           0.52 &             0.00 &             0.17 &         &              0.17 &  -9.97 \\
      37 & 08:43:34.37 & -10:25:25.55 &  0.0337 & $\sim$0.032$^{b}$ &          9.53 &            9.98 &          -0.45 &             0.70 &             0.98 &         &              0.98 &  -9.98 \\
      38 & 08:44:12.61 & -09:49:03.91 &  0.0520 &                &          9.66 &            8.53 &           1.13 &             0.00 &                  &    0.11 &          0.11         &   -9.49     \\
      39 & 08:44:31.77 & -10:01:44.75 &  0.0437 &                &          9.47 &            8.95 &           0.52 &                  &             0.05 &         &              0.05 & -10.28 \\
      40 & 08:44:17.64 & -10:24:45.39 &  0.0338 &                &          9.33 &            8.71 &           0.62 &                  &             0.05 &         &              0.05 & -10.05 \\
      41 & 08:43:13.45 & -10:07:15.54 &  0.0386 &                &          9.37 &            9.60 &          -0.23 &                  &             0.03 &         &              0.03 & -11.10 \\
      42 & 08:43:41.75 & -09:30:47.30 &  0.0384 &                &          9.46 &            8.41 &           1.05 &                  &                  &    0.05 &             0.05      &     -9.71   \\
      43 & 08:45:28.11 & -09:48:20.58 &  0.0329 &                &          9.29 &            7.88 &           1.41 &                  &                  &    0.05 &              0.05     &    -9.18    \\
      44 & 08:44:59.09 & -09:46:43.20 &  0.0383 &                &          8.99 &            9.77 &          -0.77 &             0.56 &             1.93 &         &              1.93 &  -9.48 \\
      45 & 08:42:45.50 & -09:23:59.07 &  0.0335 &                &          9.71 &            9.48 &           0.23 &             0.02 &             0.10 &         &              0.10 & -10.47 \\
      46 & 08:45:24.13 & -09:59:29.32 &  0.0874 &                &         10.21 &            9.89 &           0.32 &             0.32 &             0.15 &         &              0.15 & -10.71 \\
      47 & 08:44:38.71 & -09:45:40.14 &  0.0867 &                &          9.97 &            9.07 &           0.90 &             0.05 &                  &         &         0.05          &    -10.37    \\
      48 & 08:44:23.13 & -10:03:10.17 &  0.0194 &                &          9.18 &            8.73 &           0.45 &             0.00 &                  &         &                   &        \\
      49 & 08:45:29.87 & -09:47:54.63 &  0.0323 &                &          9.04 &            7.36 &           1.68 &                  &                  &         &                   &        \\
\hline
\hline
\label{tab:properties}
\end{tabular}
{\raggedright References: ${a}$: \cite{Jones2009}, $b$: \cite{Bilicki2014} \par}
\end{table*}

\begin{table}
\centering
\caption{Radio continuum flux densities for our \HI\ detections. All other detections have $S_{\rm 1.362 GHz}$ upper limits of 0.02~mJy.}
\begin{tabular}{lcc}
 \hline
ID & $S_{\rm 1362~MHz}$& Unresolved?\\
 & mJy & (9.1~$\times$~6.8" beam)\\
\hline

2 & 0.5 & Yes\\
3 & 7.3 & Yes\\
8 & 0.4 & Yes\\
10 & 0.1 & Yes\\
12 & 0.1 & Yes\\
18 & 0.9 & No\\
24 & 0.5 & Yes\\
25 & 0.3 & Yes\\
30 & 0.7 & No\\
33 & 0.5 & Yes\\
34 & 1.1 & Yes\\
37 & 1.5 & Yes\\
44 & 1.2 & Yes\\
45 & 0.4 & Yes\\
46 & 0.1 & Yes\\

\hline
\label{tab:radiocont}
\end{tabular}
\end{table}

\begin{figure*}
\centering
\includegraphics[width=0.8\linewidth]{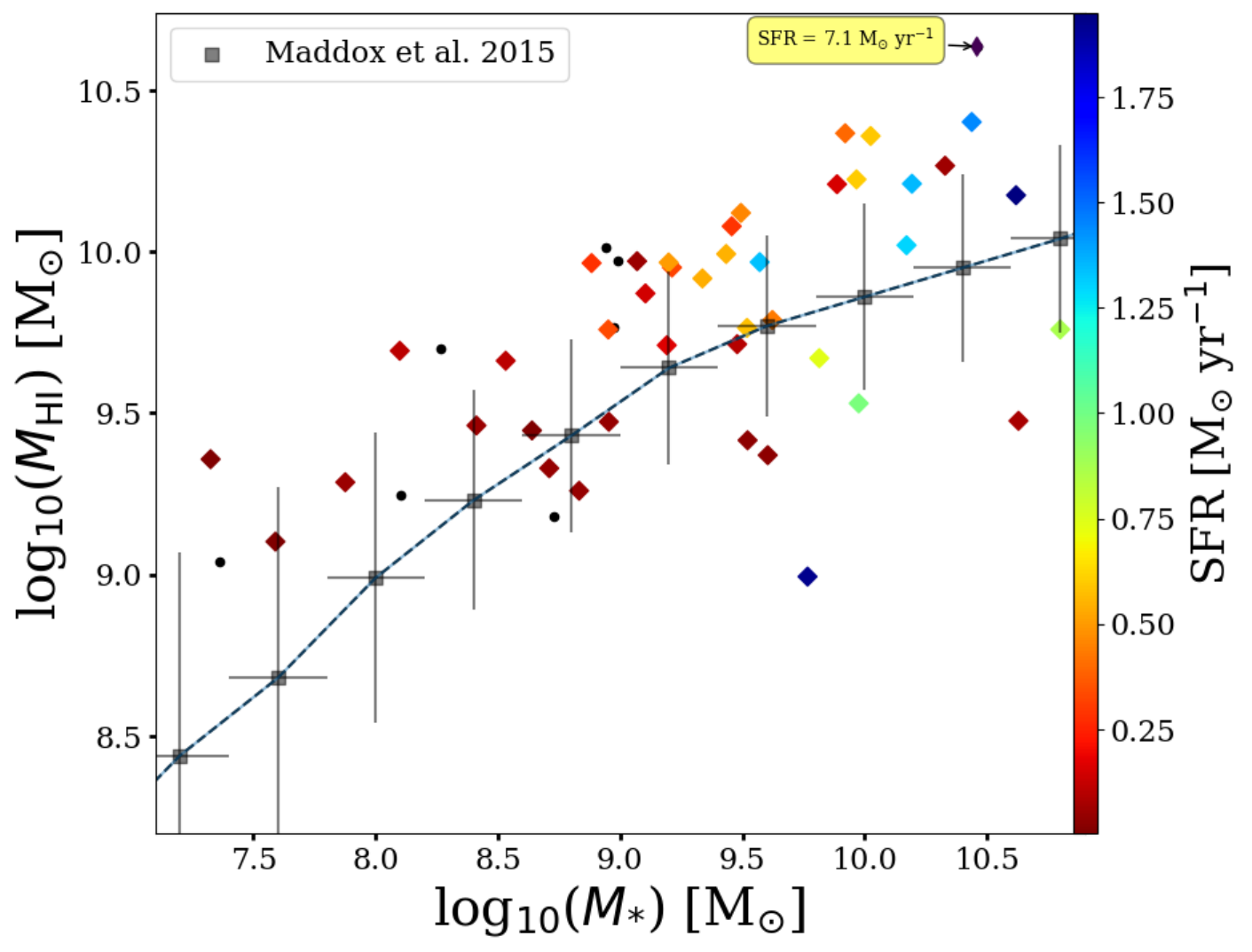}
\caption{The \HI\ mass versus the stellar mass for the 49 galaxies detected in this study. Where available, the points have been coloured by the combined SFR calculated from the mid-infrared and near-UV data, or as black circles otherwise. We note that galaxy ID3 has an enhanced SFR of more than 7~M$_{\odot}$~yr$^{-1}$. The $M_{\rm HI}$-$M_{*}$ relation from \citet{Maddox2015} for ALFALFA galaxies is overlaid for comparison in grey.}
\label{fig:MHIvMS}
\end{figure*}

\begin{figure}
\centering
\includegraphics[width=0.99\linewidth]{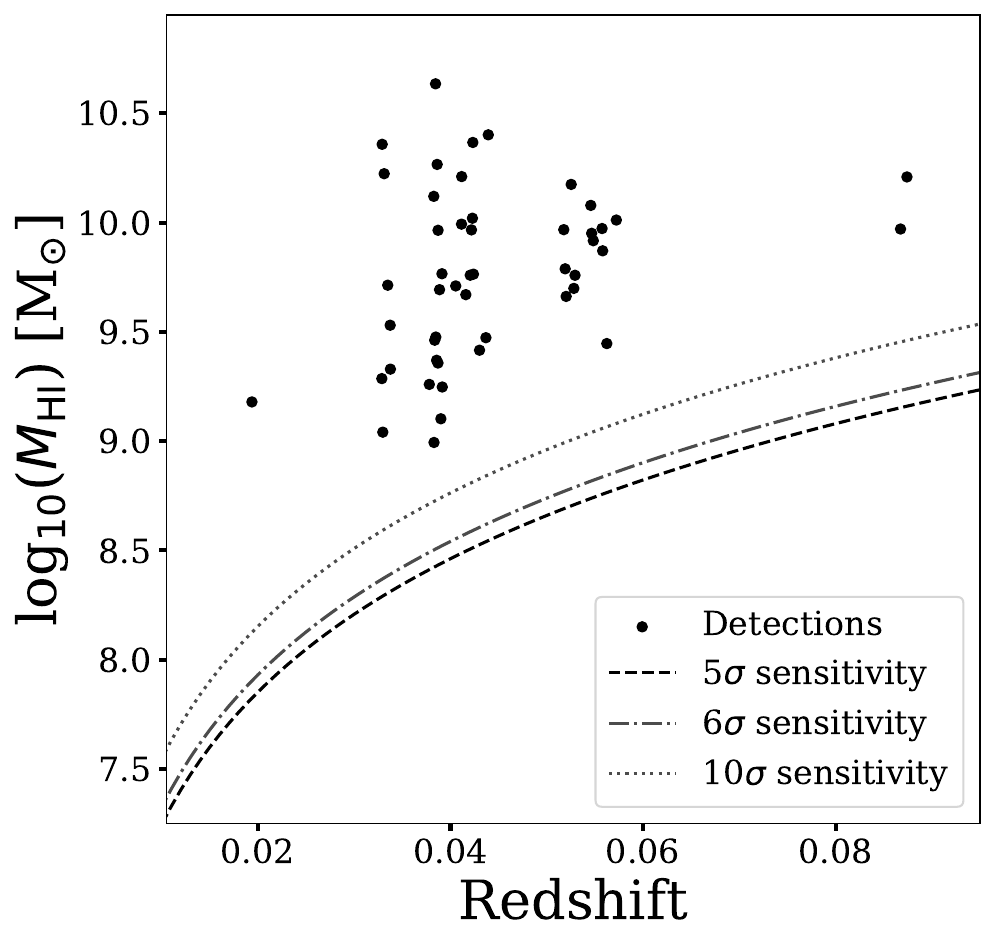}
\caption{The $M_{\rm HI}$ masses versus redshift for our sample. The dashed line gives a 5$\sigma$ sensitivity, dot-dashed line at 6$\sigma$, and dotted line at 10$\sigma$, assuming an unresolved galaxy with velocity width of 200~km\,s$^{-1}$, using equation 157 of \citet{Meyer2017}.}
\label{fig:MHIvsz}
\end{figure}

\begin{figure*}
\centering
\includegraphics[width=0.85\linewidth]{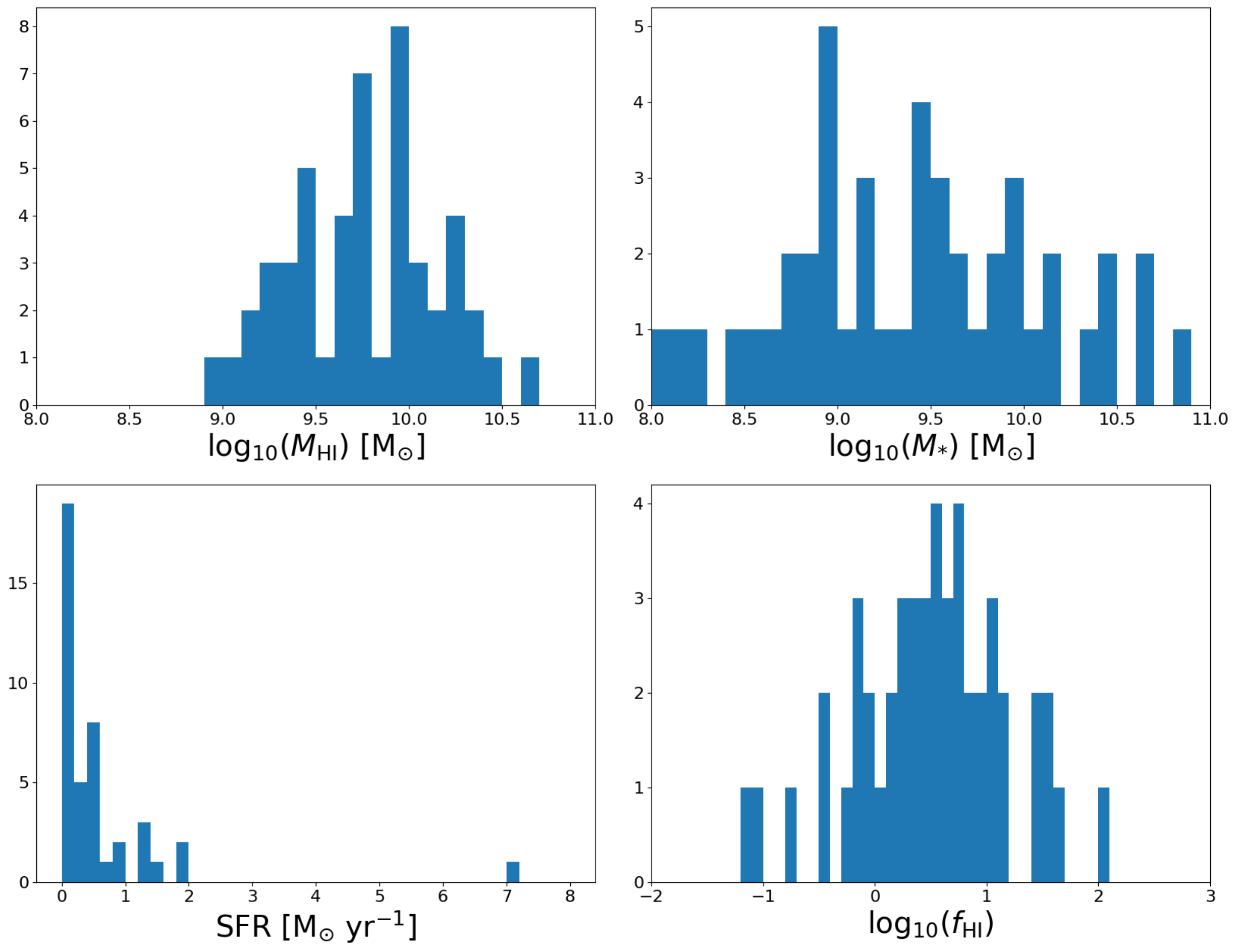}
\caption{Histograms of the \HI\ mass (top left panel), stellar mass (top right), combined SFR (bottom left), and \HI\ gas fraction (lower right) for the 49 galaxies detected in \HI\ in the 2.3~hr MeerKAT pointing.}
\label{fig:4panel}
\end{figure*}

\begin{figure}
\centering
\includegraphics[width=0.99\linewidth]{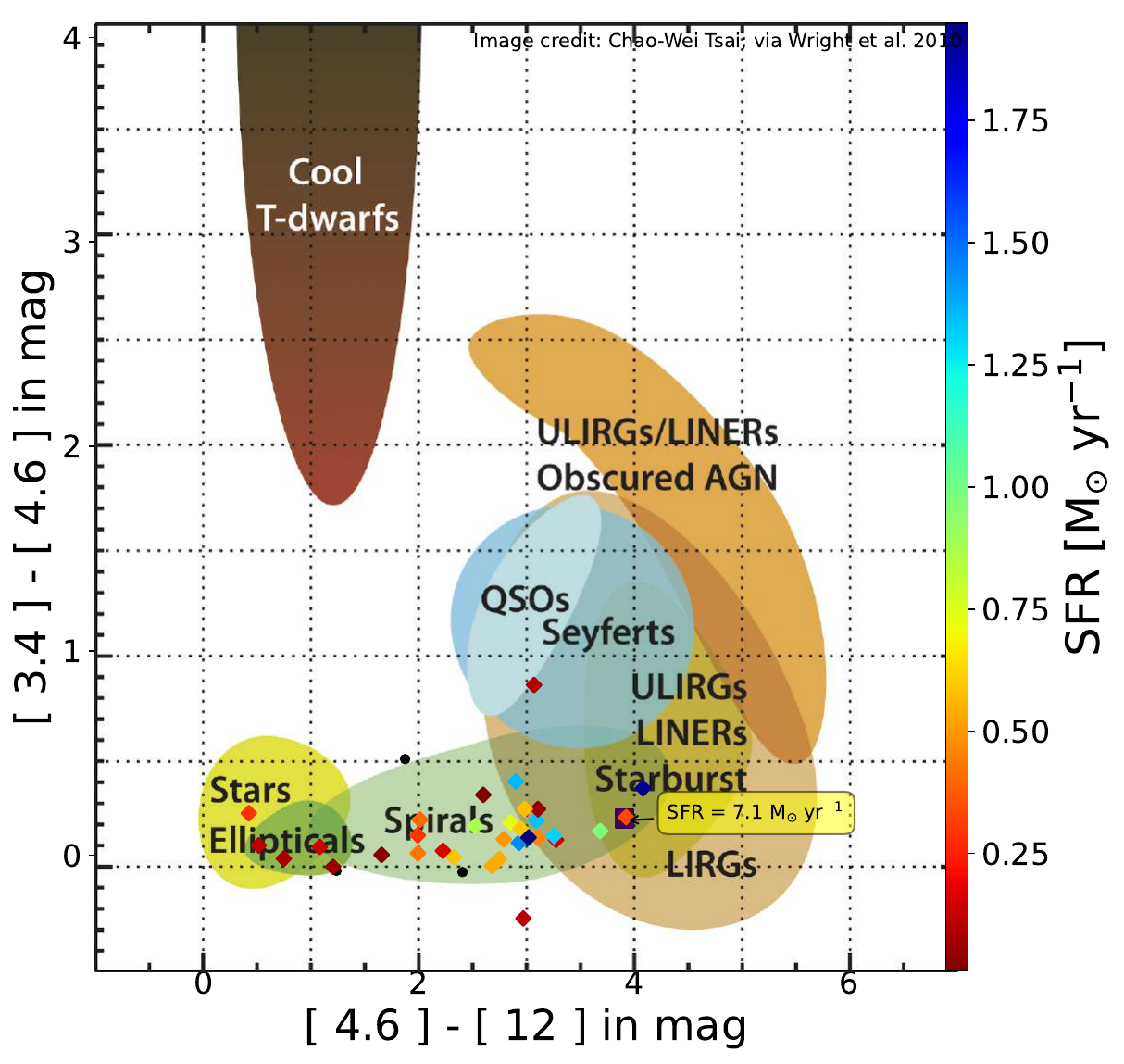}
\caption{WISE colour-colour properties of the galaxies in our sample which were detected in WISE bands W1, W2 and W3. Where available, points are coloured by the combined SFR, or given as black circles otherwise. Points are overlaid on the scheme given in fig.~10 of \citet{Wright2010}. Image generated via code by Chao-Wei Tsai.}
\label{fig:WISEcc}
\end{figure}

\subsection{Properties of the sample}

In Figure~\ref{fig:MHIvMS} we give the $M_{\rm HI}$ vs $M_{*}$ relation for the sample, with points coloured by their SFR (where available). Broadly speaking the expected correlation between the two can be seen; we compare with the relation presented by \cite{Maddox2015} in grey for the Arecibo Legacy Fast ALFA Survey\cite[ALFALFA;][]{Haynes2018}. 
There are only two significant outliers ($\sim$1 dex). One is ID43 which has the lowest \HI\ mass of our sample (just below 10$^{9}$\,M$_{\odot}$) and a higher corresponding stellar mass, and is separated below the rest of the sample in this relation by up to $\sim$1 dex (see discussion in Section~\ref{sec:ohm}). The other outlier is ID3, with a higher $M_{\rm HI}$ than other galaxies with similar stellar masses, which we discuss below and in Section~\ref{sec:interacting}. Also as expected, the higher mass galaxies, particularly in stellar mass, tend to have higher SFRs. 

In Fig.~\ref{fig:MHIvsz} we give the \HI\ mass versus redshift. Included are the  5, 6, and 10$\sigma$ sensitivity limits for our survey, using equation 157 of \cite{Meyer2017} and assuming unresolved \HI\ sources with a velocity width of 200~km\,s$^{-1}$ for 104.49 kHz-wide channels as presented here. The gap between detections and the line indicates we did not have fully searched to 5$\sigma$ (e.g. the SoFiA search had a threshold of 10 sigma to limit false detections, while the matched-filter approached used a minimum of 6$\sigma$), and/or that we missed detections, or that these sensitivity curves are too optimistic. Figure~\ref{fig:4panel} gives the distribution for \HI\ mass, stellar mass, combined SFR, and \HI\ gas fraction. 
In regards to the lower-left panel for the SFR distribution for the 42 available measurements, one clear outlier can be seen, corresponding to ID3 with a SFR greater than 7~M$_{\odot}$\,yr$^{-1}$ (the next highest being ID30 with SFR = 1.95~M$_{\odot}$\,yr$^{-1}$). We discuss this case further in Section~\ref{sec:interacting}. 

\subsubsection{Masquerading megamasers?}\label{sec:ohm}

\begin{figure}
\centering
\includegraphics[width=0.99\linewidth]{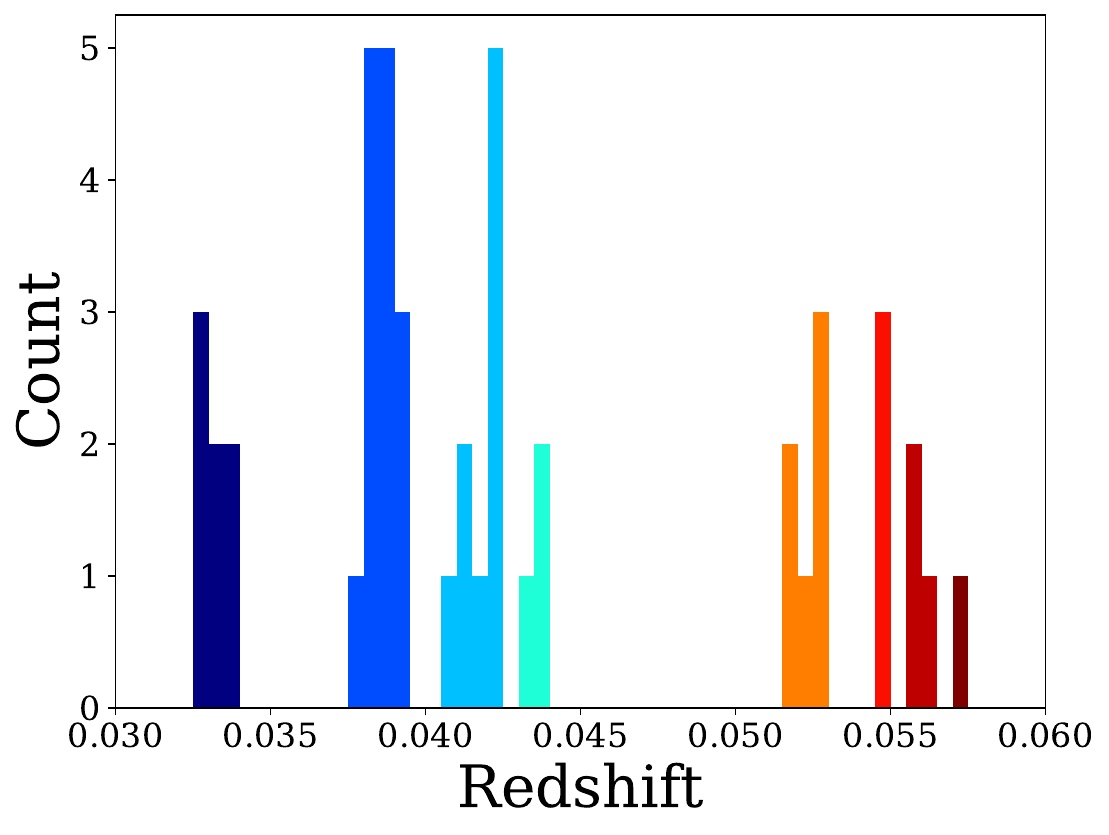}
\caption{Histograms of redshifts for 46 of the 49 galaxies in our detected sample, excluding the lowest-redshift and two highest-redshift galaxies (IDs 46--48). We note that the colouring of the bins used does not directly match that used in Fig.~\ref{fig:overlayall}.}
\label{fig:zhist}
\end{figure}

\begin{figure*}
\centering
\includegraphics[width=0.9\linewidth]{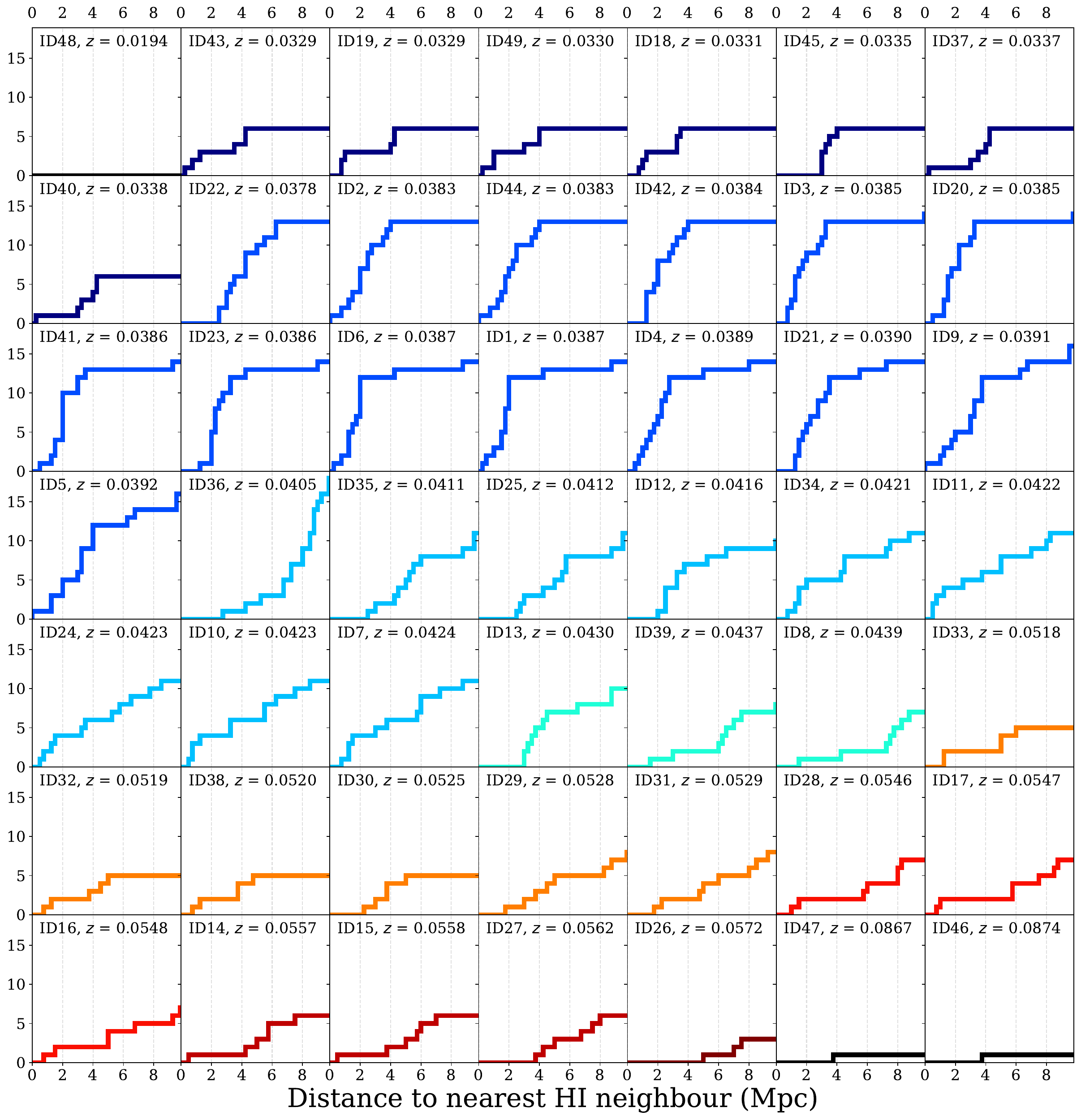}
\caption{Cumulative histograms as a function of the distance to neighbouring \HI\ galaxies, for all 49 galaxies detected by MeerKAT, ordered in increasing redshift left to right, top to bottom. The colour scheme matches that used in Fig.~\ref{fig:zhist}. Several galaxy groups can be identified - e.g. all seven galaxies at $0.030 < z < 0.035$ (dark blue) are within 5~Mpc of each other.}
\label{fig:nearneighbour}
\end{figure*}

\begin{figure*}
\centering
\includegraphics[width=0.90\textwidth]{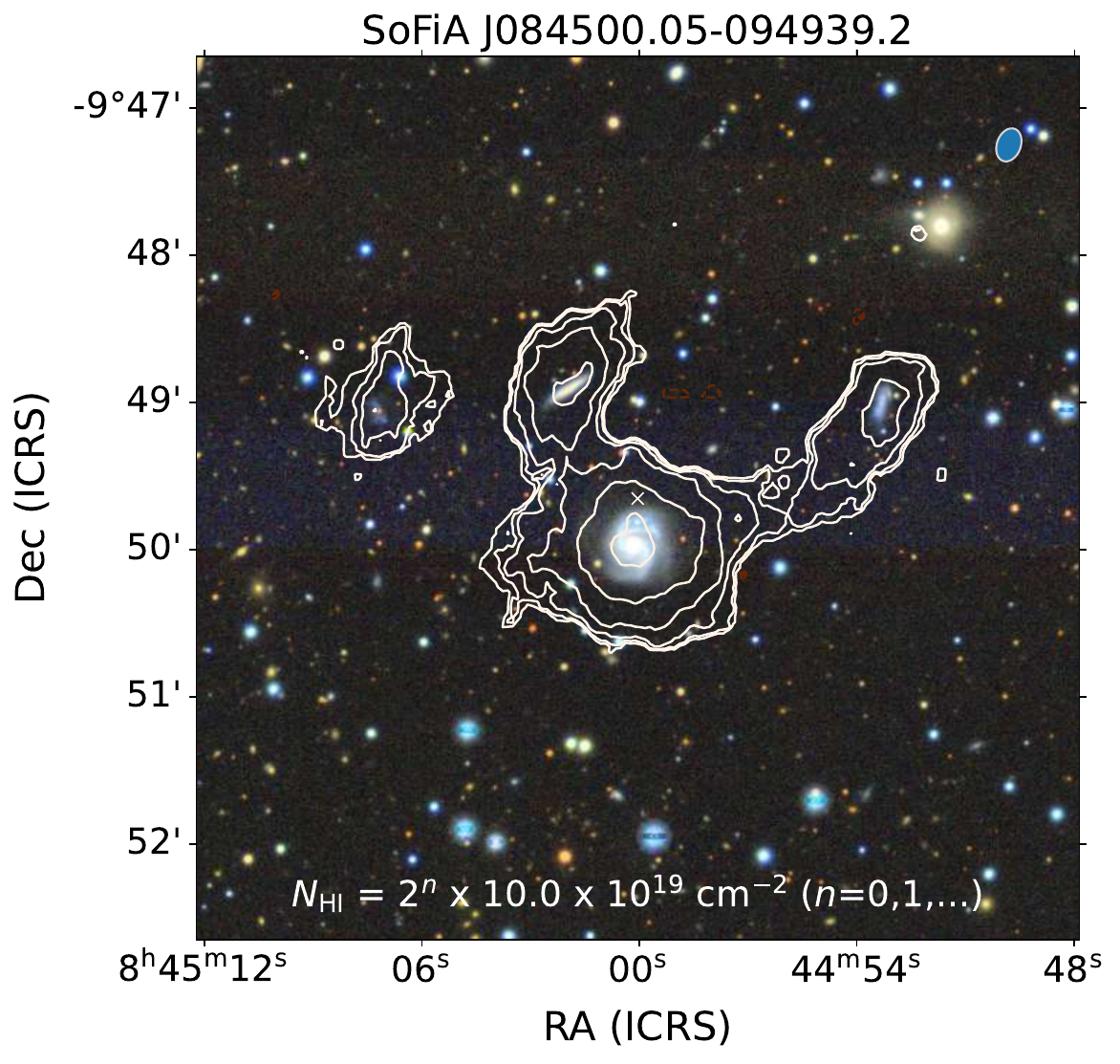}
\caption{MeerKAT \HI\ contours overlaid on a three-colour DECaLS DR10 image for galaxy IDs 1, 2, 3 and 4. There is clear interaction between the southern-most galaxy (ID3) and its neighbouring counterparts (ID1 and 2). Galaxies ID1 and 2 have been found to have relatively diminished SFRs (0.28 and 0.46~M$_{\odot}$~yr$^{-1}$) compared to the SFR of ID3 (7.11M$_{\odot}$~yr$^{-1}$). Image generated through SIP.}
\label{fig:mainset}
\end{figure*}

We have limited redshift information available in the literature for this sample. While three optical redshifts agree well with the measured \HI\ redshift from SoFiA, and a further seven photometric redshifts from WISE data also indicate that the emission seen by MeerKAT for those galaxies is \HI, that leaves another 39 galaxies without redshift information prior to this study. As such, it is possible that rather than \HI\ emission, a detection could instead be of another radio spectral line, such as hydroxyl ({\sc OH}), as demonstrated in the detection of an OH megamaser with Apertif \citep{Hess2021}, and in the LADUMA survey \citep{Glowacki2022}, both through the 1665--1667~MHz OH doublet. OH megamasers are relatively rarer with little over 100 detected to date, although the advent of SKA pathfinders such as MeerKAT and their associated spectral-line surveys are poised to improve on this space \citep{Roberts2021}.  

Short of dedicated spectral-line follow-up to verify redshifts for each source, it is possible to make a prediction of whether a detection is an OH megamaser at a higher redshift, which are typically seen in starburst galaxies, through the WISE magnitude and colour information. We give the WISE colour-colour distribution in Figure~\ref{fig:WISEcc}, overlaid on the classifications given in fig.~10 of \cite{Wright2010}. Most of our detections fall within the spiral galaxy region of the plot, with a couple falling in the starburst region (including the aforementioned ID3, which was confirmed to be detected in \HI\ from its optical spectroscopic information). One source appears to be QSO-like in its WISE colour properties, ID45. ID45 is seen to have unresolved radio continuum with flux density $S_{\rm 1362~MHz}$~=~0.4~mJy, but no other radio continuum or optical spectrum has been reported for this galaxy which could determine whether it hosts an AGN. 

Using the machine-learning approach described by \cite{Roberts2021} which uses WISE W1, W2 and W3 magnitudes and the peak frequency of the spectral-line emission with a k-Nearest Neighbours algorithm (private communication), predictions were made for all 49 galaxies on whether the emission seen corresponds to \HI\ or OH, as performed for the detection by \cite{Glowacki2022}. Only one detection, ID44, was identified as a \textit{potential} OH megamaser. The algorithm based upon the W1 vs W1--W2 information favoured the OH megamaser model ($>$99\%), but was outside any OH confidence interval for the algorithm using W1--W2 vs W2--W3 (6\% probability to be a OH megamaser). 

We note that ID44 had the lowest measured (assumed) \HI\ mass of our sample given the corresponding $z_{\rm HI}$ assumed; if it is actually an OH megamaser than its $M_{\rm OH}$ would be higher due to a higher true distance. ID44 was also the most significant outlier in the $M_{\rm HI}$ vs $M_{*}$ relation (Fig.~\ref{fig:MHIvMS}) - this galaxy being an OH~megamaser would explain this outlier result. ID44 also has the third-highest SFR of our sample (1.93~M$_{\odot}$~yr$^{-1}$), which also matches the picture of OH megamasers typically living in starburst galaxies that have undergone a recent merger or interaction. ID44 has the reddest W2--W3 colour of our sample (4.09~mag), placing it in the ULIRGs (ultra-luminous infrared galaxy) and starburst sector of Fig.~\ref{fig:WISEcc}.

Lastly, we note that ID44 is also seen in radio continuum in our MeerKAT observation, with a flux density of $S_{\rm 1362~MHz}$~=~1.2~mJy, and has near-infrared $K-$band magnitude measurements of 14.017~$\pm$~0.123 and 14.254~$\pm$~0.088 from the Two Micron All Sky Survey \cite[2MASS;][]{Skrutskie2006}. We use the photometric $K-$band-redshift relation for radio galaxies presented by \cite{Willott2003} (equation 1). Assuming $z$ = 0.0383 given by the assumption the spectral-line emission is \HI, the expected $K-$band magnitude is 10.33~mag. OH emission, corresponding to a host galaxy at redshift~$z$~$\sim$~0.219, gives the expected $K-$band magnitude of 14.24~mag, within the errors of the measured $K-$band magnitude for ID44. We repeat this analysis for all other galaxies in our sample with radio continuum (14) and $K-$band measurements in the literature (9). Two of these (IDs 2 and 12) had photometric redshift estimates that would favour OH rather than \HI, although we note no other evidence available supports the hypothesis these are OH, and ID2 is seen to be interacting with ID3 which has an optical spectroscopic redshift of 0.0385, in line with \HI\ (Section~\ref{sec:interacting}). Two other sources had photometric redshifts from the $K-$band relation that loosely support the \HI\ redshift, with the rest inconclusive, in part owing to the range of $K-$band values found in the literature for these sources (e.g. 11.975--13.008 for ID34).

We also consider the WISE information. The WISE colours for ID44 sits in the high-energy radio galaxy (HERG) distribution for the Large Area Radio Galaxy Evolution Spectroscopic Survey \cite[LARGESS;][]{Ching2017}. We applied the photometric redshift relations based upon the LARGESS sample for HERGs using the W1 and W2 bands from table 2 of \cite{Glowacki2017b}. This method provides a photometric redshift of 0.12 and 0.16 from W1 and W2 respectively, which suggests a higher-redshift host than 0.0383 expected from \HI. However, these details can only be treated as circumstantial evidence towards it being an OH megamaser at $z$~$\sim$~0.219, in addition to the predictions from methods by \cite{Roberts2021}, rather than something definitive. Further follow-up observations, namely spectroscopic optical measurements, will be necessary to properly determine whether we do have a OH megamaser within the dataset presented here. 

\subsection{Spatial distribution of the galaxies}

In Figure~\ref{fig:zhist} we give the redshift distribution for 46 of our 49 detections, omitting IDs 46--48 at $z_{\rm HI}$ of $\sim$0.087 and 0.019, to zoom into the bulk of our detections. We note that the colour scheme of the bins hence does not match that of Fig.~\ref{fig:overlayall}. Three main groups are seen: one at $z_{\rm HI}$ $\sim$ 0.033 (seven galaxies), another larger set at $\sim$0.041 (which includes three sub-groupings of 14, 9 and 3 galaxies), and another at $\sim$0.055 (also with several sub-groups, totalling 13 galaxies). This, combined with the distribution seen in Fig.~\ref{fig:overlayall}, indicates that multiple galaxy groups have been detected, potentially as part of a supergroup. The distances between members of separate galaxy groups e.g. IDs 19 and 25 at $z = 0.0329$ and $0.0412$ ranges from $\sim$15--100~Mpc. 

We present Figure~\ref{fig:nearneighbour} to further examine galaxy groups in our sample. Assuming all detections are in \HI\, for each galaxy (one galaxy focused on per panel, labelled by ID and ordered from lowest to highest redshift), a cumulative histogram is given, indicating the number of galaxies within 10~Mpc. For example, for the galaxies between $0.032 < z < 0.034$ (navy-coloured histograms), all 6 other companions are within 5~Mpc. It is evident that astronomers can use MeerKAT for a $<3$ hour on-source observation to trace \HI-rich galaxy groups and larger structures. 

\subsubsection{Interacting galaxies}\label{sec:interacting}

\begin{figure*}
\centering
\includegraphics[width=0.99\linewidth]{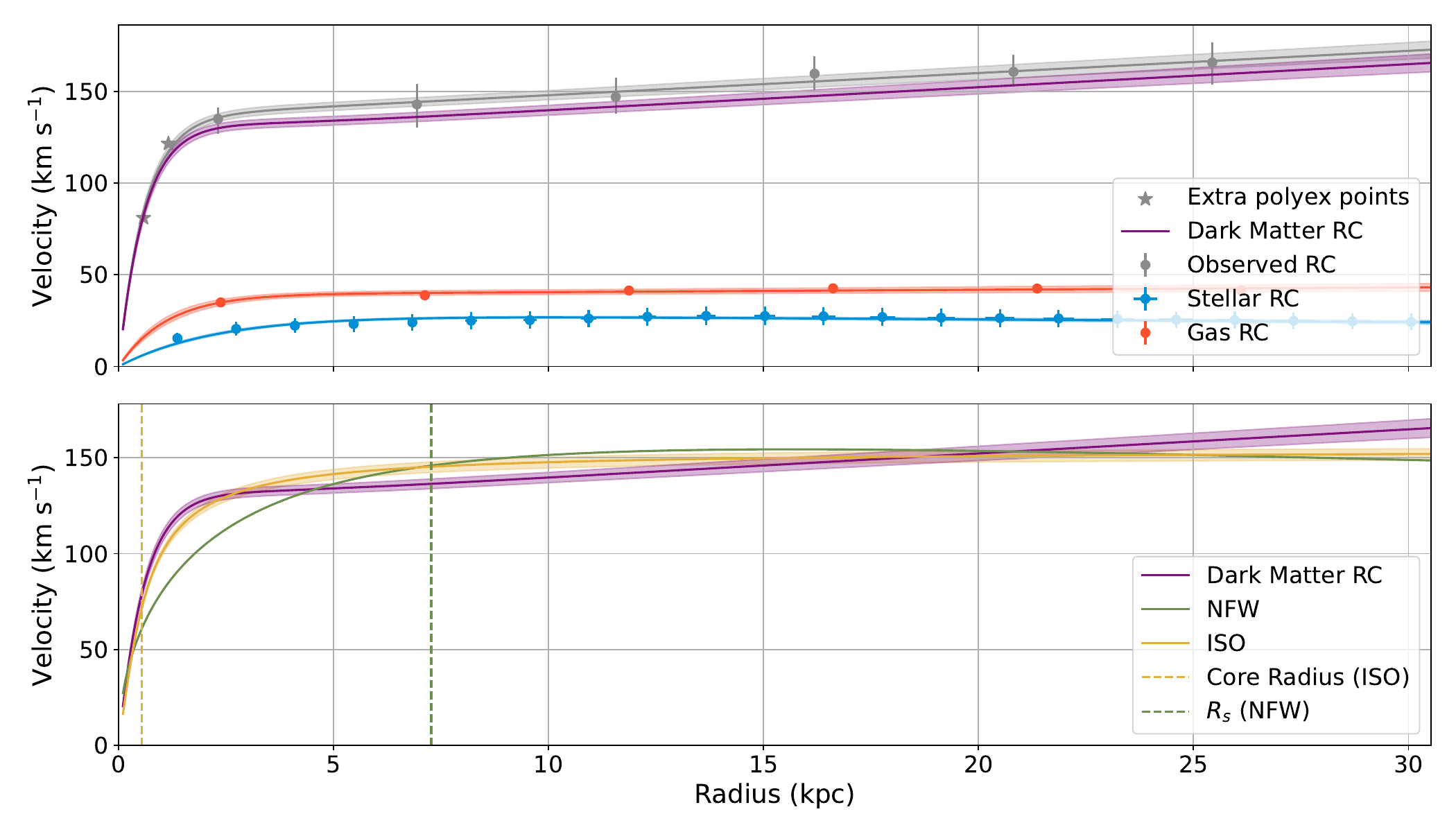}
\includegraphics[width=0.99\linewidth]{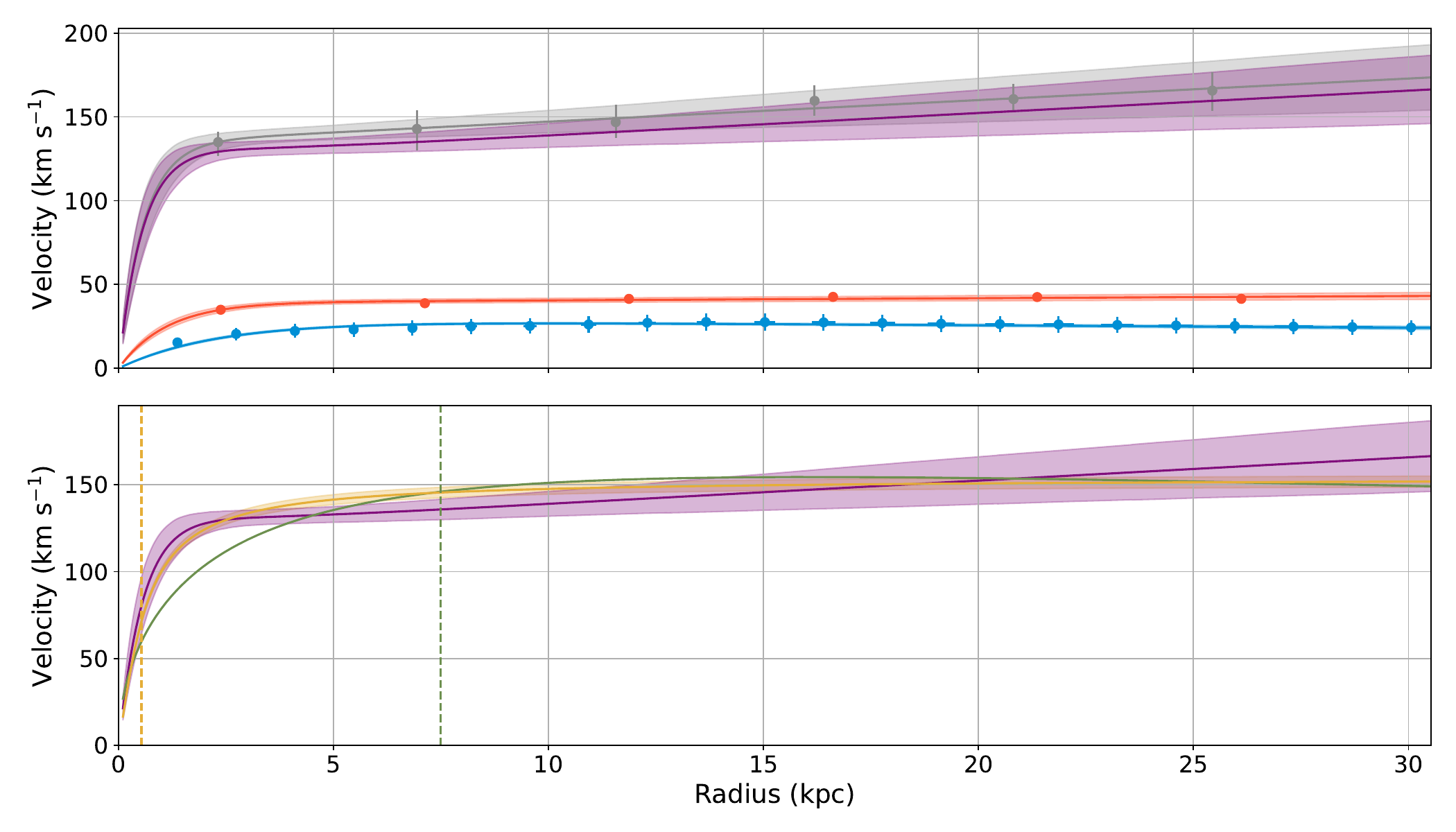}
\caption{Top: Rotation curves for galaxy ID18. Presented is the observed (total) rotation curve from the BBarolo model (grey), the stellar rotation curve derived from PanSTARRS imaging (blue), the gas rotation curve from a \HI\ surface density profile derived by BBarolo (red), and the dark matter rotation curve (purple) derived by subtracting the gas and stellar curves in quadrature from the observed. We also include two additional points (grey stars) for the BBarolo rotation curve fit to aid the observed rotation and dark matter model curve fits. 2nd panel: The ISO (orange) and the NFW (green) dark matter halo model fits to the dark matter rotation curve. Bottom two panels: as before, but without the two additional points added to the BBarolo curve. The radius values for both models decreases slightly as a result. For all panels, shaded regions indicate the 25 and 75-percentile error ranges for each fit, which is not displayed for the NFW profile due to visibility issues when displaying these larger errors.}
\label{fig:rot_massmodel}
\end{figure*}

There are a few examples in our sample of interacting galaxies. Despite the short observing time thus far obtained with MeerKAT on this field, we present the clearest example of galaxy interaction traced by \HI\ in Figure~\ref{fig:mainset}. \HI\ intensity-map contours for galaxy IDs 1, 2, 3 and 4 (top-right, middle-top, bottom, and top-left respectively) are overlaid on a DECaLS DR10 three-colour image. Note also that galaxies ID 5, 6, and 9 are located to the south (Fig.~\ref{fig:overlayall}), all which have slightly irregular velocity fields that hint at a possible interaction, and ID44 to the north (also reflected in the cumulative distributions of neighbouring \HI\ galaxies in Fig.~\ref{fig:nearneighbour} - assuming ID44 is not a hydroxyl megamaser). We also see a few galaxies with irregular \HI\ morphologies and/or disturbed velocity fields (e.g. IDs 13, 17, and 21).

Galaxy ID3, a face-on spiral galaxy (2MASS\,J08445897-0946457) at $z$~=~0.038557, is the most massive in both \HI\ and stellar content of this particular subset (and the highest \HI\ mass of all 49 galaxies presented here), and is directly interacting in \HI\ with both ID1 and ID2. It also has from its WISE and GALEX photometries the highest SFR of the sample by far, of 7~M$_{\odot}$\,yr$^{-1}$. ID3 is more $M_{\rm HI}$-massive compared to other galaxies with similar stellar masses in both this sample and in ALFALFA \cite[Fig.~\ref{fig:MHIvMS} and][]{Maddox2015}. In contrast, both ID1 and 2 have SFRs of less than 0.5~M$_{\odot}$\,yr$^{-1}$, in the lower half of the measured SFRs of the sample (Fig.~\ref{fig:4panel}), while ID4 only has a SFR$_{\rm NUV}$ of 0.11~M$_{\odot}$\,yr$^{-1}$. Our conclusion hence is that ID3 is currently stripping the \HI\ gas from its neighbours to fuel its own star formation activity. 

We note that galaxy IDs 16 and 17, as well as 43 and 49 (as noted in Section~\ref{sec:ohm}), are quite near each other, with the latter showing evidence of interaction in their spectral-line emission (see Figure A1 
of the Appendix 
available online). This is in addition to the aforementioned ID34 (Fig.~\ref{fig:combo} bottom set) with an unconfirmed \HI\ detection of a neighbouring dwarf galaxy. More sensitive radio observations will be able to further highlight the environment around these galaxies and potential pre-processing effects. 

\subsection{Mass modelling}

A few galaxies amongst our detections were sufficiently spatially resolved (spanning at least three radio beams) in \HI\ to attempt dynamical modelling to determine the stellar, the gas and the dark matter mass components of the galaxies.

\subsubsection{Rotation curves}

We used the \HI\ content to trace the total rotation curve of the galaxy. 3D-Barolo (3D-Based Analysis of Rotating Objects via Line Observations), also known as BBarolo \citep{Teodoro2015}, was used for fitting 3D tilted-ring models to the individual SoFiA subcubes for galaxy IDs 2, 3, 8, 10, 18, 19, 24, and 48. However, as ID2 was interacting with ID3, we were unable to properly produce reasonable dark matter mass models for these galaxies. We hence removed both IDs 2 and 3 from the current investigation. Outputs for the model moment maps constructed by BBarolo was compared with the SoFiA moment maps, as well as the channel-by-channel maps (see Figure A2 
for an example for ID18 in the Appendix as online material) to determine the goodness of the 3D tilted-ring fits when generating the total observed rotation curve, and the \HI\ surface density profile for each galaxy. The \HI\ surface density profile was used to determine the mass of \HI\ within each tilted ring, which was multiplied by an assumed factor of 1.4 to account for the contribution of helium \cite[as in e.g.][]{Westmeier2011, Reynolds2019}. We next converted the \HI\ surface density profile to a gas rotation curve. The `Polyex' parametric model for the rotational velocity profile was fitted to the gas and total rotation curves \cite[equation 2 of][]{Giovanelli2002}. The Polyex model is an analytic representation of rotation curves based on the amplitude of the outer rotation curve, the exponential scale length of the inner rotation curve, and the outer rotation curve slope. It has been demonstrated to work well in observation datasets \cite[e.g.][]{Elson2017} and galaxies from hydrodynamical simulations \cite[e.g.][]{Glowacki2020}. 

\begin{table*}
\centering
\caption{Derived dark matter halo quantities for the ISO and NFW models for the 8 galaxies attempted. We note the best fitting model in each case, and also give the measured total dynamical mass from BBarolo.}
\begin{tabular}{llllllllr}
 \hline
 \hline
 & ISO & & NFW & & & & &   \\
 \hline 
ID & $\rho_{\rm 0}$ & $R_{c}$ & c & $R_{\rm 200}$ & $V_{\rm 200}$ & $M_{\rm 200}$ & $M_{\rm tot}$ & Preferred\\
 & $M_{\odot}$\,pc$^{-3}$ & kpc &  & kpc & km\,s$^{-1}$ & $M_{\odot}$ & $M_{\odot}$ & model\\
\hline
8 & 0.15 & 0.7 & 5.9 & 72 & 56 &  5.4$\times$10$^{10}$ & 6.6$\times$10$^{10}$ & ISO\\
10 & 0.35 & 1.0 & 9.6 & 143 & 120 & 4.8$\times$10$^{11}$ & 2.2$\times$10$^{11}$ & ISO\\
18 & 1.5 & 0.5 & 17.8 & 129 & 111 & 3.7$\times$10$^{11}$ & 1.6$\times$10$^{11}$ & ISO\\
19 & 2.9 & 0.3 & 22.6 & 107 & 87 & 1.9$\times$10$^{11}$ & 1.1$\times$10$^{11}$ & ISO\\
24 & 2.0 & 0.4 & 25.0 & 108 & 89 & 2.0$\times$10$^{11}$ & 1.1$\times$10$^{11}$ & ISO\\
48 & 2.3 & 0.2 & 26.5 & 63 & 51 & 3.8$\times$10$^{10}$ & 1.7$\times$10$^{10}$ & ISO\\
\hline
\hline
\label{tab:modelling}
\end{tabular}
\end{table*}

The stellar rotation curve was calculated from the isophotes derived from PanSTARRS $r$-band imaging (see Section~\ref{sec:optical}). Equation~\ref{eqn:stellarmass} was applied to obtain the stellar mass in each isophote radius. Similarly, the stellar rotation curve was then calculated from the stellar mass profile. In the top panel of Figure~\ref{fig:rot_massmodel} we give the total observed, gas, and stellar rotation curves for ID18. We subtracted the gas and stellar rotation curves in quadrature from the observed rotation curve to obtain a dark matter rotation curve. The rotation curves for other galaxies are given in Figure~A4 within the Appendix 
as online material.

Owing to a poor sampling of the inner rising parts of the rotation curves, which resulted in high uncertainties for the Polyex fits, we tested fitting for additional points. Relative to the datapoint in the rotation curve  at ($r_{\rm 1}$, $v_{\rm 1}$) in the innermost part of the galaxy, we added two more datapoints at ($r_{\rm 1}$/2, $v_{\rm 1}\times0.9$) and ($r_{\rm 1}$/4, $v_{\rm 1}\times0.6$). These two points were found to agree well with the original Polyex fit. We then repeated the Polyex fitting, as shown for ID18 in the top panel of Figure~\ref{fig:rot_massmodel}. We find that this was not necessary for reasonable rotation curves and corresponding mass model fits for IDs 10 and 18, but did reduce the error in the Polyex and mass model curve fits, with more significant improvements observed for the remaining four galaxies (see Figures in Appendix). We note that including these two additional points also decreases the $\rho_{\rm 0}$ values and slightly decreases the radii parameter values found from the mass modelling fits, although parameter values prior to this were consistent with values for other galaxies in the literature (expanded on below).

\subsubsection{Dark matter halo}

We adopted two models for the dark matter halo. The first is the $\Lambda$CDM Navarro-Frenk-White (NFW) cusp-dominated model \citep{Navarro1996}, with a density profile of

\begin{equation}
    \rho_{\rm NFW}(r) = \frac{\rho_{\rm crit}}{(r/R_{s})(1 + r/R_{s})^{2}},
\end{equation}

where $R_{s}$ is the scale radius, and $\rho_{\rm crit}$ is the critical density of the universe ($\frac{3H^{2}}{8 \pi G}$). This results in the velocity profile 

\begin{equation}
    V_{\rm NFW}(r) = V_{\rm 200}\sqrt{\frac{ln(1+cx)-cs/(1+cx)}{x[ln(1+c) - c/(1+c)]}},
\end{equation}

where $R_{\rm 200}$ is the point where the dark matter halo density is greater than the critical density, $V_{\rm 200}$ is the velocity at this point, c = $R_{\rm 200}$/$R_{s}$, and $x$ = $r/R_{\rm 200}$.

The second dark matter halo model considered was the spherical pseudo-isothermal ISO core-dominated dark matter halo model, which assumes a constant density within the galaxy core \citep{Blok2008}. It is described by the density profile

\begin{equation}
    \rho_{\rm iso}(r) = \frac{\rho_{0}}{1 + r/R_{c})^{2}},
\end{equation}

where $\rho_{\rm 0}$ is the core density, and $R_{c}$ is the core radius, both treated as free parameters within physical bounds from the literature. The corresponding rotational velocity is then:

\begin{equation}
    V_{\rm iso}(r) = \sqrt{4 \pi G \rho_{\rm 0}R_{c}^{2}(1 -\frac{R_{c}}{r}{\rm tan^{-1}}(\frac{r}{R_c}))}.
\end{equation}

In the bottom panel of Figure~\ref{fig:rot_massmodel} we show the best fits to the dark matter halo rotation curves from the NFW and ISO models for ID18, with other galaxies in the bottom panels of Figure~A4 in the Appendix as online material. In Table~\ref{tab:modelling} we give the corresponding ISO and NFW values, indicate which model was preferred (determined via a $\chi^{2}$ test), and the dark matter halo mass derived from BBarolo. ISO was the preferred model in all six galaxies investigated here, with large errors found for the NFW curve fits (as such, they are not plotted for visibility).

The majority of our galaxies are dark matter dominated, with dark matter halo masses exceeding 1$\times$10$^{11}$~M$_{\odot}$, and M$_{\rm 200}$ greater still. We note that the inner, rising parts of the rotation curves are not well sampled, particularly in our \HI\ data. 
As four of our six galaxies had noticeably improved dark matter rotation curves and corresponding mass models from the inclusion of two additional points at inner radii, we conclude that just using the existing dataset results in a higher uncertainty on the values derived from the fitted mass models.
Deeper and higher resolution \HI\ data will be able to improve on this aspect. The values we obtain for the ISO and NFW model parameters (Table~\ref{tab:modelling}) are however consistent with those found for other galaxies. In table 3 of \cite{Blok2008}, values ranged for the following parameters in the examined sample of THINGS galaxies: $\rho_{\rm 0}$: 0.9--298.7~$M_{\odot}$\,pc$^{-3}$; $R_{\rm c}$: 0.01---45.63~kpc; $c$: $<$0.1--30.9; $V_{\rm 200}$: 35.2--$>$500~km\,s$^{-1}$ (noting that \cite{Blok2008} also presents ISO and NFW parameter values from other implementations in further tables). \cite{Swaters2011} found values of $\rho_{\rm 0}$: 0.5--570~$M_{\odot}$\,pc$^{-3}$ and $R_{\rm c}$: 0.5---2.7~kpc for dwarf galaxies, and \cite{Oh2015} found values of $\rho_{\rm 0}$: 0.008--2.132~$M_{\odot}$\,pc$^{-3}$ and $R_{\rm c}$: 0.15---8.4~kpc for LITTLE THINGS galaxies.

\section{Serendipity}\label{sec:serendipity}

We examine whether our serendipitous discovery of 49 \HI-rich galaxies is unexpected. While \cite{Ranchod2021} reported on the discovery of an \HI\ group with 23 members in a MIGHTEE pointing, this is to date the first reported discovery in an observation not intending to search for such galaxies in the first place. But is this unusual? This is not straightforward to tell, as larger spectral-line surveys are ongoing, so truly untargeted surveys with the likes of MeerKAT or other SKA pathfinder telescopes are not yet available. For example, while Pre-Pilot and Pilot survey results are now available for WALLABY, these observations were all in pre-selected fields, with full survey observations only recently started. MIGHTEE-HI also targets specific fields with pre-existing ancillary datasets, including ``10--15 lower-mass clusters and galaxy groups per square degree" \citep{Maddox2021}. 

In the work by \cite{Jones2020}, four optical galaxy group catalogues were cross-matched with ALFALFA to measure the \HI\ mass function (HIMF) for group galaxies. They concluded there was no single group galaxy HIMF; while two were found to be similar, the other two included either more lower-mass, \HI-rich galaxies or no low-mass galaxies due to selection. They also concluded that due to the far greater number of
field galaxies in ALFALFA than group galaxies, there was no region of their environment parameter space where group galaxies were the dominant population. \cite{Cautun2014} applied a `NEXUS+' model and found 77\% of the total volume fraction of the two high resolution Millennium simulations was occupied by voids, and 18\% by walls, and only 6\% by filaments (fig.~8). Similarly, \cite{Falck2015} found agreement for the Planck and \textit{WMAP} simulations (fig.~3). These results suggest that to discover several groups in a relatively short observation in a single MeerKAT pointing (where the sensitivity of ALFALFA was not reached) is not likely. 

In early examination of five-hour tracks undertaken with MeerKAT with the MeerKAT HI Observations of Nearby Galactic Objects - Observing Southern Emitters survey \cite[MHONGOOSE;][]{Blok2016}, which targets 30 nearby disk and dwarf galaxies, additional \HI\ detections have been discovered with a median of $\sim$20 detections per pointing within a 50~MHz chunk, with an upper bound of 50 (private communication). Given this observation has less than half the on-source time of MHONGOOSE pointings, 49 individual detections is more than is typically expected to be detected. However, optical spectroscopic redshift information of the field is required to better ascertain whether this pointing was simply fortunate enough to target a dense area of the sky in terms of large scale structure. Nonetheless, if one assumes there are several \HI-rich galaxies to be found within any one MeerKAT pointing based on this discovery, findings from MIGHTEE-HI, and preliminary MHONGOOSE results, then there are undoubtedly a large number of galaxies awaiting detections within existing and upcoming MeerKAT Open Time observations. 

The sensitivity of MeerKAT in this frequency space is currently unparalleled, which enables the detection of lower \HI-mass galaxies than those detected in the HIPASS and ALFALFA all-sky surveys in less observing time. MeerKAT can also achieve greater sensitivity than the ongoing WALLABY all-sky survey with ASKAP, albeit with a far smaller field of view. Observations with MeerKAT hence will be able to detect galaxy group members with lower \HI\ mass than any other survey until the SKA. Below $z~<~0.1$ (above $\sim$1300~MHz) there is no significant radio frequency interference (RFI), owing to the radio-quiet site of the MeerKAT site. This redshift/frequency space is included in L-band spectral-line observations with MeerKAT - typically zoom-mode observations with MeerKAT will be focused on \HI\ rather than e.g. OH. We encourage Open Time observers to investigate their datasets for \HI-rich galaxies (as well as potentially OH megamasers). Such findings will collectively increase the number of galaxy groups we can study in \HI\, and aid investigations into e.g. the effect of pre-processing on galaxy group members, and more broadly galaxy and galaxy group/cluster evolution, driven by the large science surveys of SKA pathfinder telescopes.

\section{Conclusions}\label{sec:conclusions}

We present the serendipitous discovery of 49 \HI-rich galaxies within a single 2.3 hour pointing with MeerKAT in Open Time data. With the use of ancillary data from PanSTARRS, WISE, 2MASS, and GALEX, we obtained stellar masses and star formation rates to complement the \HI\ information we present (\HI\ masses, spectra, intensity and velocity maps). 

Within our sample we find examples of galaxy interactions, including a case of one galaxy stripping and fuelling star formation (to greater than 7~M$_{\odot}$~yr$^{-1}$) from two neighbouring galaxies. Multiple \HI-rich galaxy groups are identified, with three major groupings at $z~\sim$~0.033, 0.041, and 0.055 - therefore it is possible that we have also detected a supergroup or filament in this MeerKAT pointing. We also find one potential OH megamaser within our sample, although this is not confirmed due to a lack of optical spectroscopic redshifts available. We generate rotation curves and dark matter mass models for six galaxies in our sample that were sufficiently spatially resolved and not found to be significantly interacting with another galaxy. 

We considered the probability of detecting 49 \HI-rich galaxies in this relatively short MeerKAT pointing. It appears that 49 detections are unlikely, prompted by the study by \cite{Jones2020} which demonstrated that there is no single group galaxy HIMF and a far greater number of field galaxies than group galaxies within ALFALFA. Given the demonstrated ability of MeerKAT in both major science surveys and another Open Time proposal \citep{Healy2021} to detect new galaxies in \HI\ and OH, which is attributed to an RFI-quiet environment and high sensitivity in this space, we encourage other users of spectral-line Open Time data to investigate their datasets. With sufficient effort, significant datasets can be obtained to complement large science surveys aiming to address outstanding questions in galaxy evolution, including the formation of galaxy groups and their transition into supergroups and clusters. 

\section*{Acknowledgements}

We thank the anonymous referee for the helpful feedback provided that has improved this paper. We thank Erwin de Blok, Barbara Catinella, Luca Cortese, Nathan Deg, Rory Hackett, Scott Haydon, Julia Healy, Dane Kleiner, Ferry Lanter, Filippo Maccagni, John Forbes, and Andrew Sullivan for useful discussions and feedback on the paper. MG is supported by the Australian Government through the Australian Research Council's Discovery Projects funding scheme (DP210102103).

The MeerKAT telescope is operated by the South African Radio Astronomy Observatory (SARAO; \url{www.sarao.ac.za}), which is a facility of the National Research Foundation (NRF), an agency of the Department of Science and Innovation. The MeerKAT data presented in this paper were processed using the ilifu cloud computing facility  (\url{www.ilifu.ac.za}), which is operated by a consortium that includes the University of Cape Town (UCT), the University of the Western Cape, the University of Stellenbosch, Sol Plaatje University, the Cape Peninsula University of Technology and the South African Radio Astronomy Observatory. The ilifu facility is supported by contributions from the Inter-University Institute for Data Intensive Astronomy (IDIA, which is a partnership between the UCT, the University of Pretoria and the University of the Western Cape), the Computational Biology division at UCT, and the Data Intensive Research Initiative of South Africa (DIRISA). This work was carried out using the data processing pipelines developed at the Inter-University Institute for Data Intensive Astronomy (IDIA) and available at \url{https://idia-pipelines.github.io}. IDIA is a partnership of the University of Cape Town, the University of Pretoria and the University of the Western Cape. This work made use of the CARTA (Cube Analysis and Rendering Tool for Astronomy) software (\url{https://cartavis.github.io}), and iDaVIE-v \cite[Immersive Data Visualisation Interactive Explorer for Volumetric Rendering;][]{Comrie2021b,Jarrett2021}. This research made use of hips2fits,\footnote{https://alasky.cds.unistra.fr/hips-image-services/hips2fits} a service provided by CDS. This research has made use of the NASA/IPAC Extragalactic Database (NED),
which is operated by the Jet Propulsion Laboratory, California Institute of Technology, under contract with the National Aeronautics and Space Administration. Part of this research was conducted through the International Centre for Radio Astronomy Research (ICRAR) summer studentship program. 

\section*{Data Availability}

Data will be made available upon reasonable request. The Open Time MeerKAT data used in this survey is already publicly available via the SARAO archive (SBID 1617470178). 



\bibliographystyle{mnras}
\bibliography{bibliography} 




\label{appendixpage}
\appendix

\section{Images and notes on individual sources}\label{app:comboimages}

We present further SIP output images as in Fig.~1, and give notes for each detection. We also present rotation curves and dark matter model fits as in Fig.~10. 

\textbf{\textit{ID1:}} Host is a blue galaxy in a group environment. Interacting with ID3, where it appears to be having its \HI\ gas stripped by ID3. This may explain its relatively diminished SFR of 0.28~M$_{\odot}$\,yr$^{-1}$ and disturbed velocity map.

\textbf{\textit{ID2:}} Host is an edge-on disk galaxy. Like ID2, this galaxy is interacting with ID3, with more redshifted gas on its south side being stripped. Small radio continuum point source seen with flux density $S_{\rm 1362~MHz}$~=~0.4~mJy.

\textbf{\textit{ID3:}} Host is a large (log$_{\rm 10}$($M_{*}$) = 10.46~M$_{\odot}$) face-on spiral galaxy (2MASS\,J08445897-0946457) with an optical spectroscopic redshift of $z$~=~0.038557. It has a high SFR of 7.11M$_{\odot}$~yr$^{-1}$. This is attributed to the active stripping of gas from both ID1 and 2. Large radio continuum point source seen with flux density $S_{\rm 1362~MHz}$~=~7.3~mJy.

\textbf{\textit{ID4:}} Host appears to be a faint blue galaxy. It is close to ID2, but does not yet show clear evidence of interaction with the given sensitivity of the MeerKAT observation, although its velocity map is not {\myedit very regular}. 

\textbf{\textit{ID5:}} Faint optical source identified as the host, with log$_{\rm 10}$($M_{*}$) = 8.10~M$_{\odot}$. To the south of IDs 1, 2, 3 and 4.

\textbf{\textit{ID6:}} Faint optical source identified as the host, with log$_{\rm 10}$($M_{*}$) = 7.32~M$_{\odot}$. To the south of IDs 1, 2, 3 and 4.

\textbf{\textit{ID7:}} Host is a larger galaxy relative to ID5/6. To the south of IDs 1, 2, 3 and 4.

\textbf{\textit{ID8:}} Host is a large face-on spiral galaxy, with a potential nearby companion visible to the right in the optical, with a minor peak in the \HI\ intensity map aligning with the companion. No evidence of interaction in its velocity map or spectrum - higher resolution imaging is required. Small radio continuum point source seen with flux density $S_{\rm 1362~MHz}$~=~0.4~mJy. 

\textbf{\textit{ID9:}} Host is a faint blue galaxy in the optical. Disturbed velocity map. To the south of IDs 1, 2, 3 and 4.

\textbf{\textit{ID10:}} Host is a large blue galaxy in the optical, with an apparent nearby red companion. \HI\ spectrum is slightly asymmetric with more flux observed at the redshifted (higher velocity) end. Part of a sub-group with IDs 11, 12 and 13, to the south of lower-ID galaxies. Potential faint radio continuum emission with flux density $S_{\rm 1362~MHz}$~$\sim$~0.1~mJy.

\textbf{\textit{ID11:}} Host {\myedit actually appears as a blue galaxy in DeCALs}. Asymmetric \HI\ spectrum observed, and with a potential \HI\ feature extending to the upper left, as seen in the intensity and velocity map. Part of a sub-group with IDs 10, 12 and 13, to the south of lower-ID galaxies. 

\textbf{\textit{ID12:}} Host is a mostly face-on disk galaxy, with log$_{\rm 10}$($M_{*}$) = 9.81~M$_{\odot}$. Part of a sub-group with IDs 10, 11 and 13, to the south of lower-ID galaxies. Potential faint radio continuum emission with flux density $S_{\rm 1362~MHz}$~$\sim$~0.1~mJy.

\textbf{\textit{ID13:}} Host is a faint blue galaxy. Disturbed intensity and particularly velocity map. Part of a sub-group with IDs 10, 11 and 12, to the south of lower-ID galaxies. 

\textbf{\textit{ID14:}} Host identified in the optical albeit required to use PanSTARRS imaging with noticeable streaks seen in the image. Not part of the aforementioned group galaxies (higher redshift), instead tracing out a separate filament primarily in the right hand side of the MeerKAT pointing (IDs 14--17, 26--32, 38). 

\textbf{\textit{ID15:}} Host is an edge-on disk galaxy. 

\textbf{\textit{ID16:}} Host is a blue spiral galaxy. Close to ID17. 

\textbf{\textit{ID17:}} Host is difficult to distinguish - either of two similarly-sized blue galaxies in the optical. Close to ID16. Disturbed velocity map and a possible extended feature to the south.  

\textbf{\textit{ID18:}} Host is an edge-on disk galaxy, with a bright counterpart near it spatially ({\myedit very likely} a star). Close to ID19. Lopsided \HI\ spectrum with more flux on its blueshifted (lower velocity) side. Elongated radio continuum source seen with flux density $S_{\rm 1362~MHz}$~=~0.9~mJy.

\textbf{\textit{ID19:}} Two galaxies nearby, with the \HI\ centred on the leftmost galaxy. {\myedit The peak of the \HI\ however aligns with the rightmost galaxy, although this peak contour is smaller than the radio beam.} Fairly symmetric spectrum. Close to ID18. 

\textbf{\textit{ID20:}} Host is a {\myedit possible S0} galaxy, with the \HI\ extend somewhat perpendicular to the optical alignment. Close to IDs 21 and 22. Spectroscopic redshift of $z$~=~0.03886.

\textbf{\textit{ID21:}} Very faint optical host identified (log$_{\rm 10}$($M_{*}$) = 7.59~M$_{\odot}$), which drops out in the PanSTARRS $z$ and $y$ bands. Disturbed \HI\ maps. Between IDs 20 and 22 - a potential interaction at play?  

\textbf{\textit{ID22:}} A faint edge-on host identified, with no coverage in DECaLS DR10. To the north and close to IDs 20 and 21. 

\textbf{\textit{ID23:}} Host is a large and spectacular edge-on spiral galaxy. \HI\ gas distribution appears to have a deficiency in the centre by the host galaxy bulge; prior to viewing optical imaging had been identified as two separate galaxies, including by SoFiA. ID45 identified as a nearby galaxy. 

\textbf{\textit{ID24:}} Host is a large bright galaxy (log$_{\rm 10}$($M_{*}$) = 10.17~M$_{\odot}$). Small radio continuum point source seen with flux density $S_{\rm 1362~MHz}$~=~0.5~mJy.

\textbf{\textit{ID25:}} Host is a large spiral face-on galaxy, with log$_{\rm 10}$($M_{*}$) = 10.19~M$_{\odot}$. Optical spectroscopic redshift of $z$~=~0.04080. Small radio continuum point source seen with flux density $S_{\rm 1362~MHz}$~=~0.3~mJy.

\textbf{\textit{ID26:}} Faint optical host identified, with log$_{\rm 10}$($M_{*}$) = 8.94~M$_{\odot}$.

\textbf{\textit{ID27:}} Narrow \HI\ spectrum, with a faint optical galaxy also identified as the host. Velocity map is curious in that the centre is redshifted - worth further scrutiny in any follow-up observations. 

\textbf{\textit{ID28:}} Hos{\myedit t} is an edge-on spiral galaxy. Possible \HI\ feature emerging in the north end. Close to ID 29.

\textbf{\textit{ID29:}} Quite unusual \HI\ distribution evident. Optical host is hard to identify in PanSTARRS but clearer in the north side as a faint blue galaxy in DECaLS DR10, suggesting the southern \HI\ content is a tail or other interaction feature. Close to ID 28. 

\textbf{\textit{ID30:}} Host is a bright, large galaxy in the optical, with log$_{\rm 10}$($M_{*}$) = 10.62~M$_{\odot}$. Extended, diffuse radio continuum source seen with flux density $S_{\rm 1362~MHz}$~=~0.7~mJy.

\textbf{\textit{ID31:}} Host is a faint blue galaxy. \HI\ is extended, with redshifted components evident in the velocity map to the north and south. 

\textbf{\textit{ID32:}} Host is an edge-on spiral galaxy. Lopsided \HI\ spectrum {\myedit and the velocity field indicates a strong warp in the disk}. 

\textbf{\textit{ID33:}} Host is a large spiral galaxy. Lopsided \HI\ spectrum. Small radio continuum point source seen with flux density $S_{\rm 1362~MHz}$~=~0.5~mJy.

\textbf{\textit{ID34:}} Host is a large optical galaxy with the largest stellar mass in our sample (log$_{\rm 10}$($M_{*}$) = 10.80~M$_{\odot}$). Evidence of a nearby companion and faint tail connecting it to the main galaxy, although deeper \HI\ observations are required to confirm this. Small radio continuum point source seen with flux density $S_{\rm 1362~MHz}$~=~1.1~mJy.

\textbf{\textit{ID35:}} Host is a {\myedit blue} galaxy in the optical, at the edge of DECaLS DR10 coverage. 

\textbf{\textit{ID36:}} Host appears as a red galaxy in  {\myedit PanSTARRs}. Most of the \HI\ emission is to the south. 

\textbf{\textit{ID37:}} Host is a face-on spiral galaxy. Disturbed \HI\ velocity map. Small radio continuum point source seen with flux density $S_{\rm 1362~MHz}$~=~1.5~mJy.

\textbf{\textit{ID38:}} Host is a faint blue galaxy. Diffuse extended \HI\ emission seen. 

\textbf{\textit{ID39:}} Host is a faint blue galaxy. Diffuse extended \HI\ emission seen.

\textbf{\textit{ID40:}} Host is a rather faint edge-on galaxy. Extended \HI\ emission to the south with a hole seen.

\textbf{\textit{ID41:}} Faint optical host identified with log$_{\rm 10}$($M_{*}$) = 9.60~M$_{\odot}$.

\textbf{\textit{ID42:}} Host is an edge-on blue galaxy. Extended \HI\ emission to the south. 

\textbf{\textit{ID43:}} Host is a faint blue galaxy. Interacting with ID49.

\textbf{\textit{ID44:}} Host is a small but bright galaxy. \HI\ poor, but also potentially identified as an OH megamaser. Optical spectroscopic redshift data is required to verify the emission type. Small radio continuum point source seen with flux density $S_{\rm 1362~MHz}$~=~1.2~mJy.

\textbf{\textit{ID45:}} Host is a compact galaxy. Small radio continuum point source seen with flux density $S_{\rm 1362~MHz}$~=~0.4~mJy.

\textbf{\textit{ID46:}} A higher redshift detection ($z_{\rm HI}$~=~0.08740). Host is a blue galaxy. Small radio continuum point source seen with flux density $S_{\rm 1362~MHz}$~=~0.1~mJy.

\textbf{\textit{ID47:}} A higher redshift detection ($z_{\rm HI}$~=~0.08674). Host is a blue galaxy. Disturbed and diffuse \HI\ seen.

\textbf{\textit{ID48:}} A lower redshift detection ($z_{\rm HI}$~=~0.01938). Host is a blue galaxy. 

\textbf{\textit{ID49:}} A faint blue galaxy is identified as the host. Seen to be interacting with ID44. Disturbed gas kinematics.

\begin{landscape}
\begin{figure}
\includegraphics[width=1.16\textwidth]{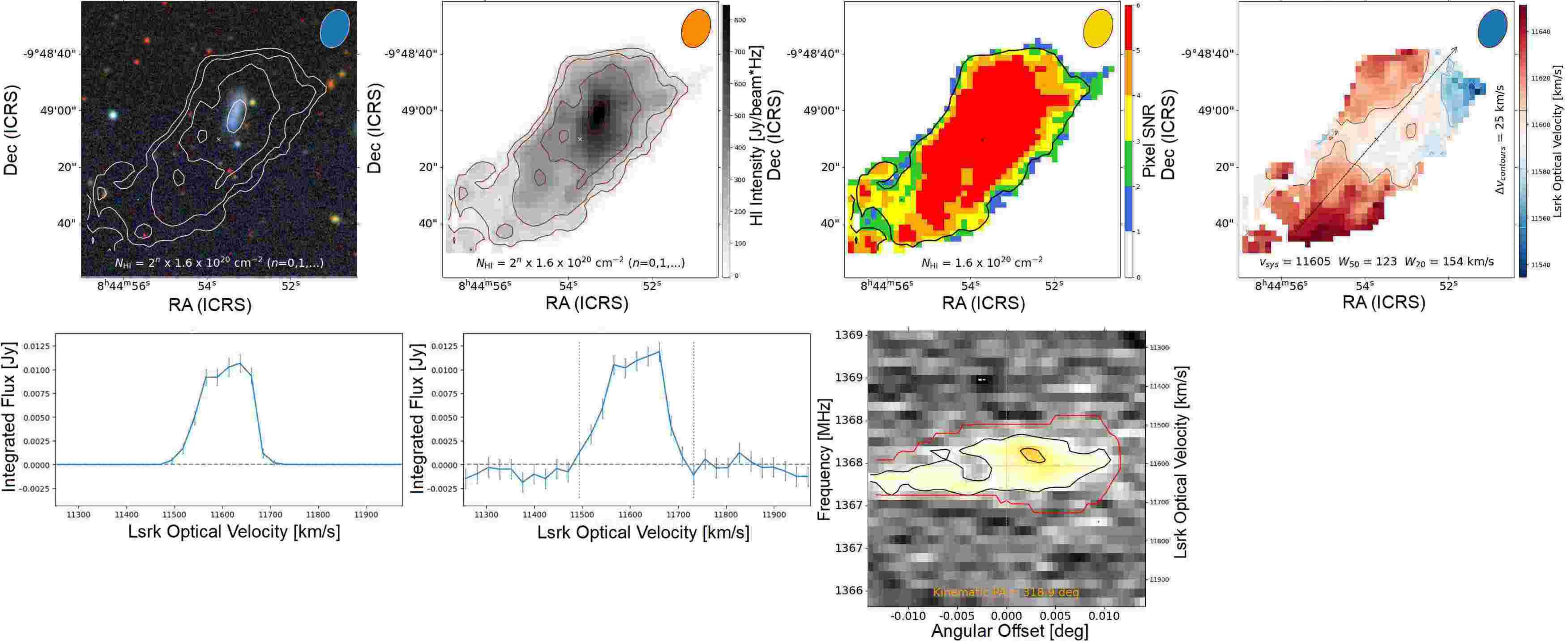}
\includegraphics[width=1.16\textwidth]{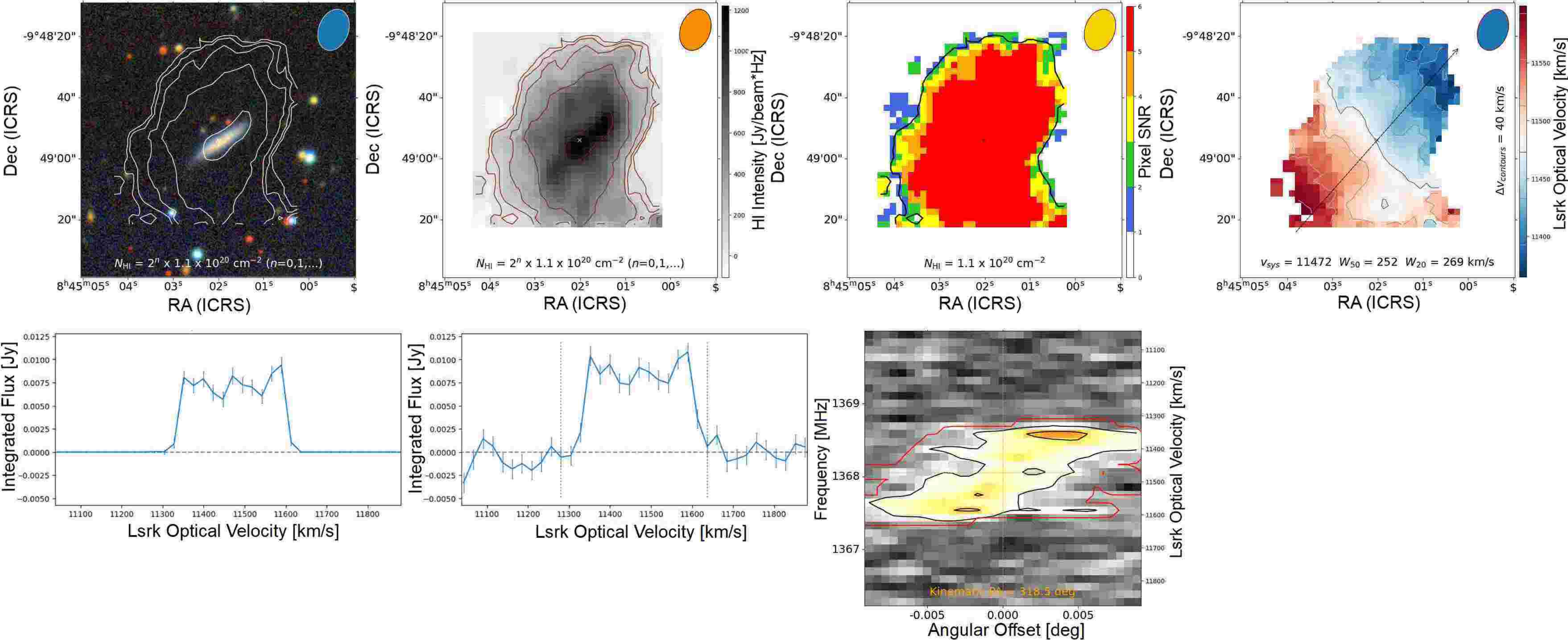}
\caption{SIP outputs for galaxy IDs 1 and 2, as in ~Fig.~1. Contours are overlaid on either DECaLS DR10 or PanSTARRS three-colour images. These two galaxies are interacting with ID3. Images created via SIP.}
\label{fig:comboappendix}
\end{figure}
\end{landscape}

\begin{landscape}
\begin{figure}
\ContinuedFloat
\includegraphics[width=1.16\textwidth]{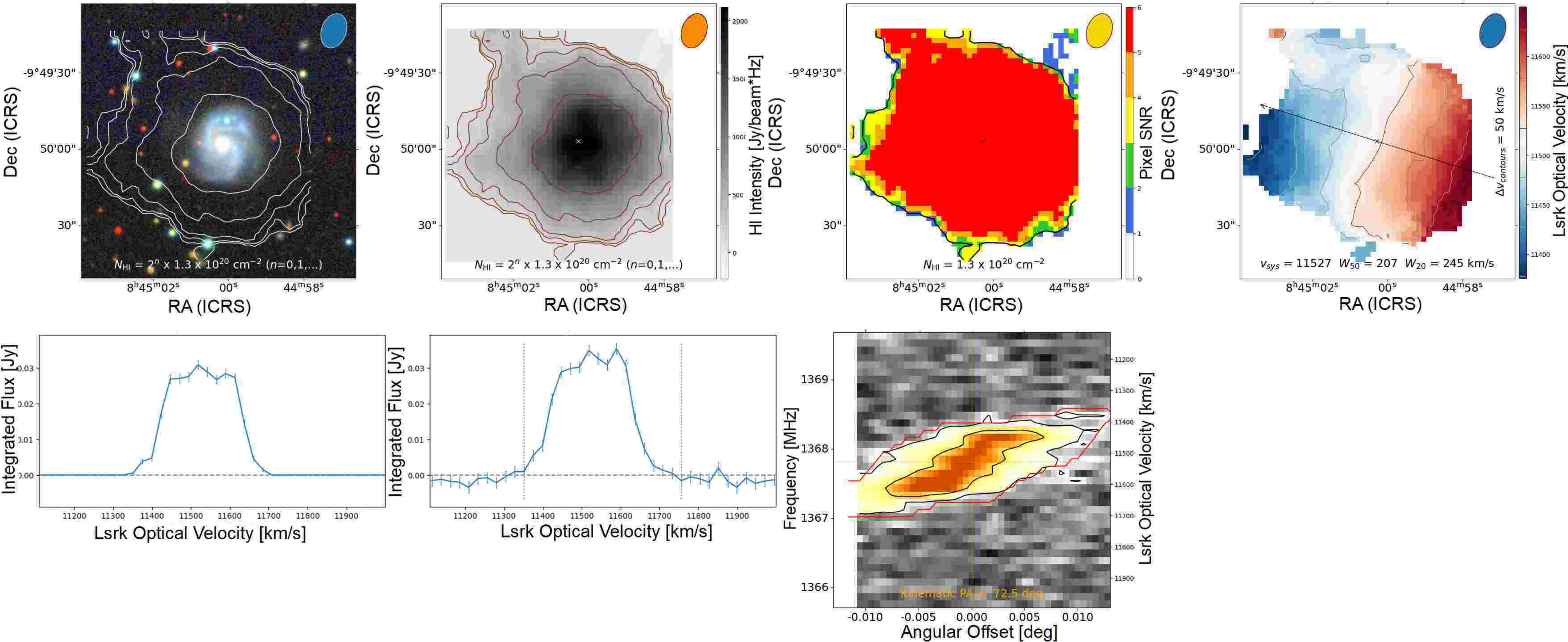}
\includegraphics[width=1.16\textwidth]{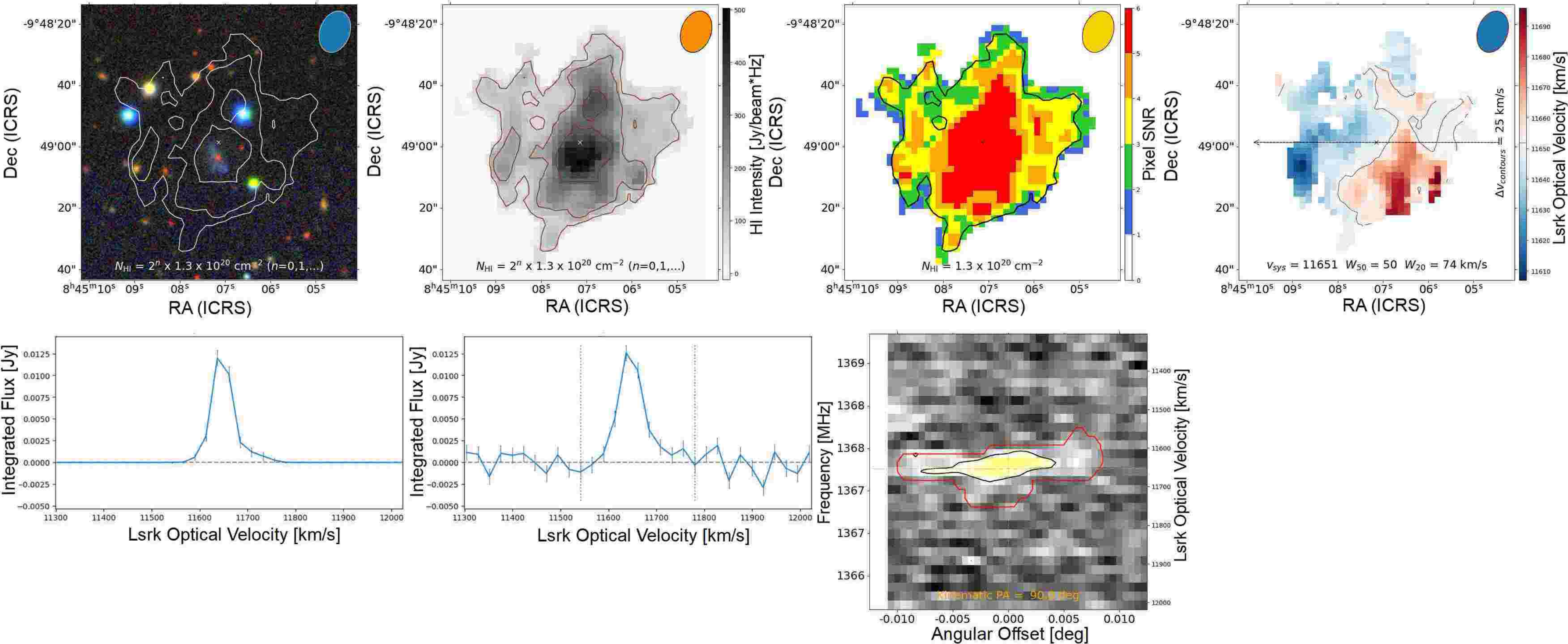}
\caption{Continued. SIP outputs for galaxy IDs 3 and 4. ID3 is interacting with ID 1 and 2.}
\end{figure}
\end{landscape}

\begin{landscape}
\begin{figure}
\ContinuedFloat
\includegraphics[width=1.16\textwidth]{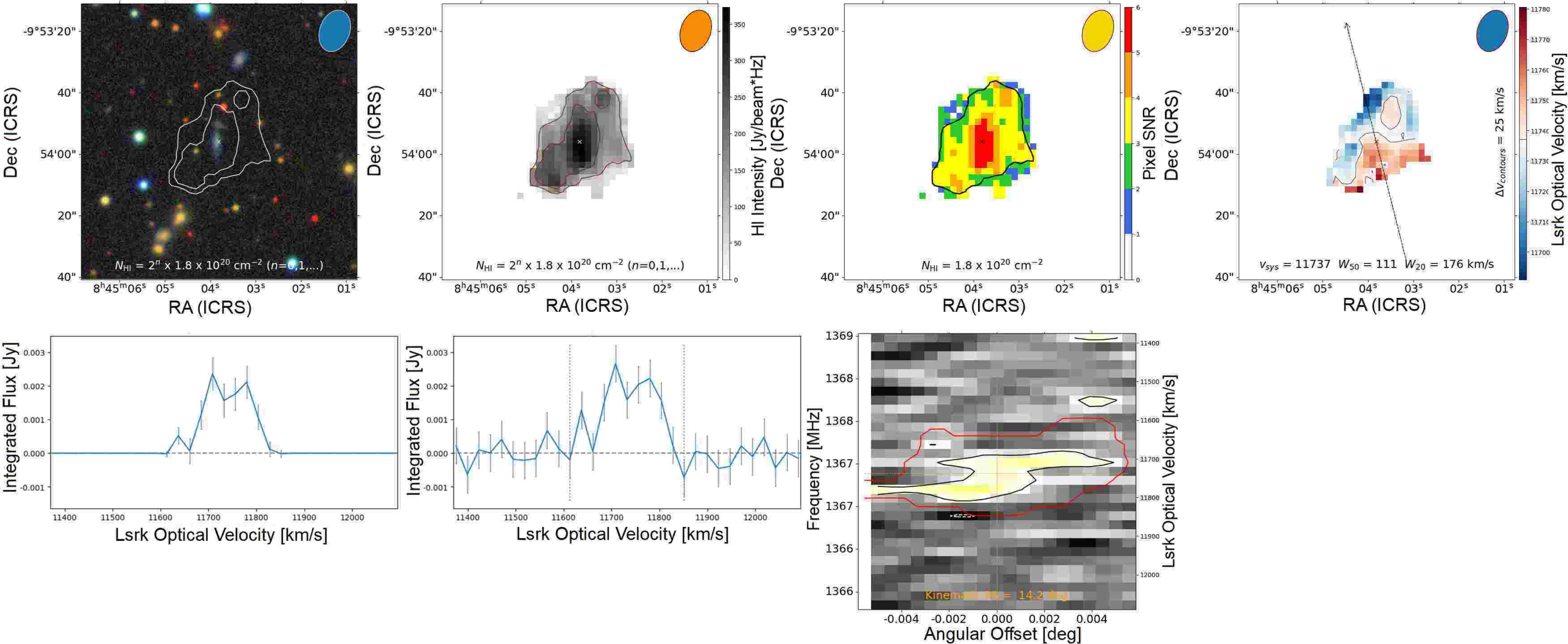}
\includegraphics[width=1.16\textwidth]{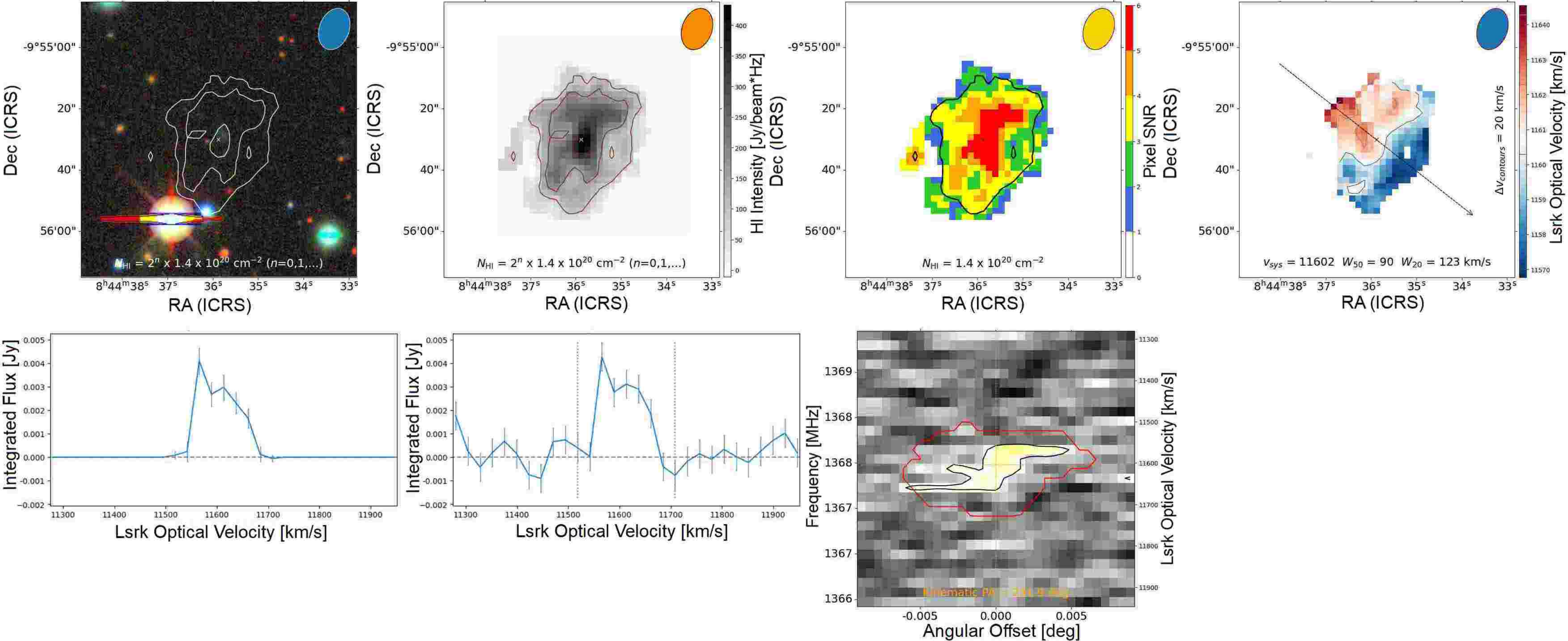}
\caption{Continued. SIP outputs for galaxy IDs 5 and 6.}
\end{figure}
\end{landscape}

\begin{landscape}
\begin{figure}
\ContinuedFloat
\includegraphics[width=1.16\textwidth]{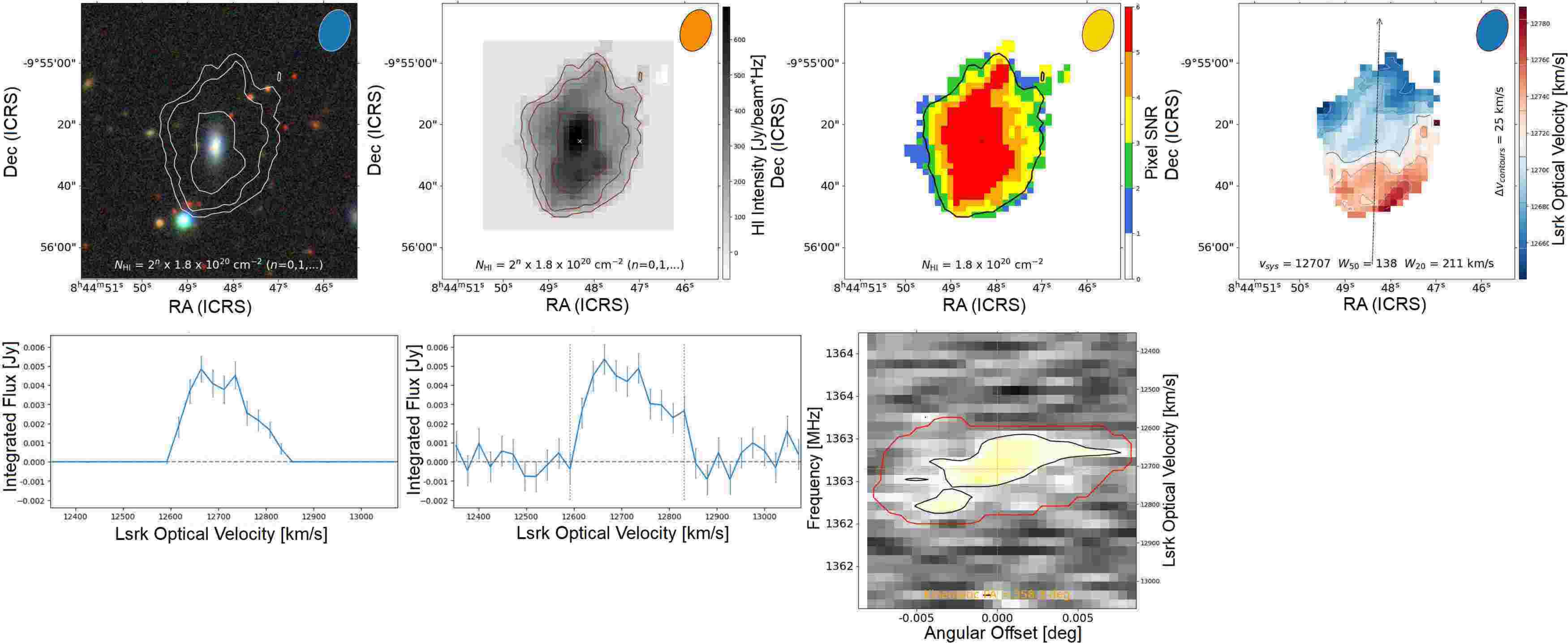}
\includegraphics[width=1.16\textwidth]{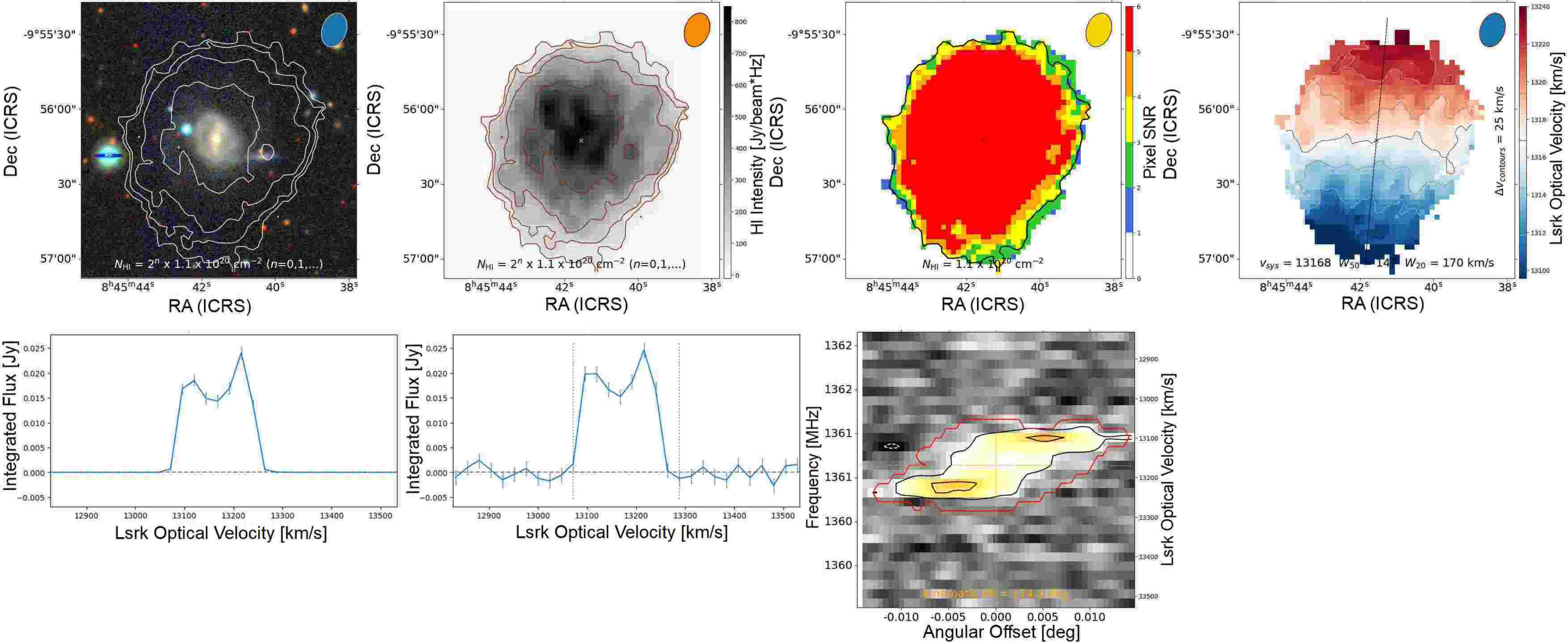}
\caption{Continued. SIP outputs for galaxy IDs 7 and 8.}
\end{figure}
\end{landscape}

\begin{landscape}
\begin{figure}
\ContinuedFloat
\includegraphics[width=1.16\textwidth]{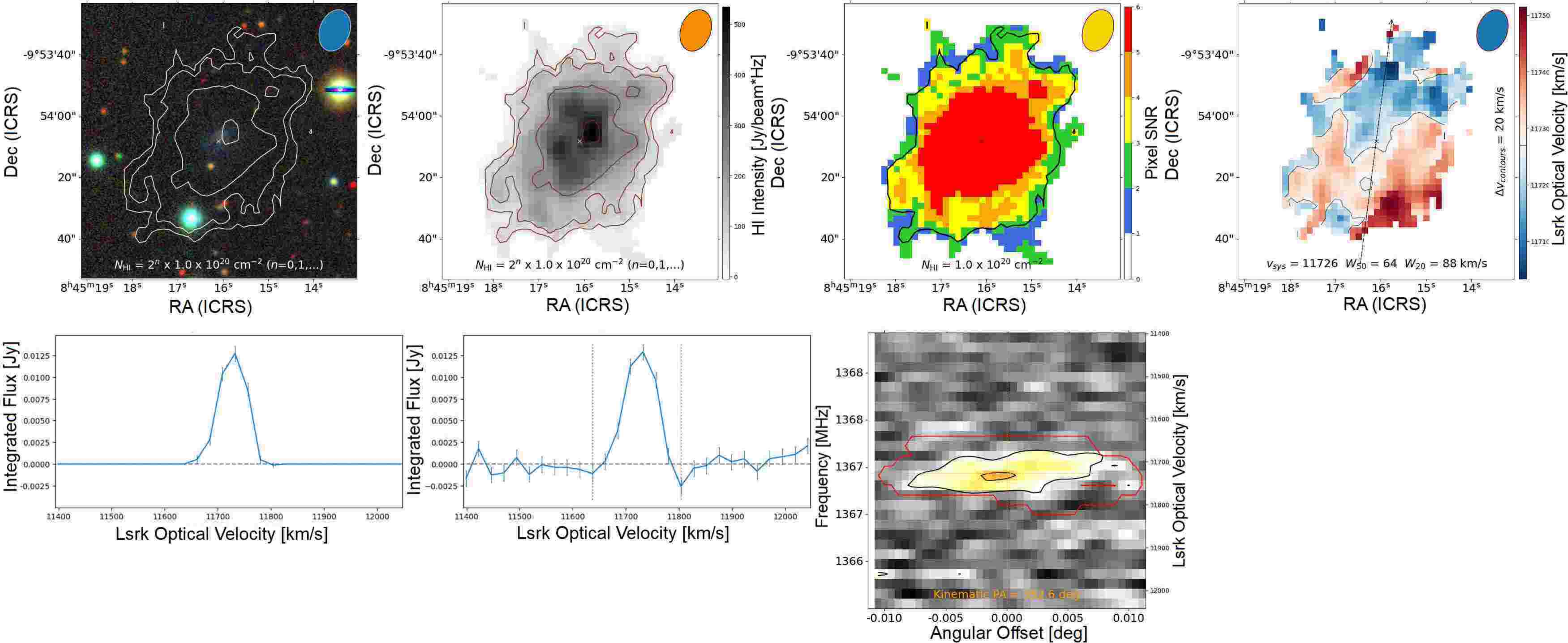}
\includegraphics[width=1.16\textwidth]{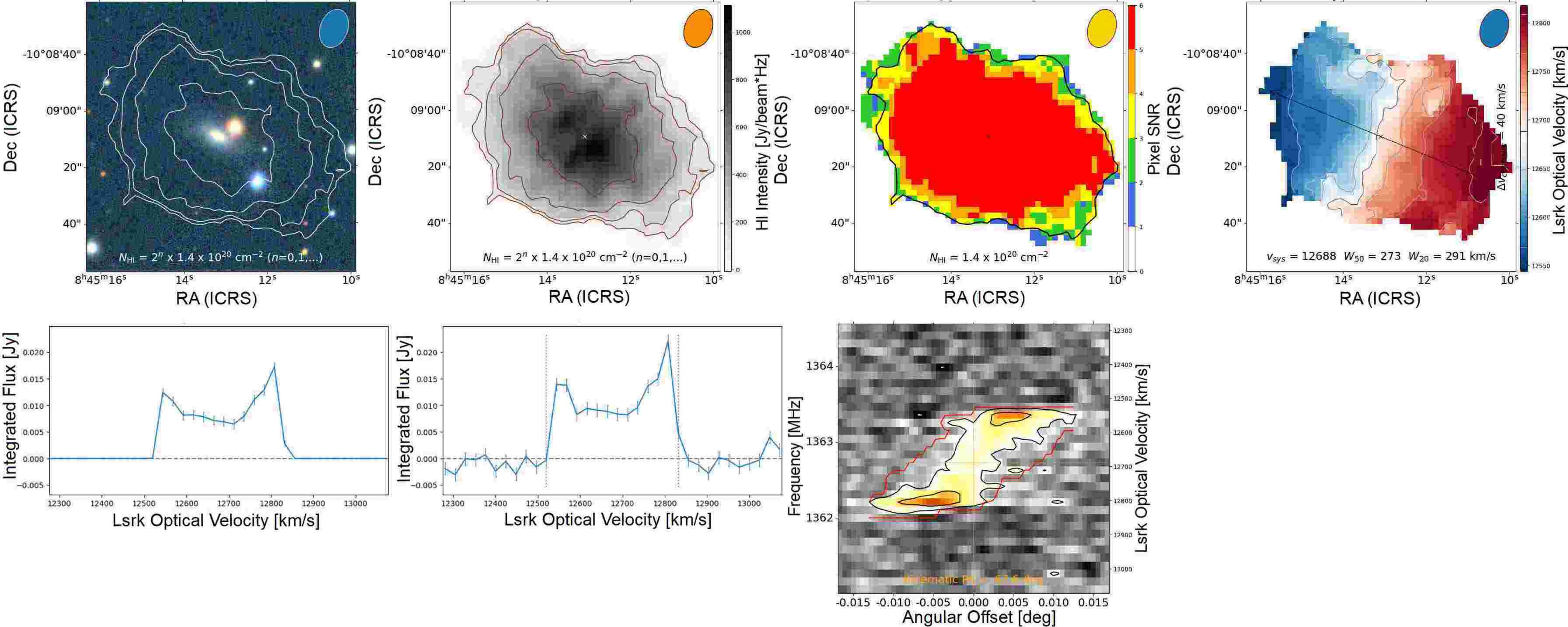}
\caption{Continued. SIP outputs for galaxy IDs 9 and 10.}
\end{figure}
\end{landscape}

\begin{landscape}
\begin{figure}
\ContinuedFloat
\includegraphics[width=1.16\textwidth]{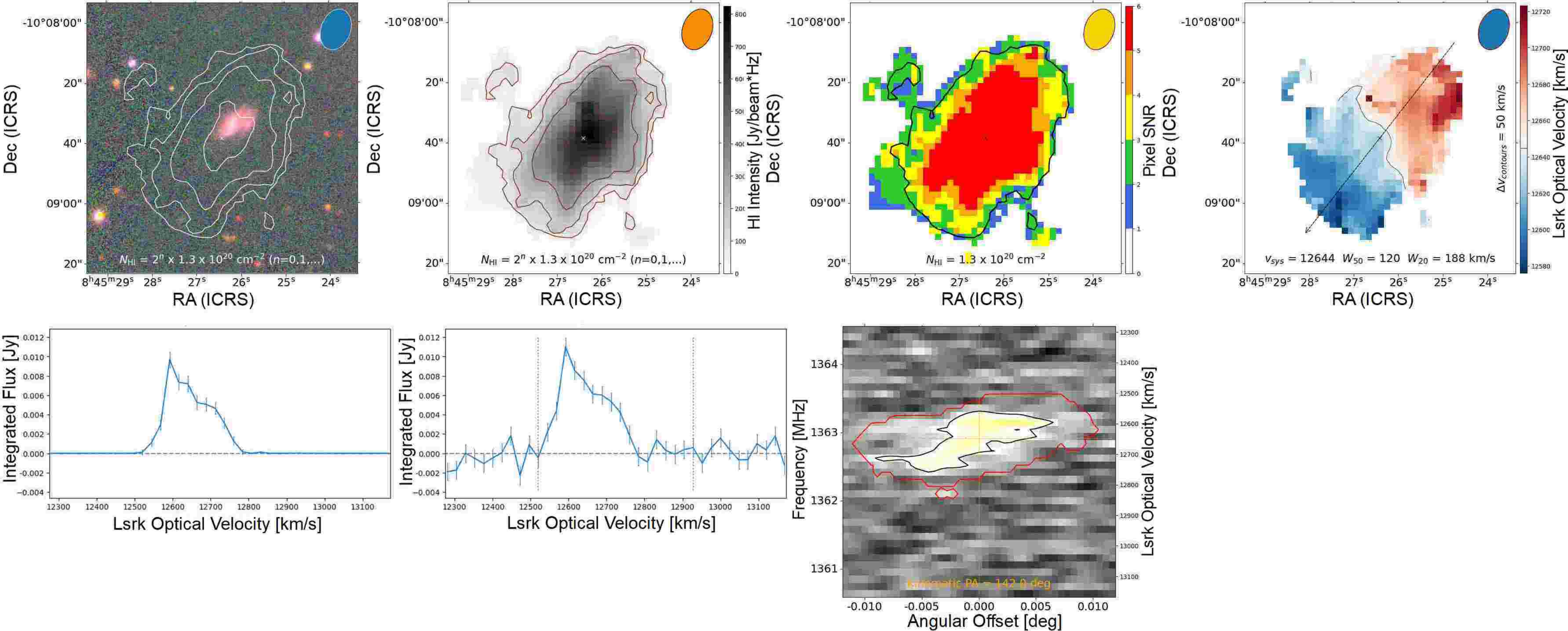}
\includegraphics[width=1.16\textwidth]{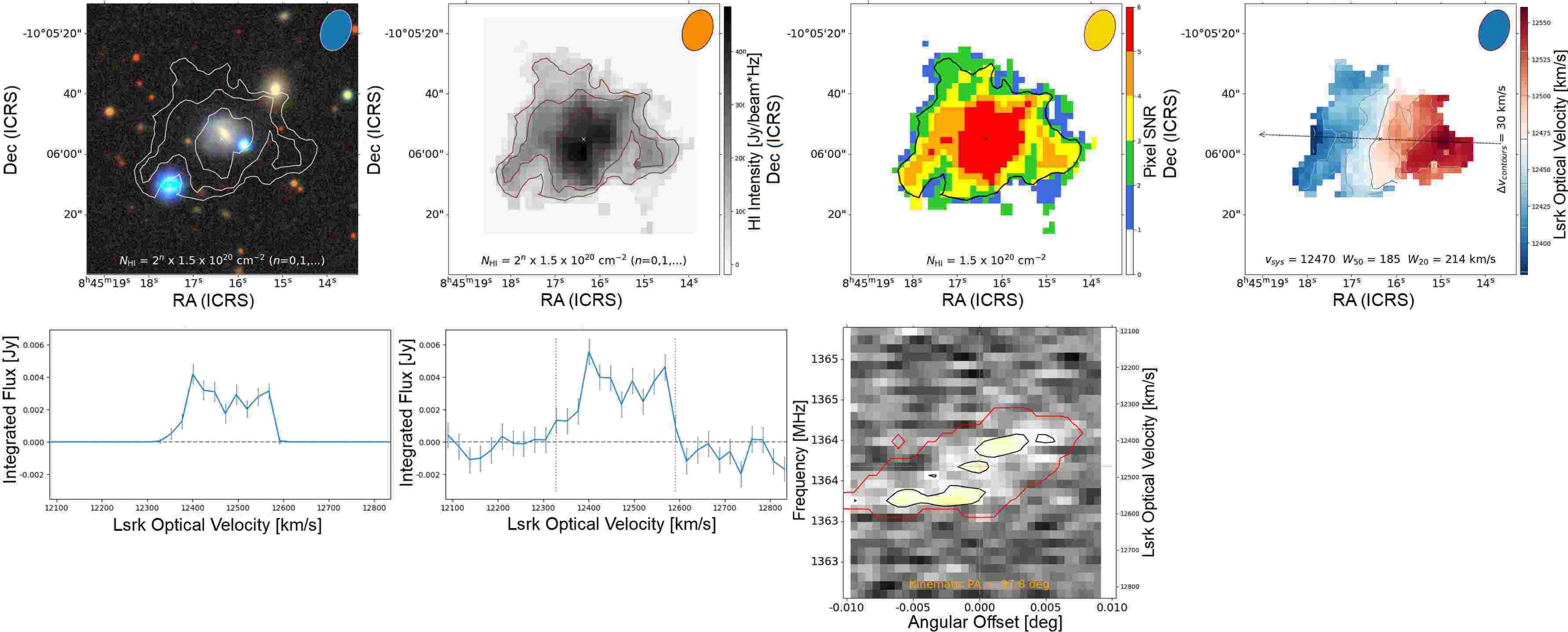}
\caption{Continued. SIP outputs for galaxy IDs 11 and 12.}
\end{figure}
\end{landscape}

\begin{landscape}
\begin{figure}
\ContinuedFloat
\includegraphics[width=1.16\textwidth]{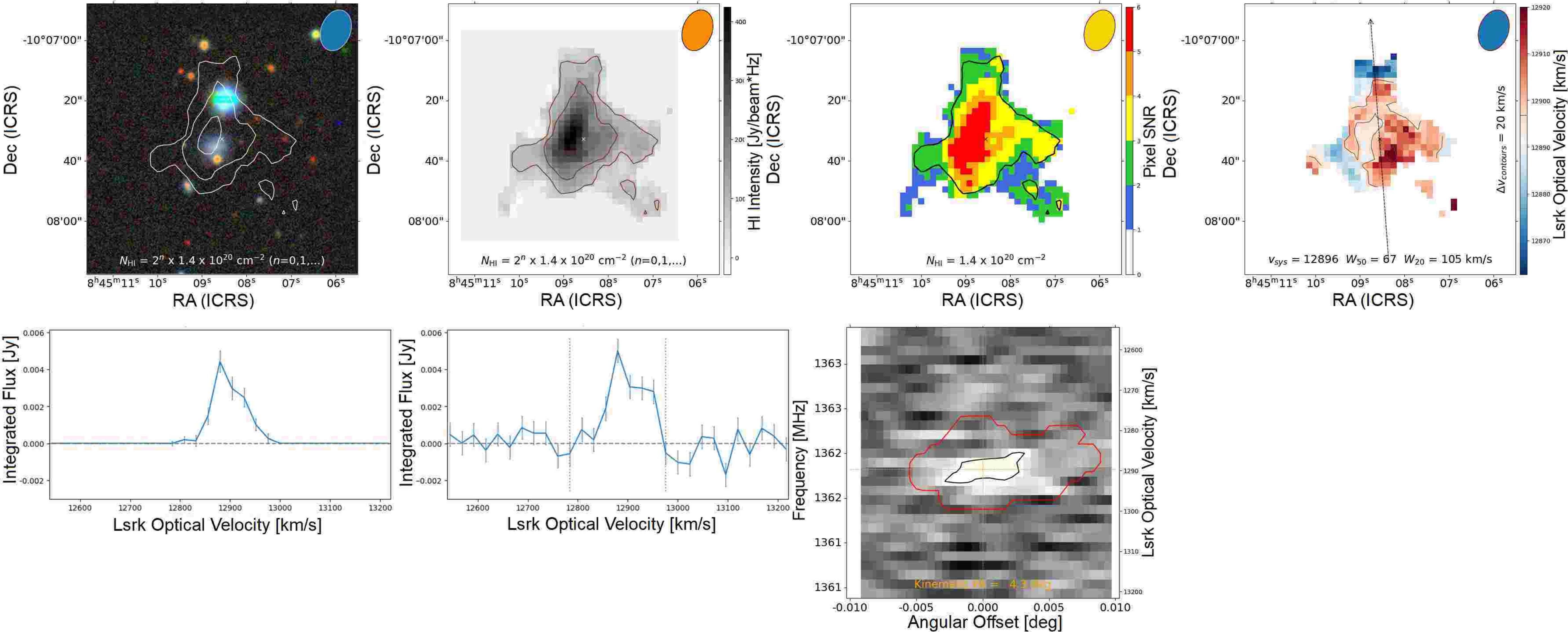}
\includegraphics[width=1.16\textwidth]{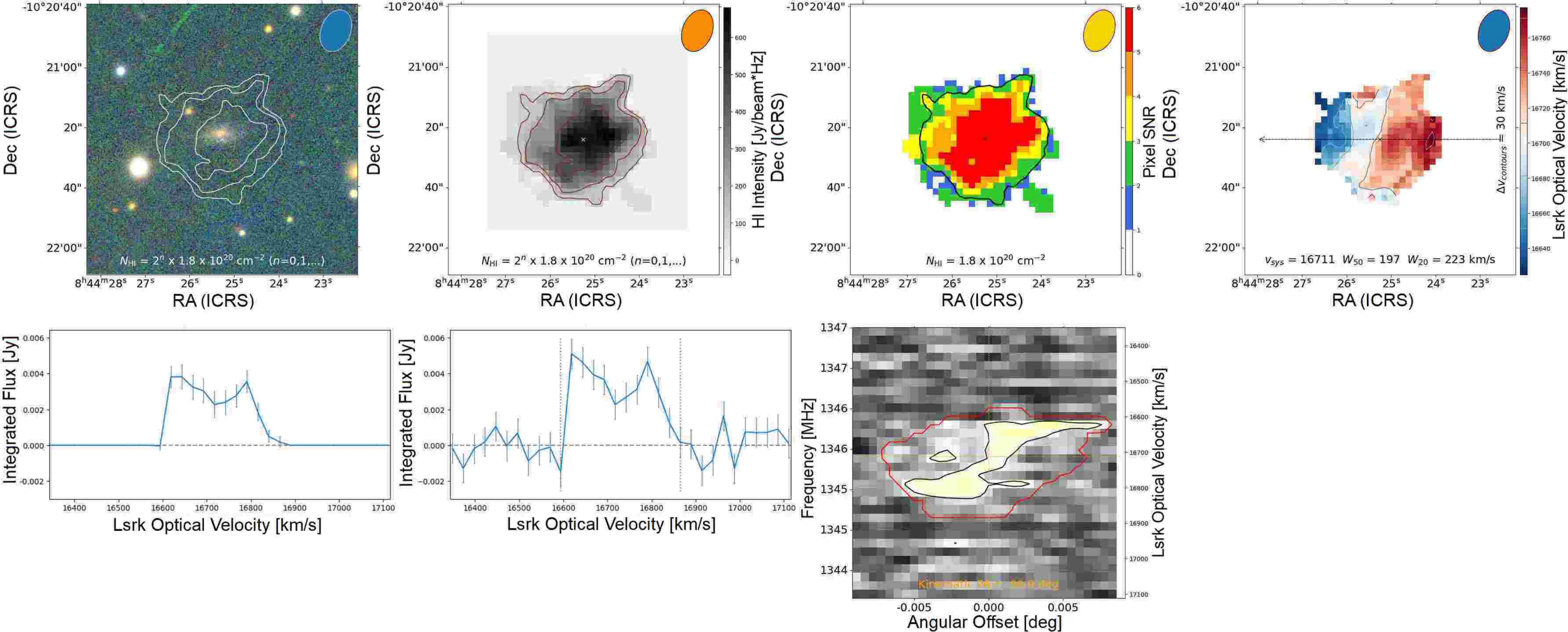}
\caption{Continued. SIP outputs for galaxy IDs 13 and 14.}
\end{figure}
\end{landscape}

\begin{landscape}
\begin{figure}
\ContinuedFloat
\includegraphics[width=1.16\textwidth]{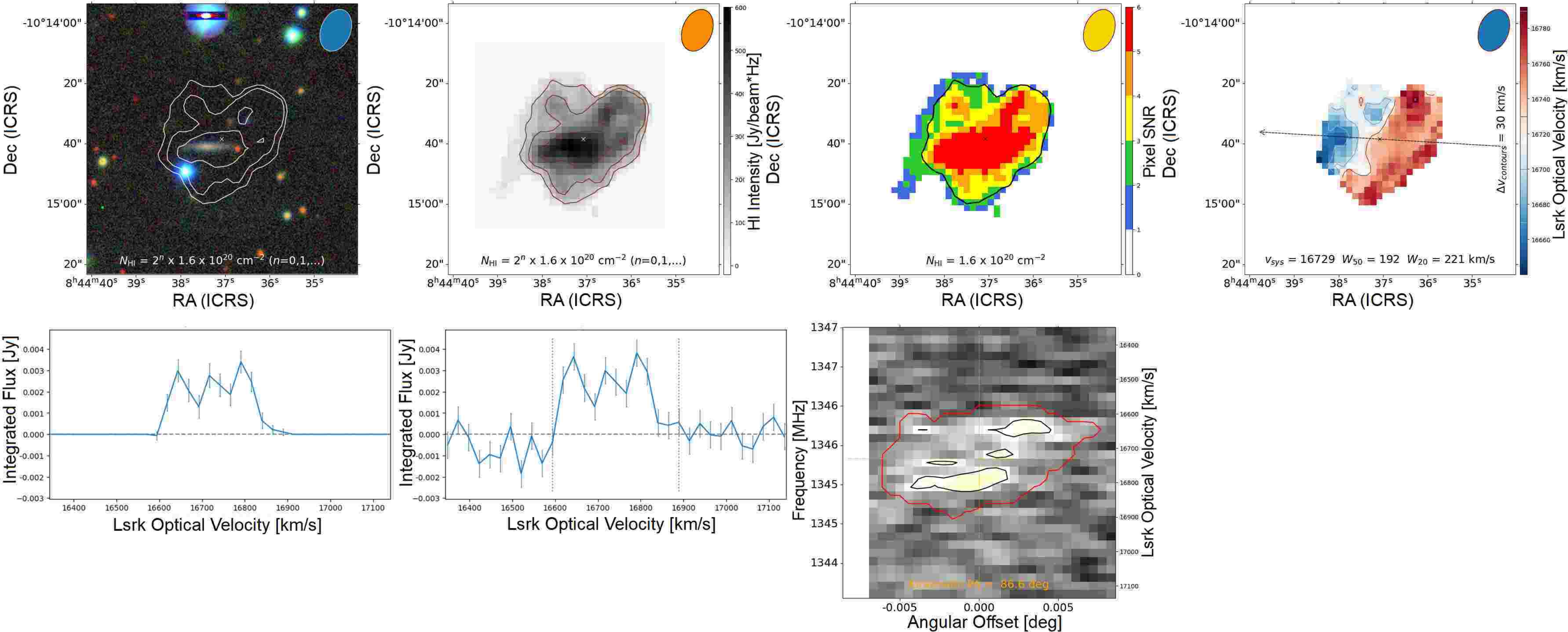}
\includegraphics[width=1.16\textwidth]{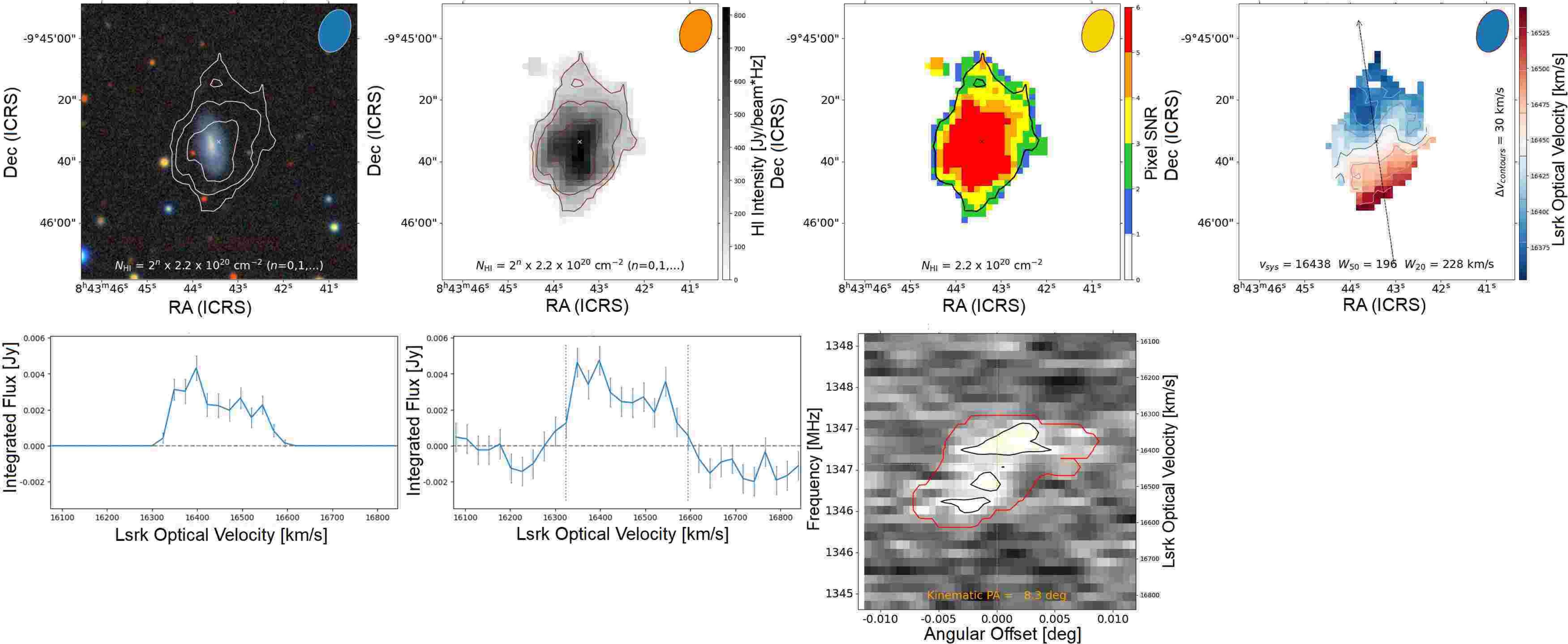}
\caption{Continued. SIP outputs for galaxy IDs 15 and 16.}
\end{figure}
\end{landscape}

\begin{landscape}
\begin{figure}
\ContinuedFloat
\includegraphics[width=1.16\textwidth]{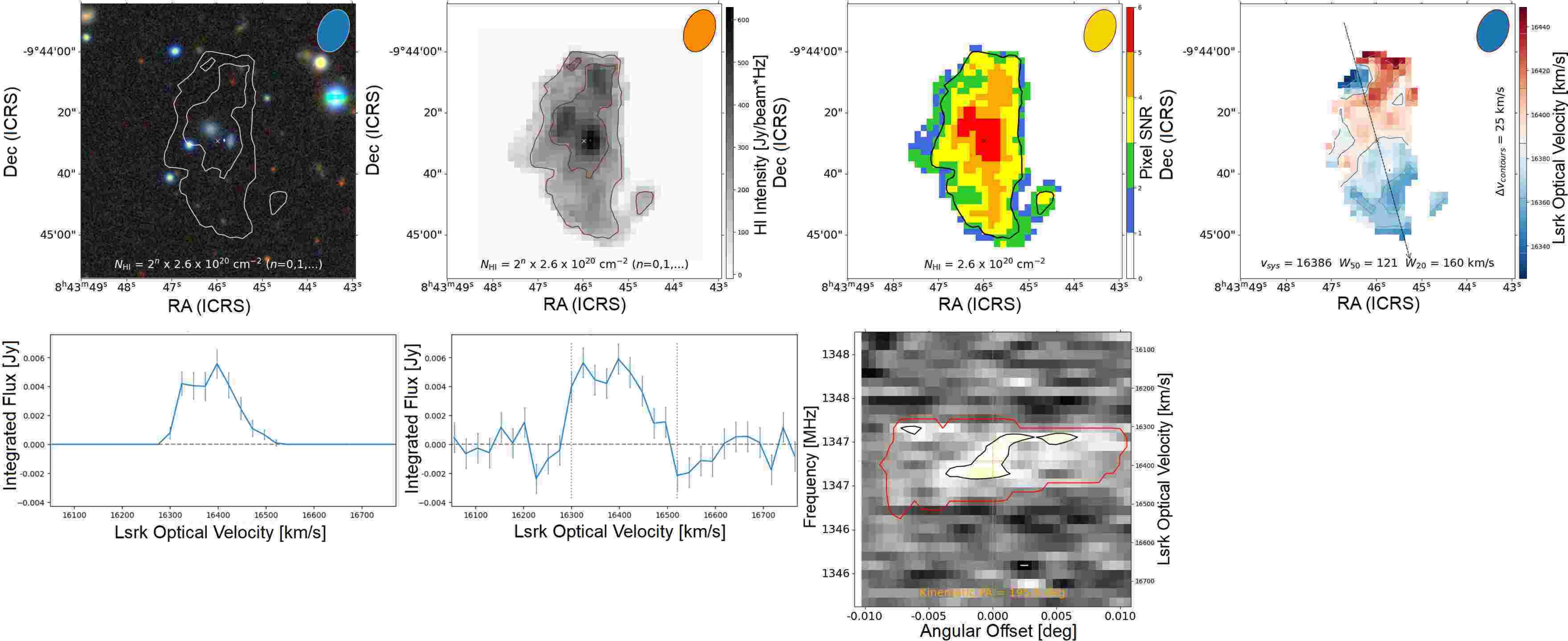}
\includegraphics[width=1.16\textwidth]{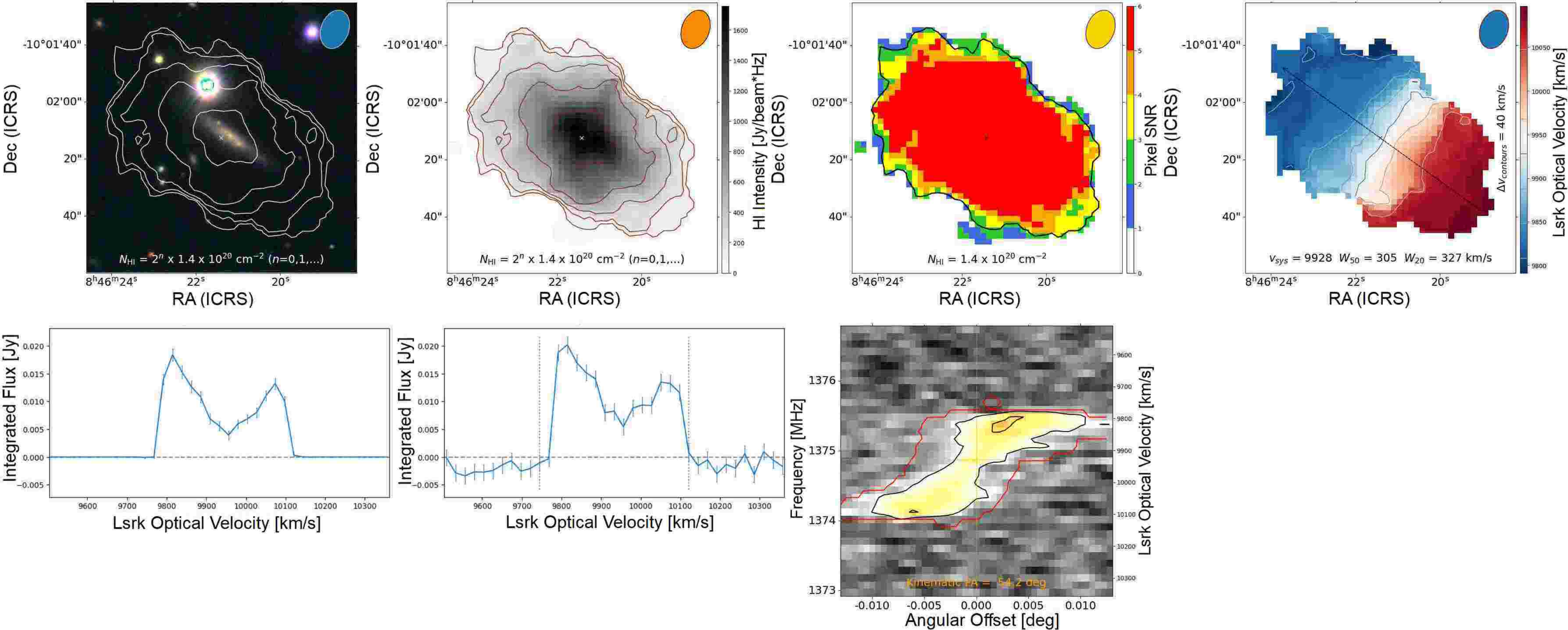}
\caption{Continued. SIP outputs for galaxy IDs 17 and 18.}
\end{figure}
\end{landscape}

\begin{landscape}
\begin{figure}
\ContinuedFloat
\includegraphics[width=1.16\textwidth]{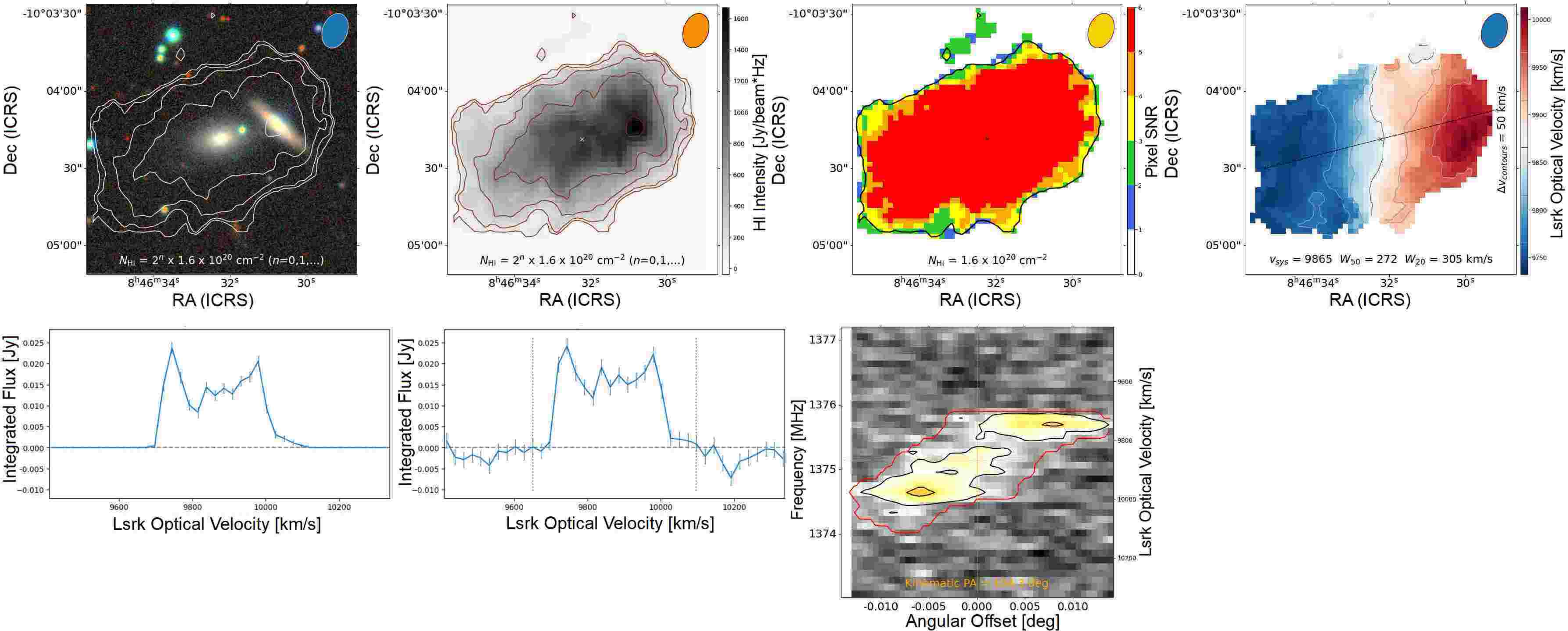}
\includegraphics[width=1.16\textwidth]{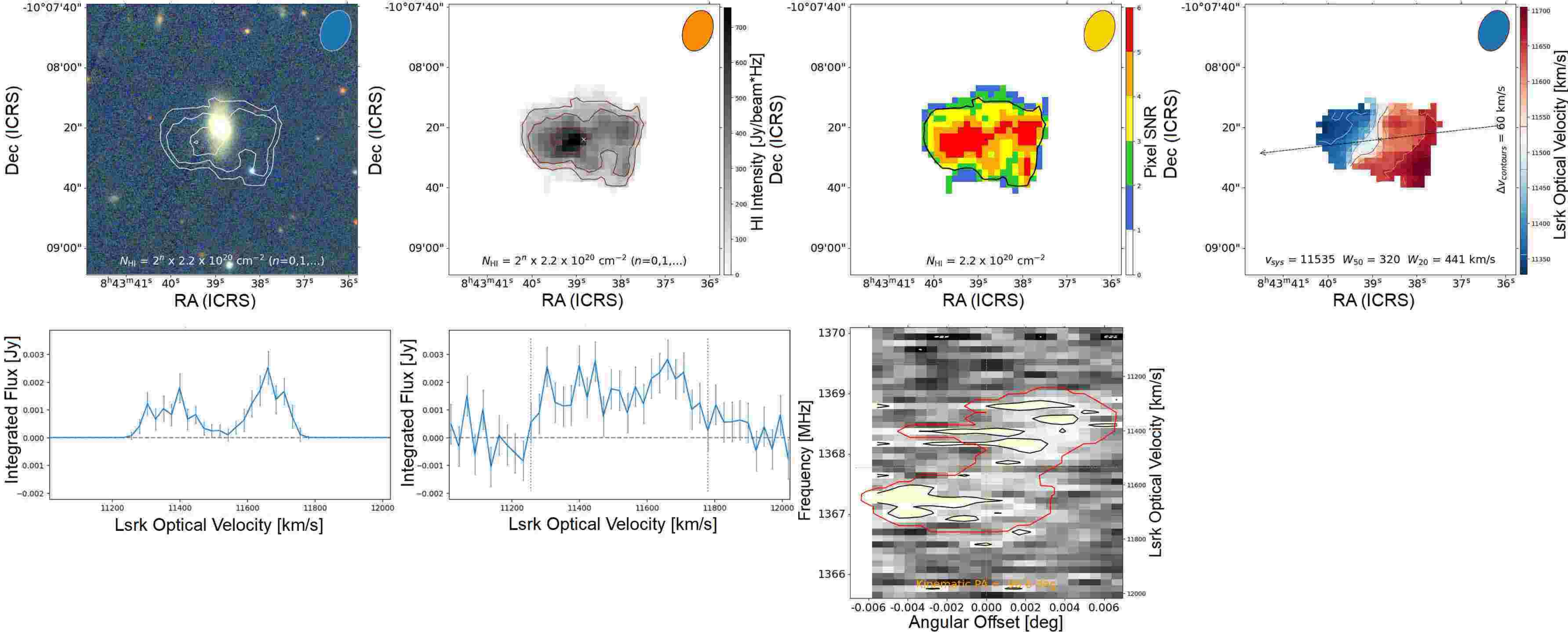}
\caption{Continued. SIP outputs for galaxy IDs 19 and 20.}
\end{figure}
\end{landscape}

\begin{landscape}
\begin{figure}
\ContinuedFloat
\includegraphics[width=1.16\textwidth]{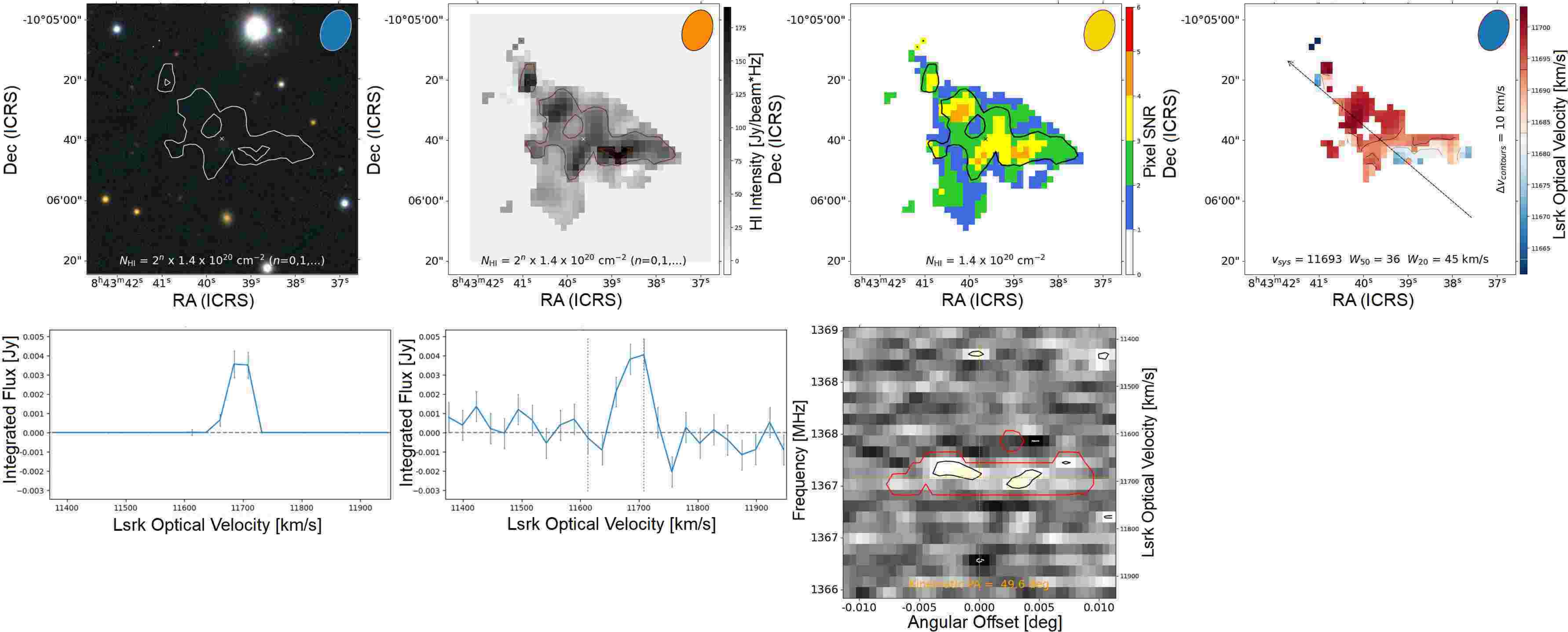}
\includegraphics[width=1.16\textwidth]{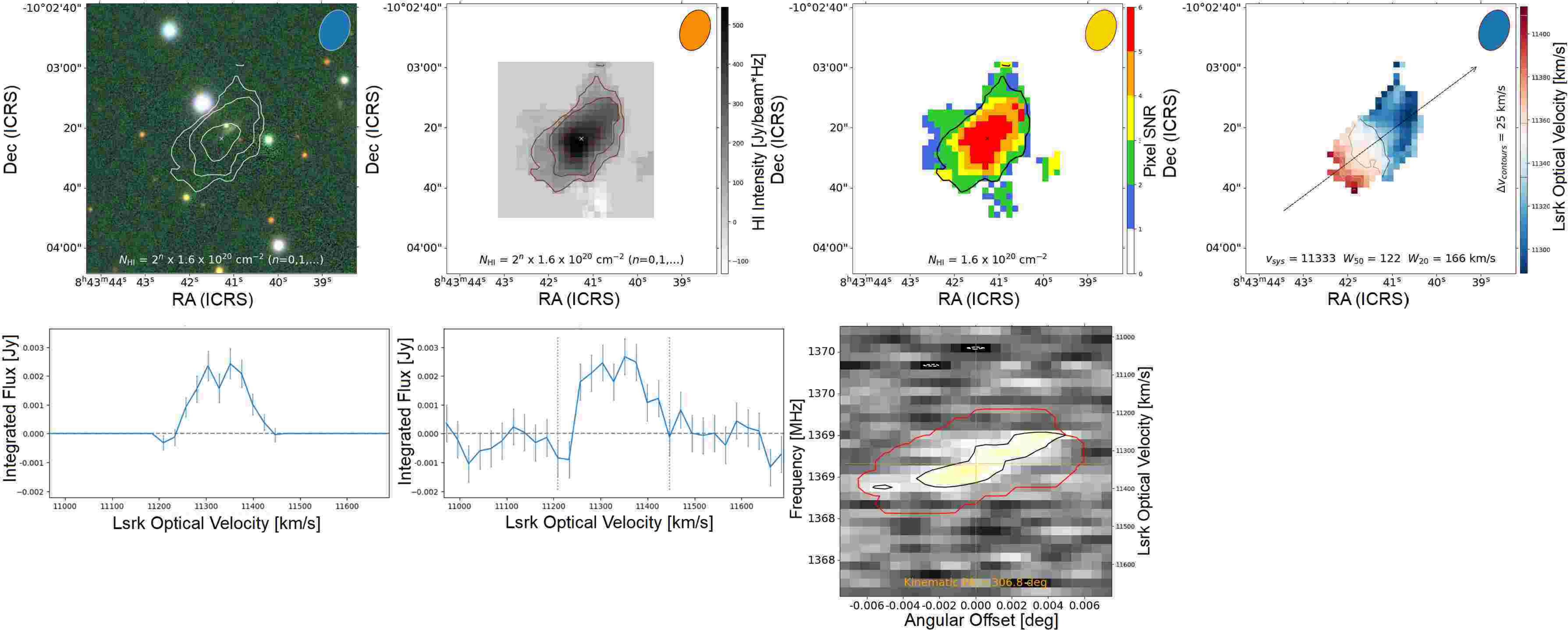}
\caption{Continued. SIP outputs for galaxy IDs 21 and 22.}
\end{figure}
\end{landscape}

\begin{landscape}
\begin{figure}
\ContinuedFloat
\includegraphics[width=1.16\textwidth]{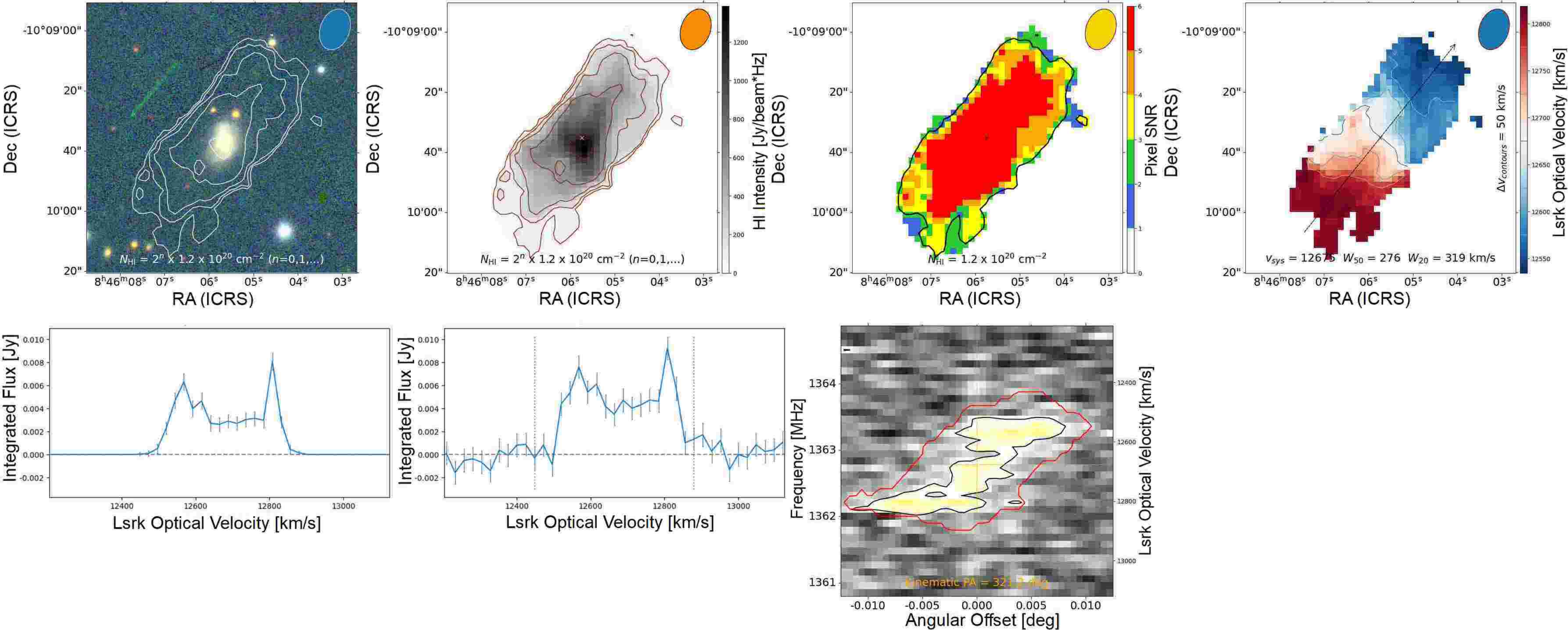}
\includegraphics[width=1.16\textwidth]{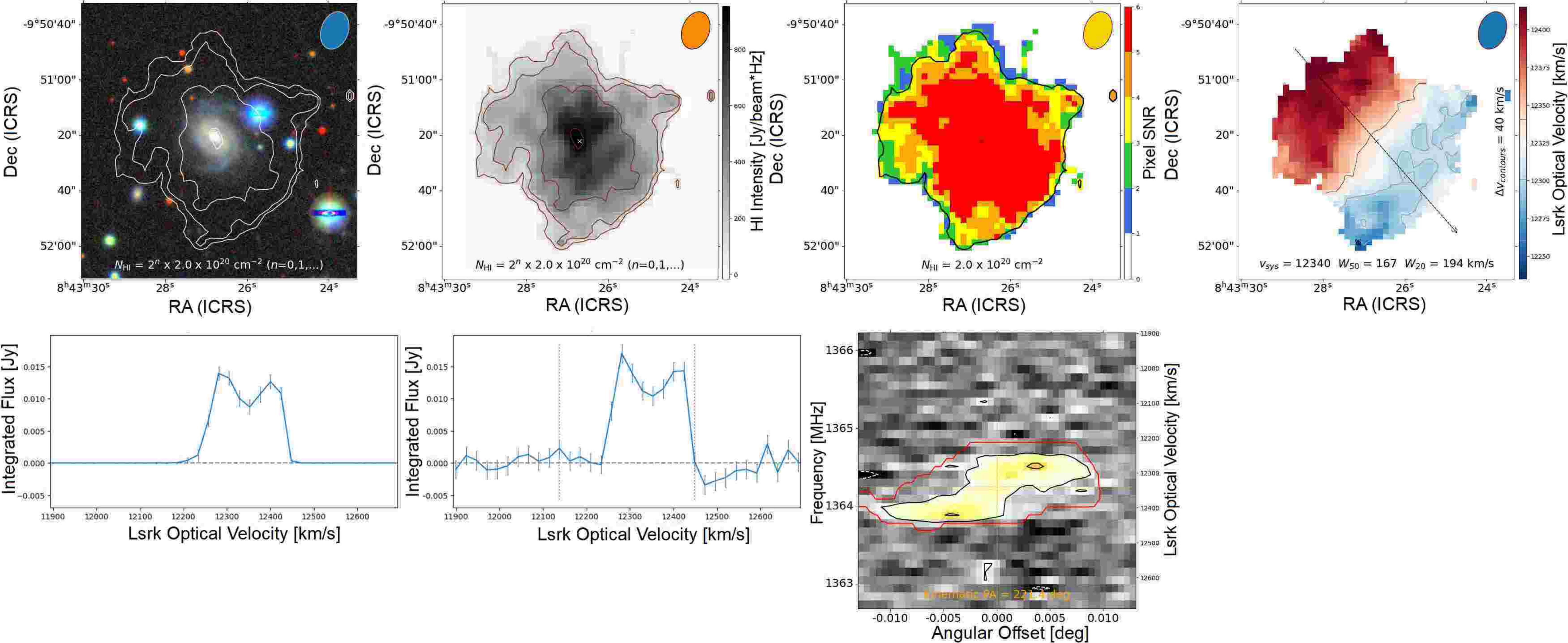}
\caption{Continued. SIP outputs for galaxy IDs 24 and 25.}
\end{figure}
\end{landscape}

\begin{landscape}
\begin{figure}
\ContinuedFloat
\includegraphics[width=1.16\textwidth]{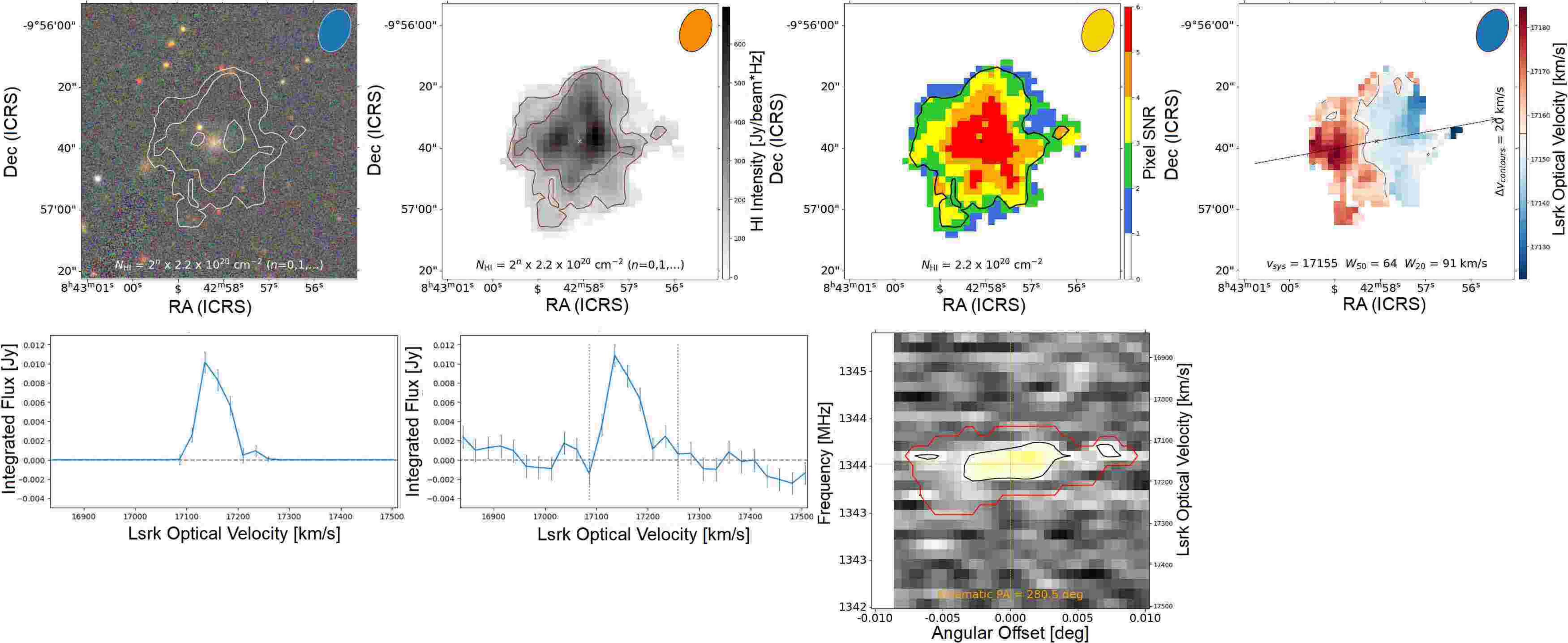}
\includegraphics[width=1.16\textwidth]{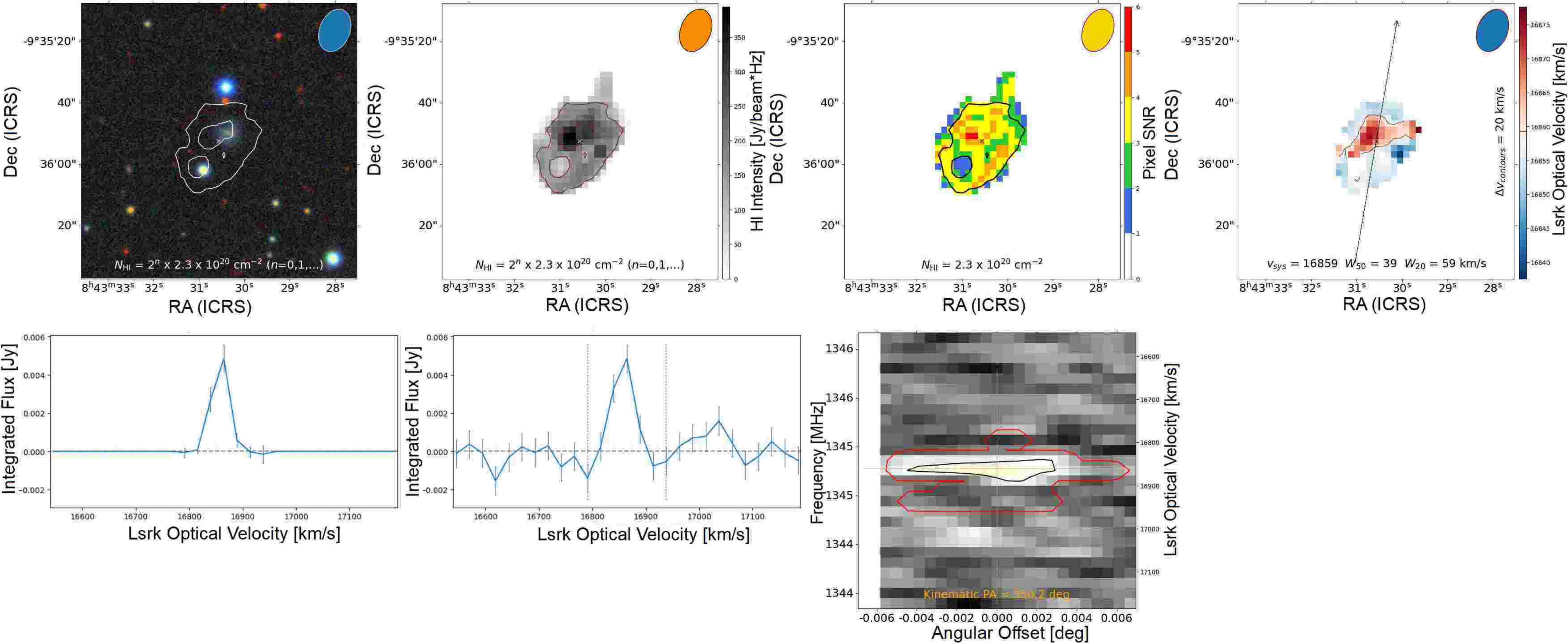}
\caption{Continued. SIP outputs for galaxy IDs 26 and 27.}
\end{figure}
\end{landscape}

\begin{landscape}
\begin{figure}
\ContinuedFloat
\includegraphics[width=1.16\textwidth]{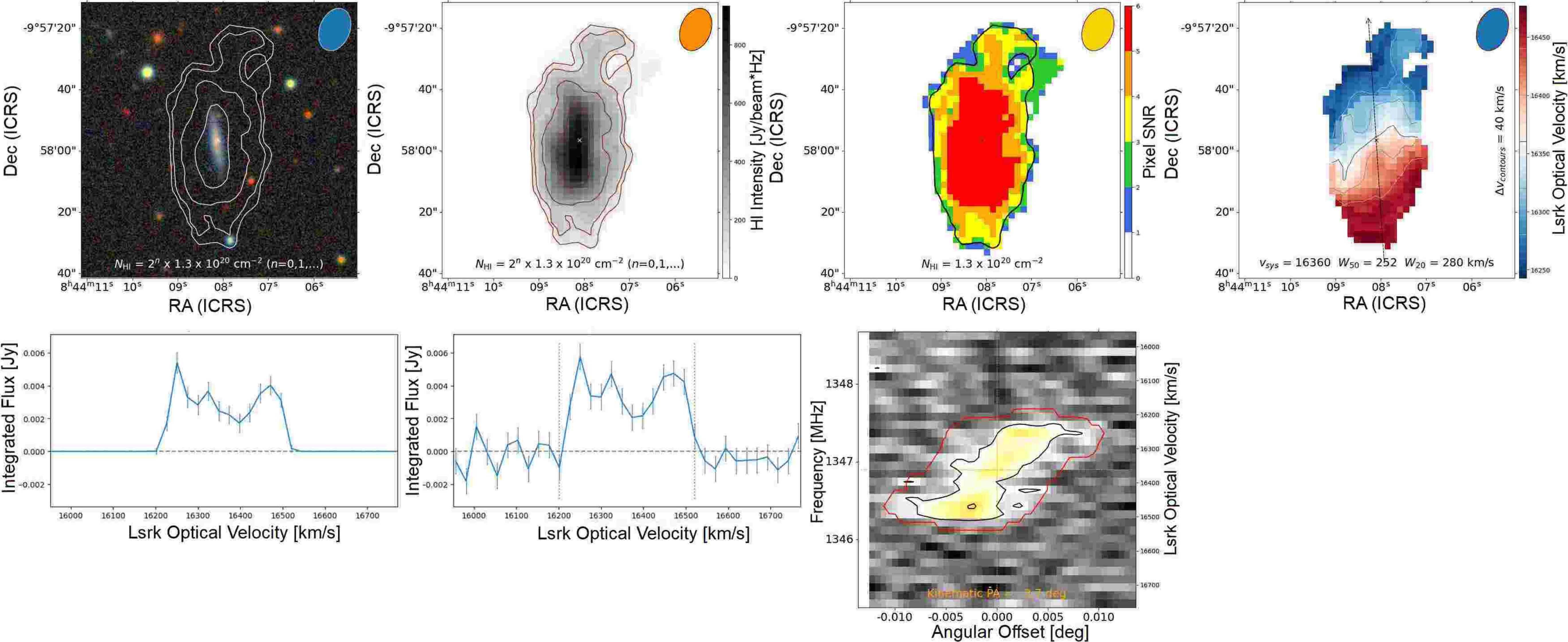}
\includegraphics[width=1.16\textwidth]{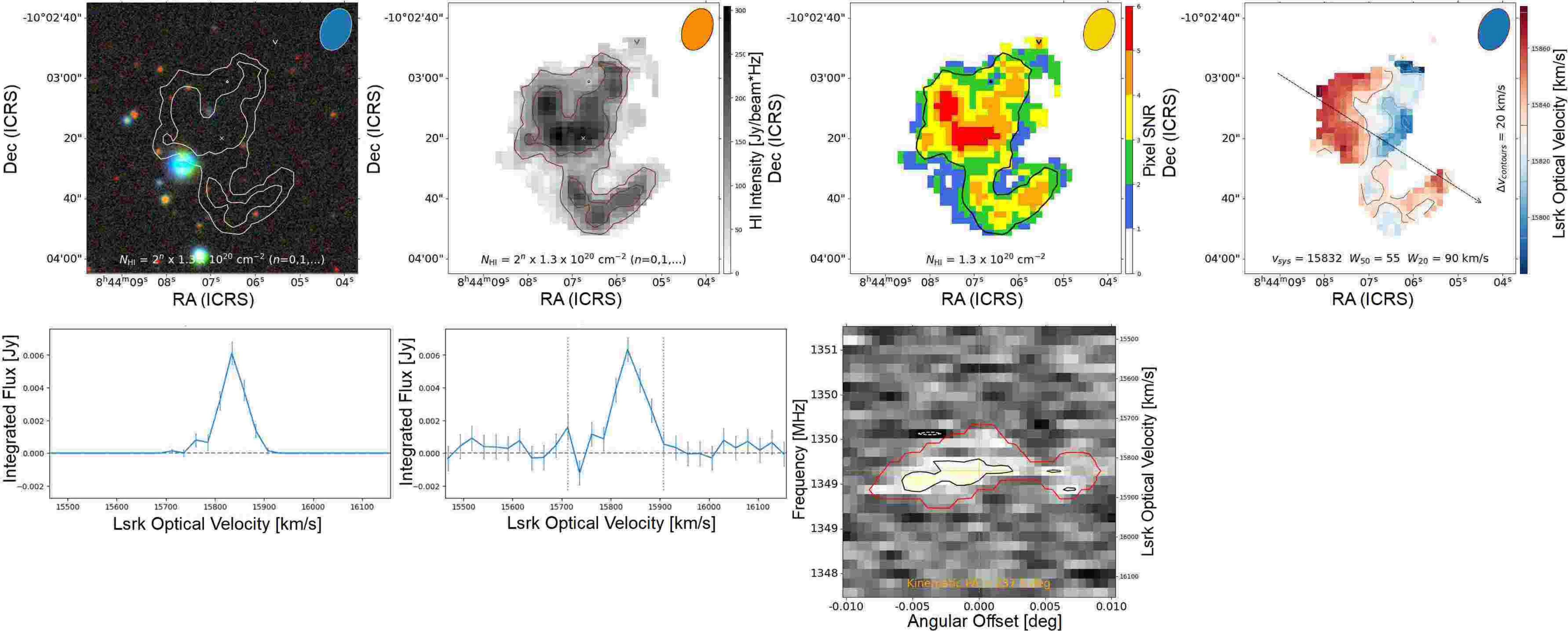}
\caption{Continued. SIP outputs for galaxy IDs 28 and 29.}
\end{figure}
\end{landscape}

\begin{landscape}
\begin{figure}
\ContinuedFloat
\includegraphics[width=1.16\textwidth]{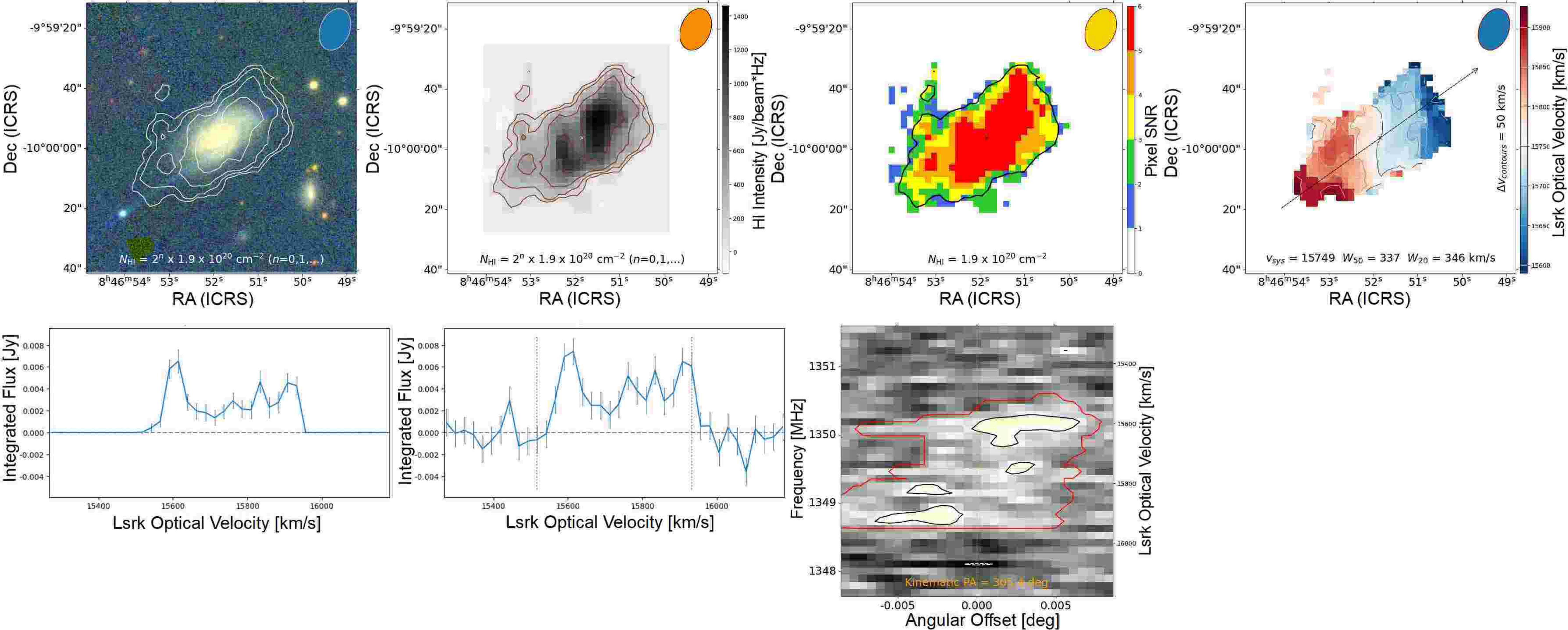}
\includegraphics[width=1.16\textwidth]{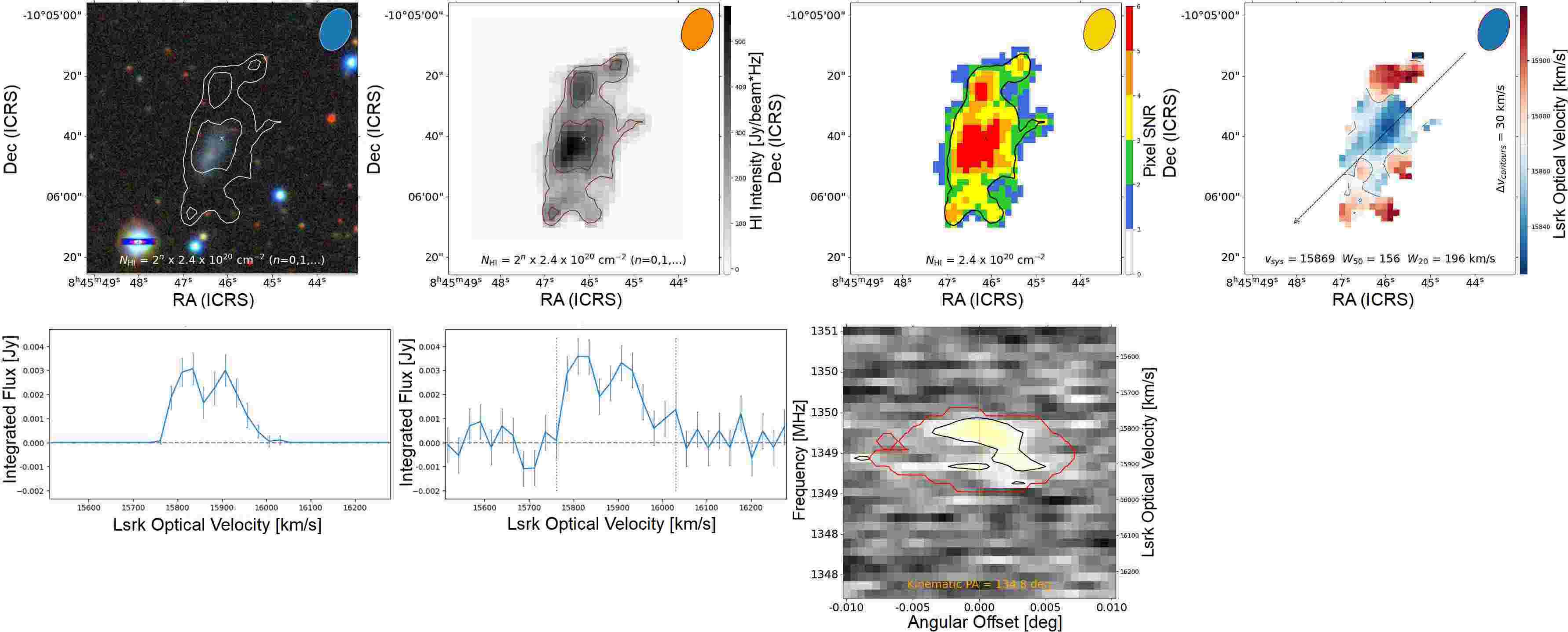}
\caption{Continued. SIP outputs for galaxy IDs 30 and 31.}
\end{figure}
\end{landscape}

\begin{landscape}
\begin{figure}
\ContinuedFloat
\includegraphics[width=1.16\textwidth]{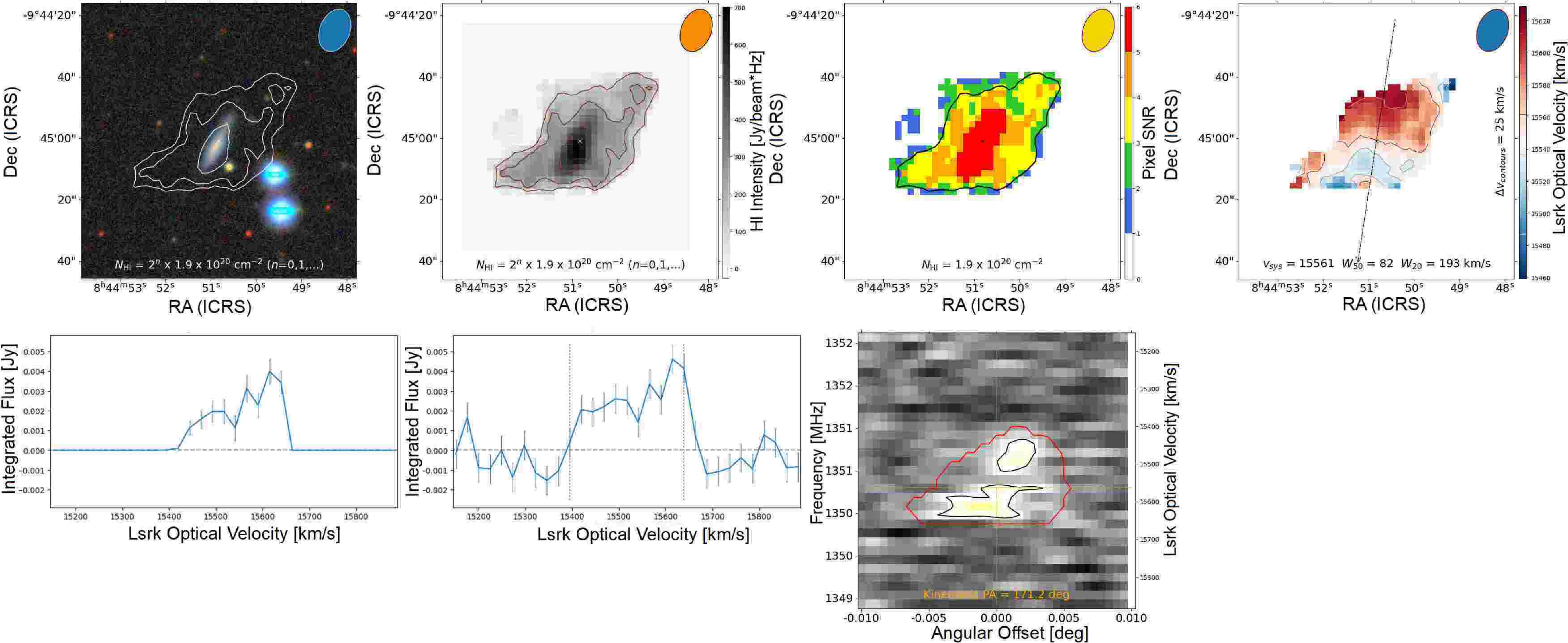}
\includegraphics[width=1.16\textwidth]{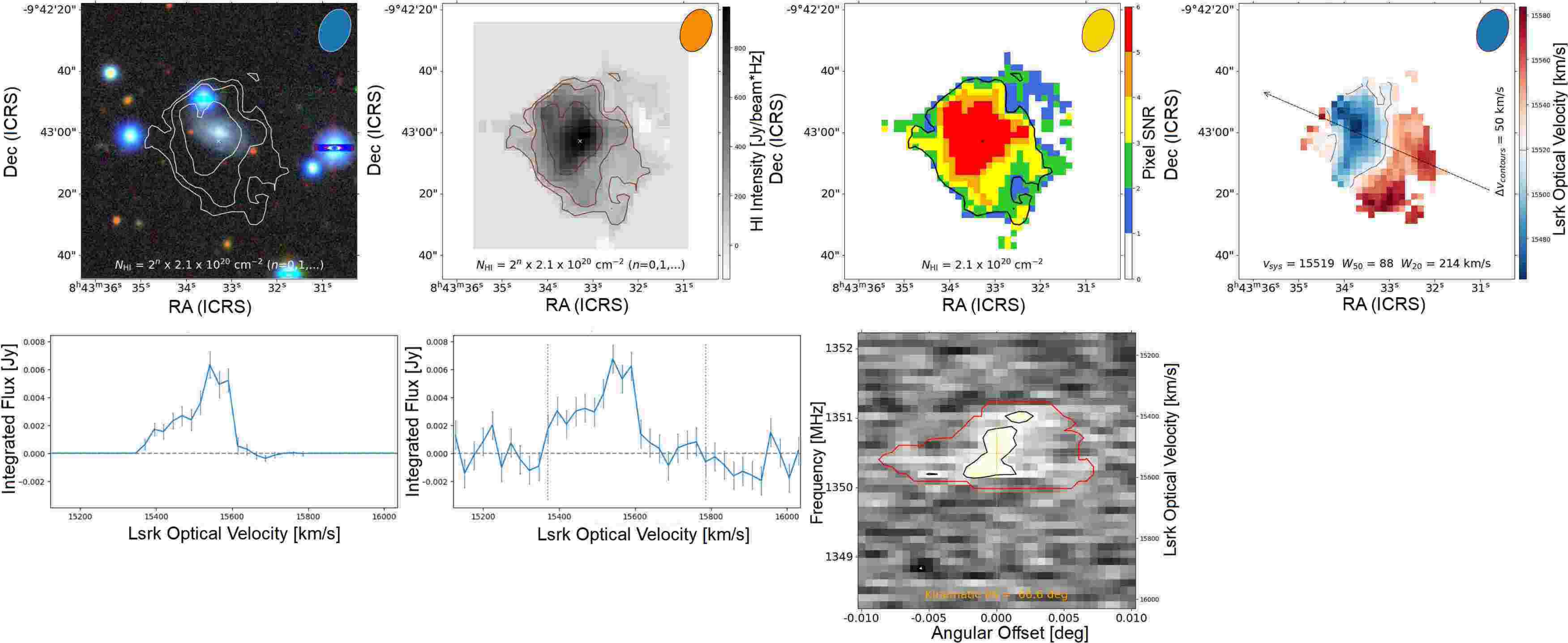}
\caption{Continued. SIP outputs for galaxy IDs 32 and 33.}
\end{figure}
\end{landscape}

\begin{landscape}
\begin{figure}
\ContinuedFloat
\includegraphics[width=1.16\textwidth]{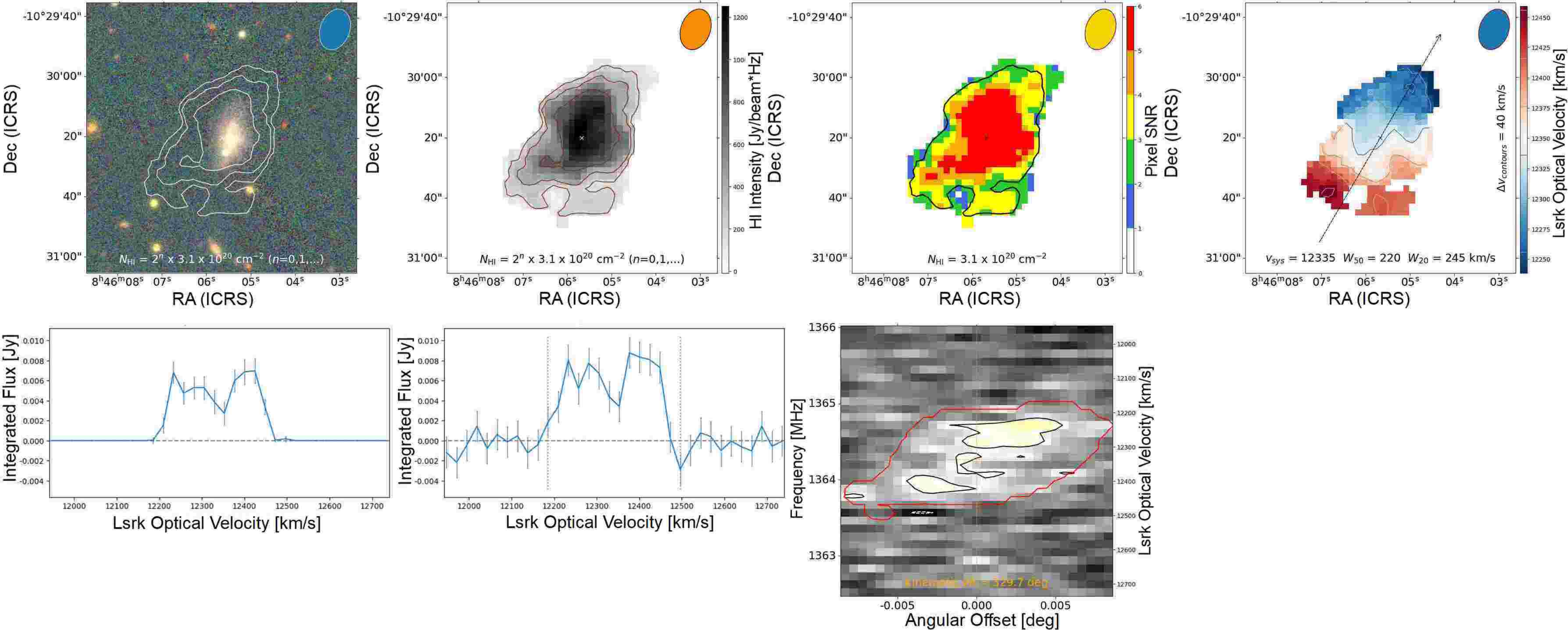}
\includegraphics[width=1.16\textwidth]{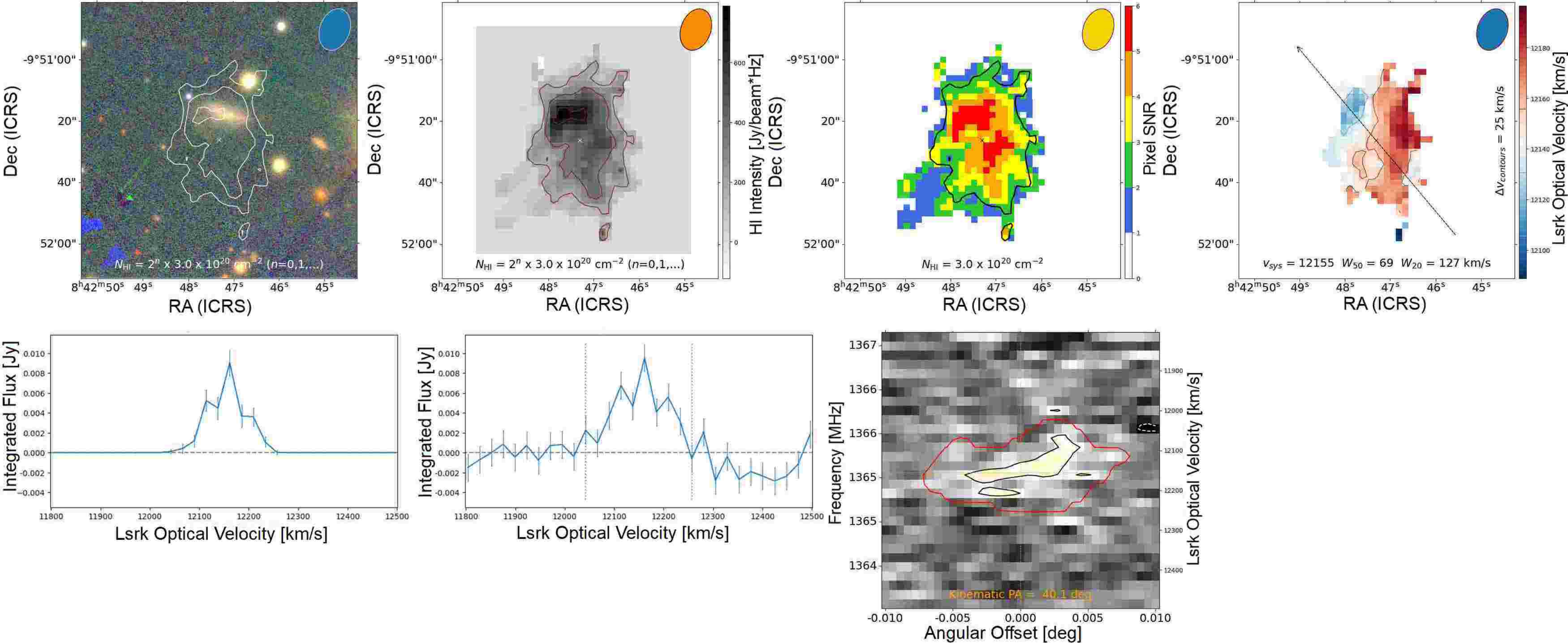}
\caption{Continued. SIP outputs for galaxy IDs 35 and 36.}
\end{figure}
\end{landscape}

\begin{landscape}
\begin{figure}
\ContinuedFloat
\includegraphics[width=1.16\textwidth]{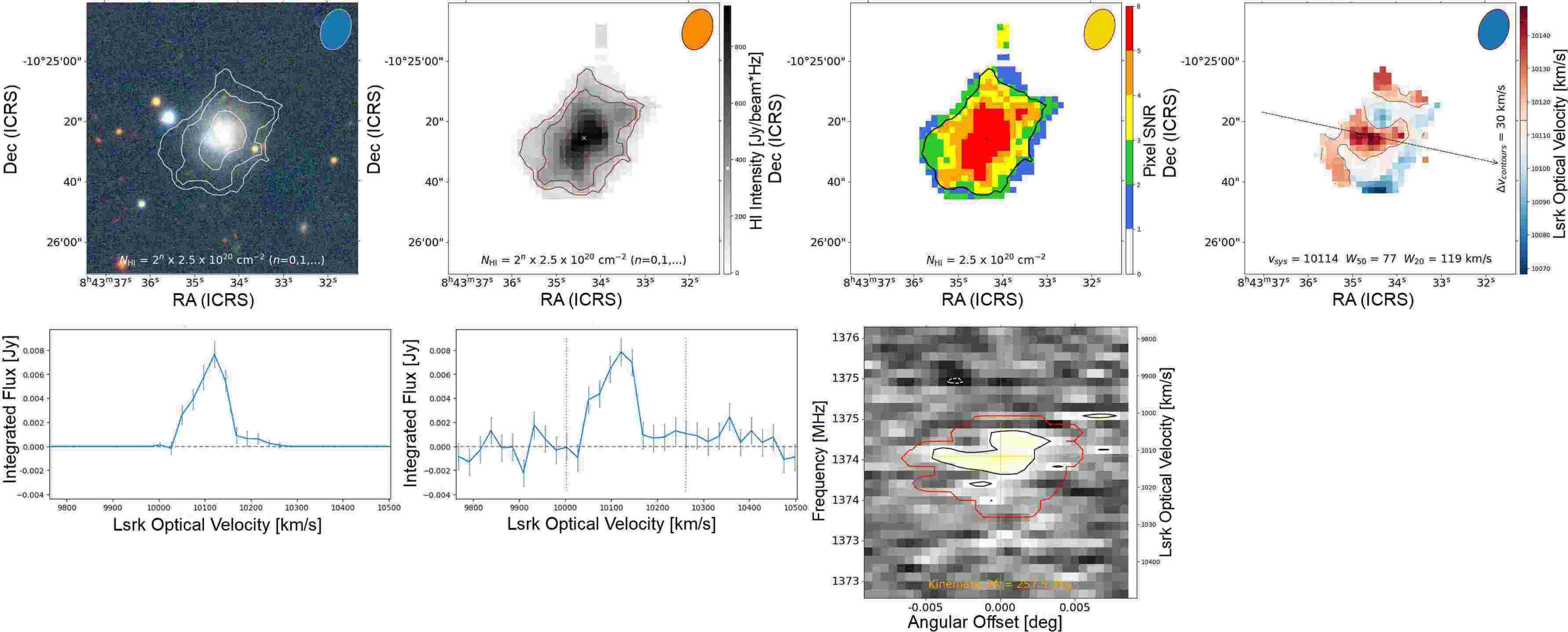}
\includegraphics[width=1.16\textwidth]{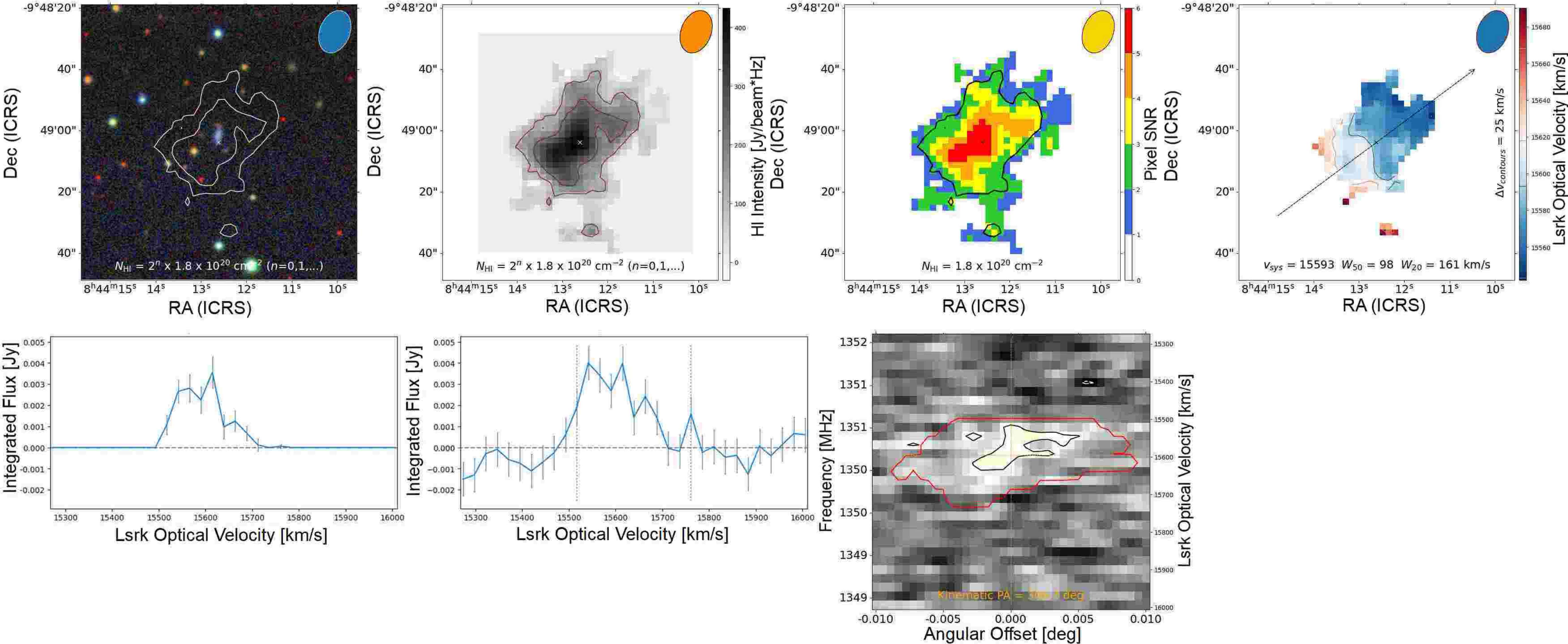}
\caption{Continued. SIP outputs for galaxy IDs 37 and 38.}
\end{figure}
\end{landscape}

\begin{landscape}
\begin{figure}
\ContinuedFloat
\includegraphics[width=1.16\textwidth]{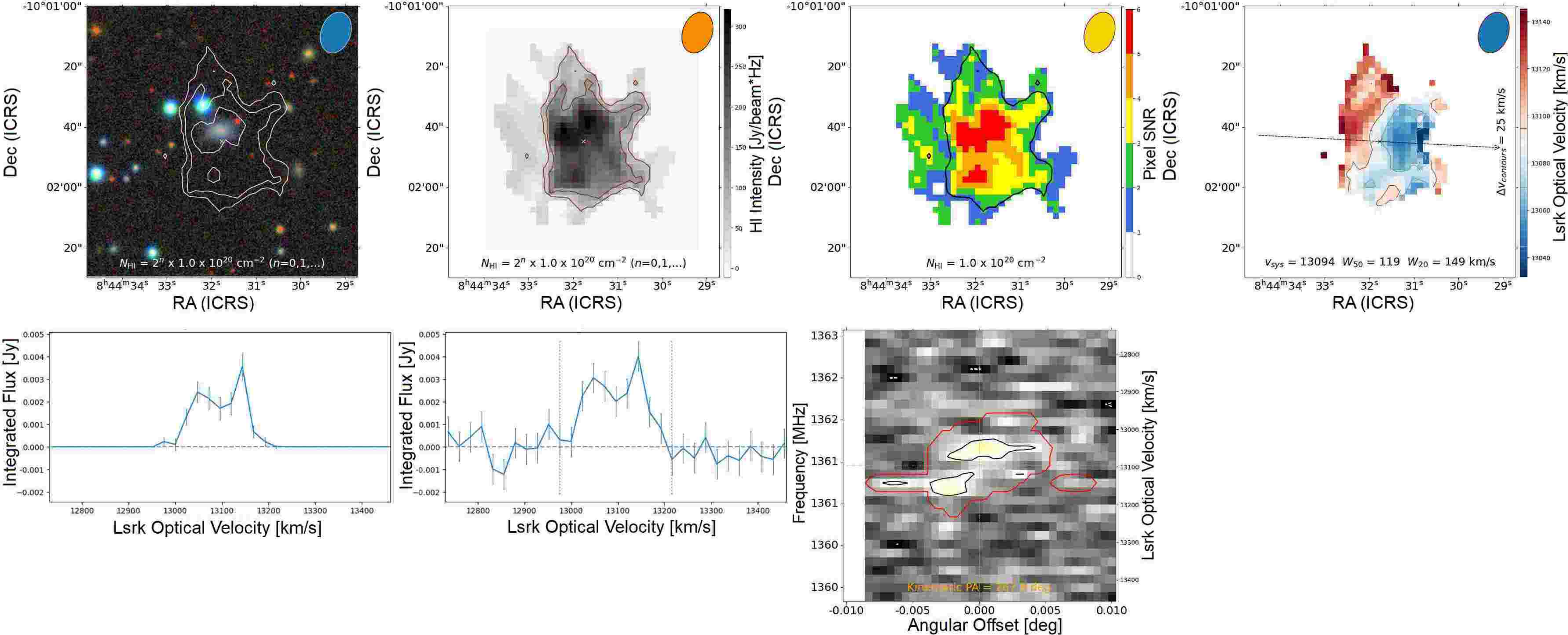}
\includegraphics[width=1.16\textwidth]{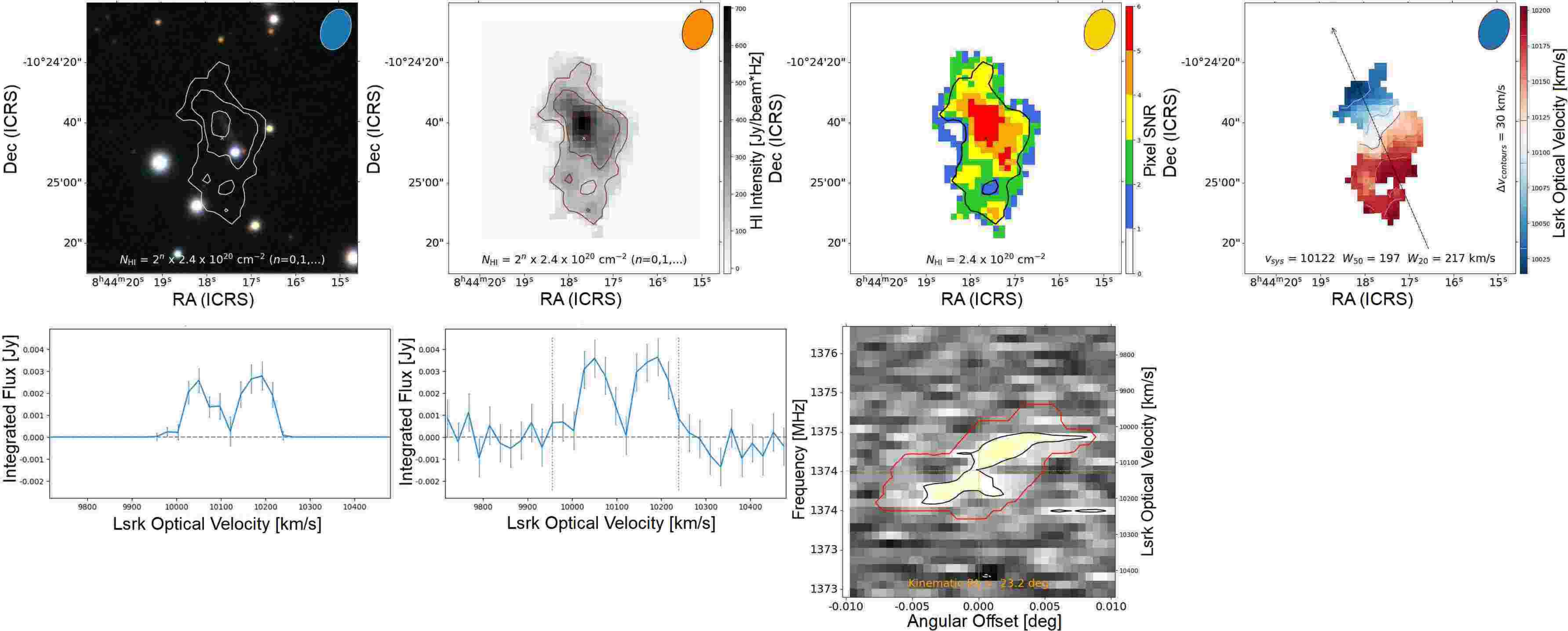}
\caption{Continued. SIP outputs for galaxy IDs 39 and 40.}
\end{figure}
\end{landscape}

\begin{landscape}
\begin{figure}
\ContinuedFloat
\includegraphics[width=1.16\textwidth]{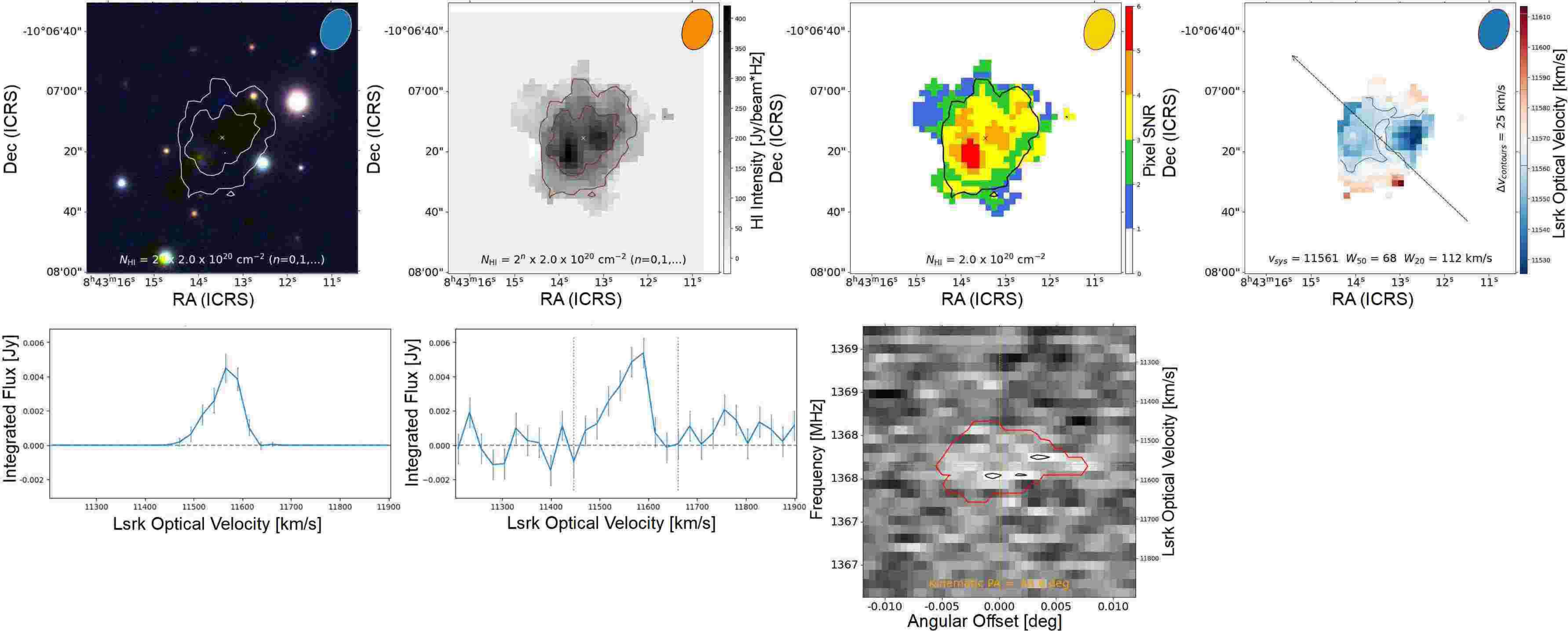}
\includegraphics[width=1.16\textwidth]{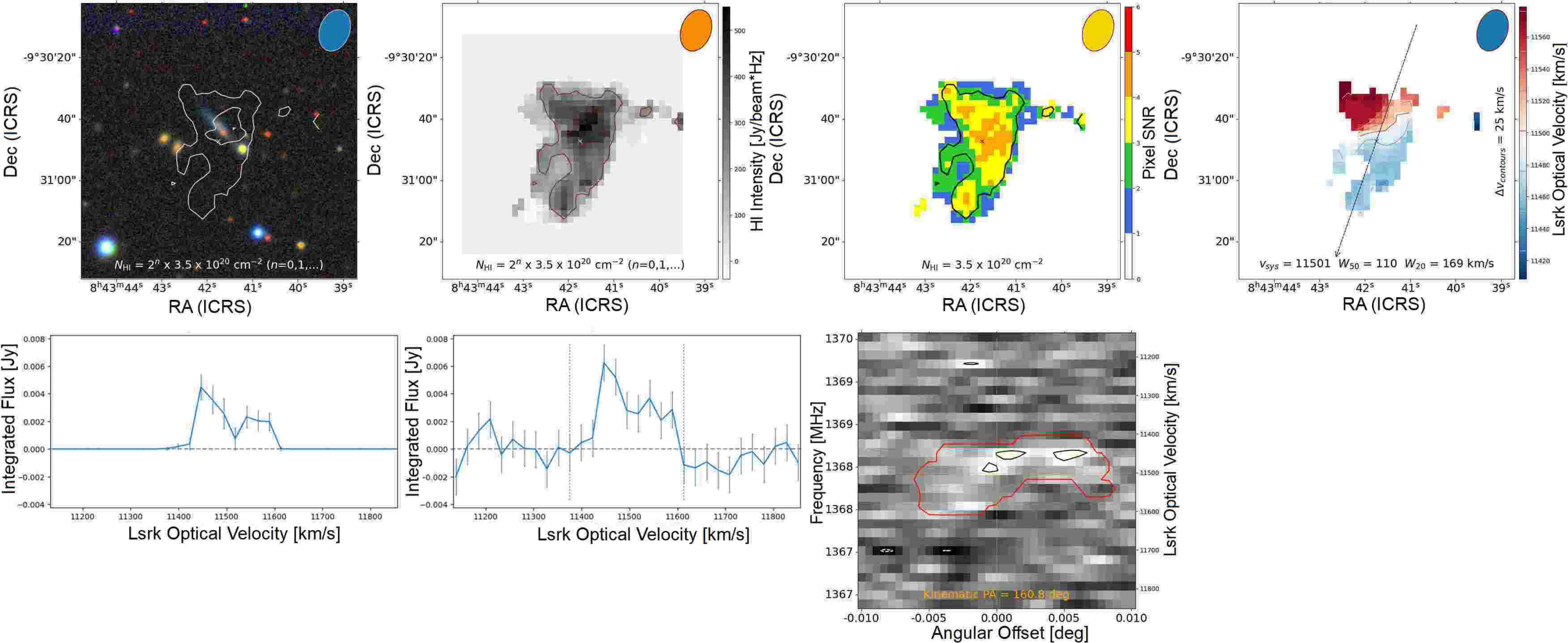}
\caption{Continued. SIP outputs for galaxy IDs 41 and 42.}
\end{figure}
\end{landscape}

\begin{landscape}
\begin{figure}
\ContinuedFloat
\includegraphics[width=1.16\textwidth]{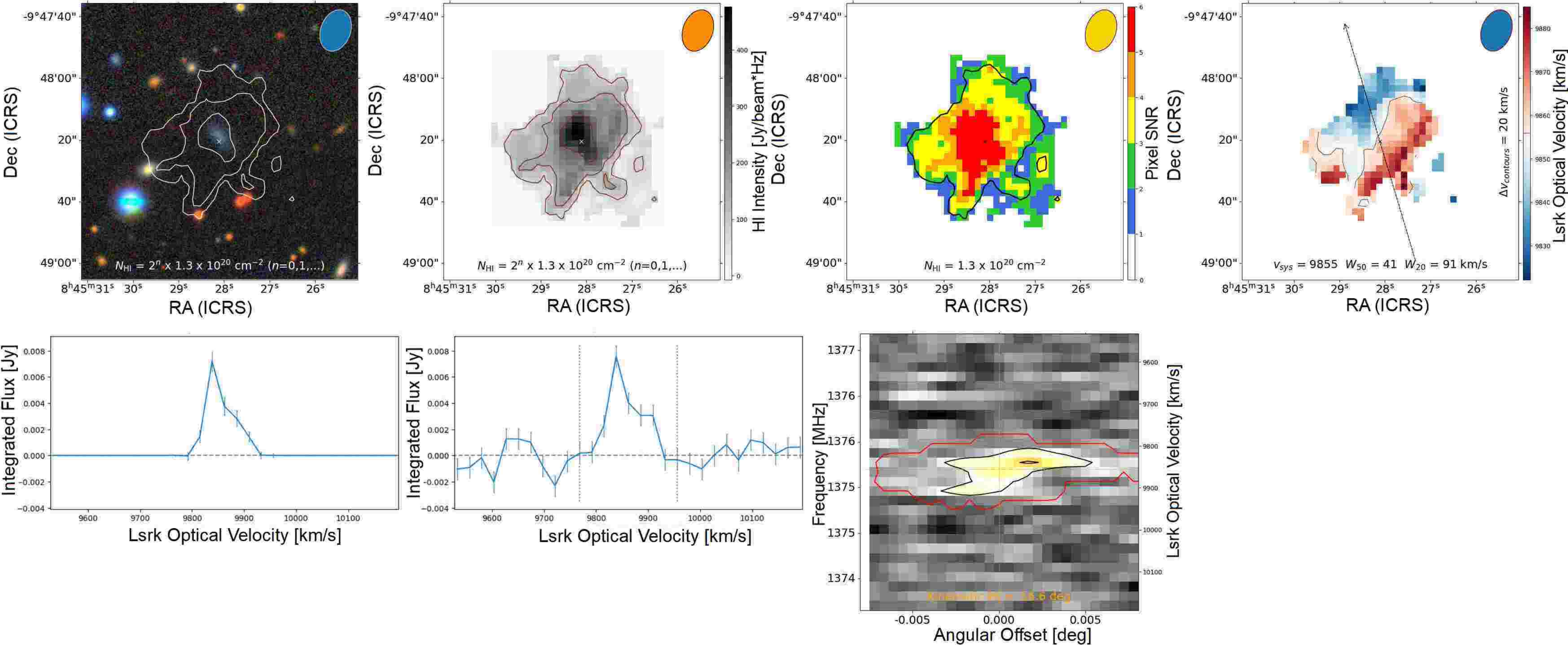}
\includegraphics[width=1.16\textwidth]{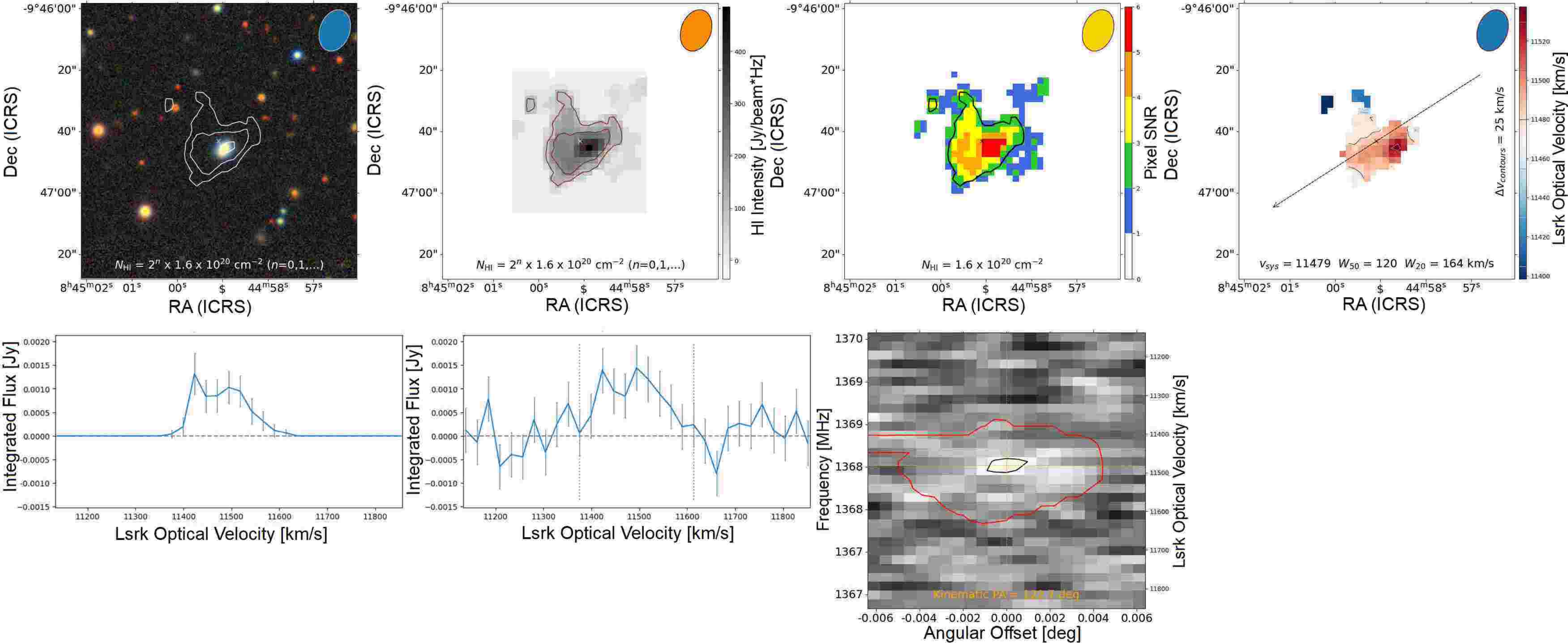}
\caption{Continued. SIP outputs for galaxy IDs 43 and 44. Note ID43 is interacting with ID49 (see final set of plots for this figure).}
\end{figure}
\end{landscape}

\begin{landscape}
\begin{figure}
\ContinuedFloat
\includegraphics[width=1.16\textwidth]{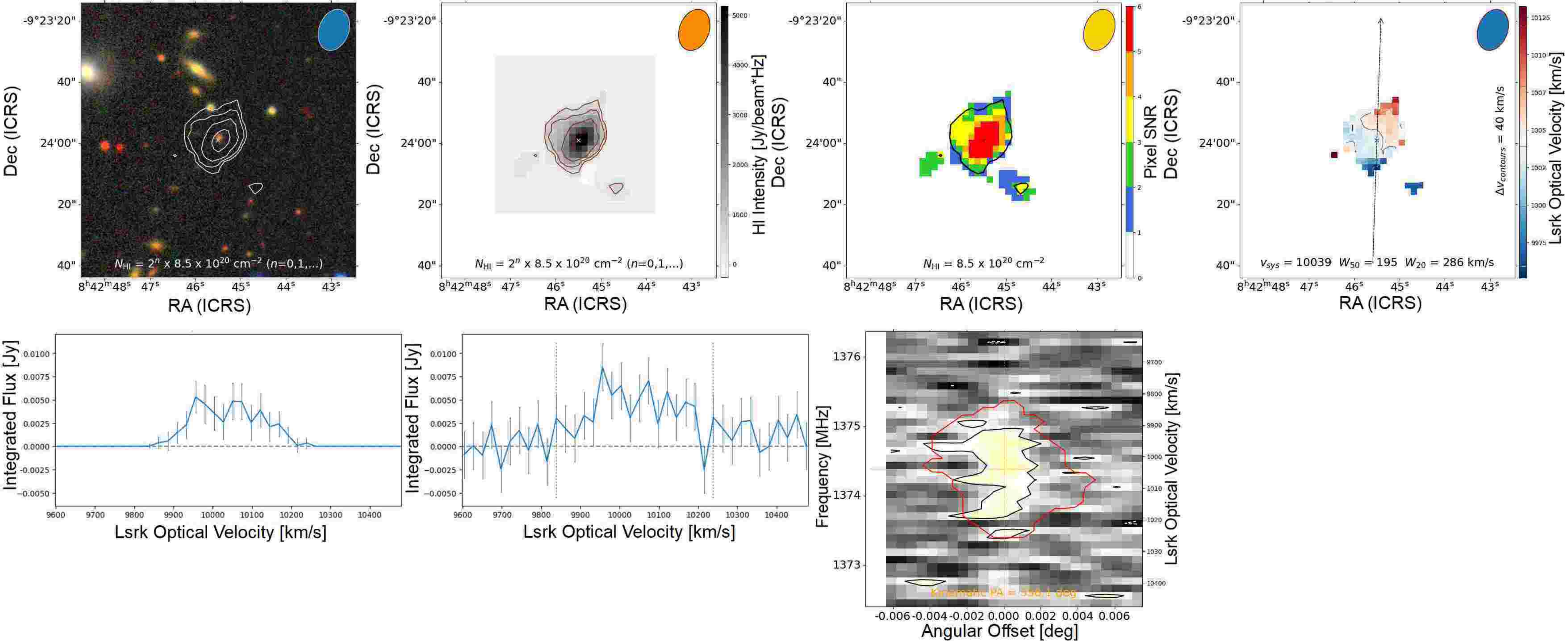}
\includegraphics[width=1.16\textwidth]{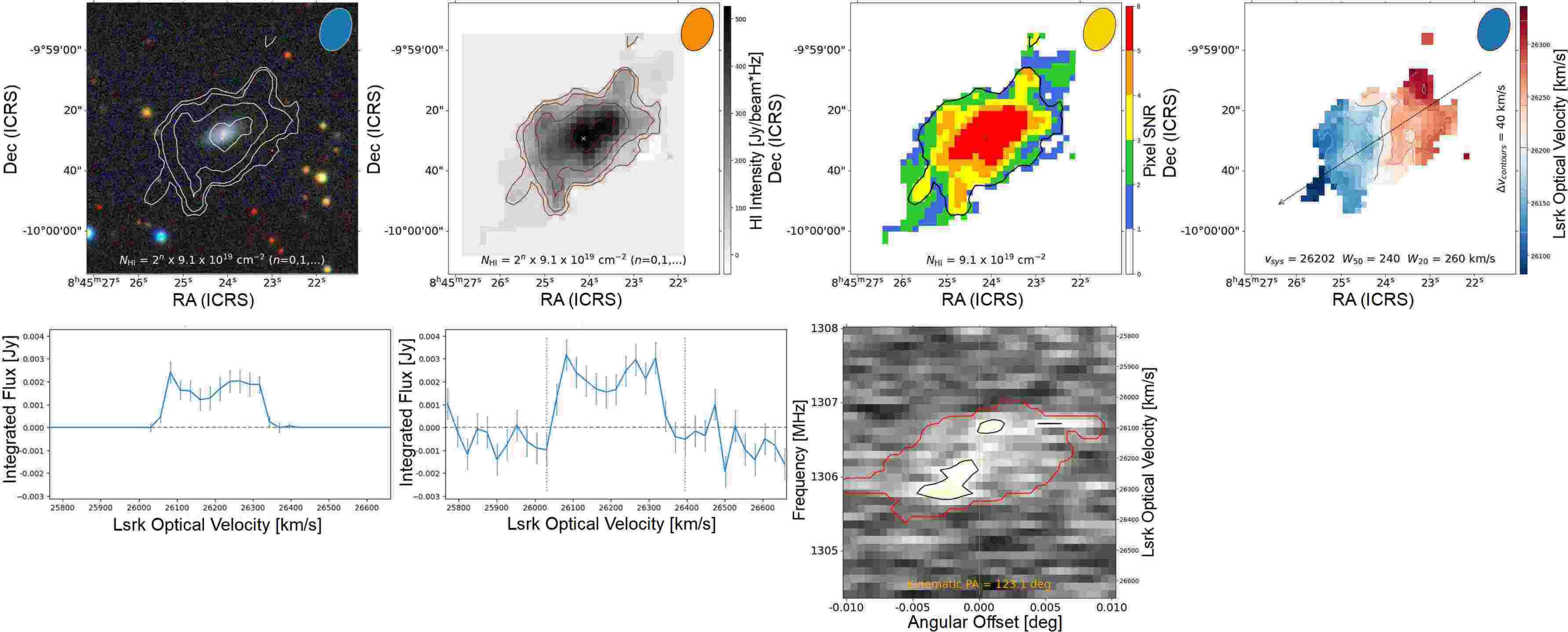}
\caption{Continued. SIP outputs for galaxy IDs 45 and 46.}
\end{figure}
\end{landscape}

\begin{landscape}
\begin{figure}
\ContinuedFloat
\includegraphics[width=1.16\textwidth]{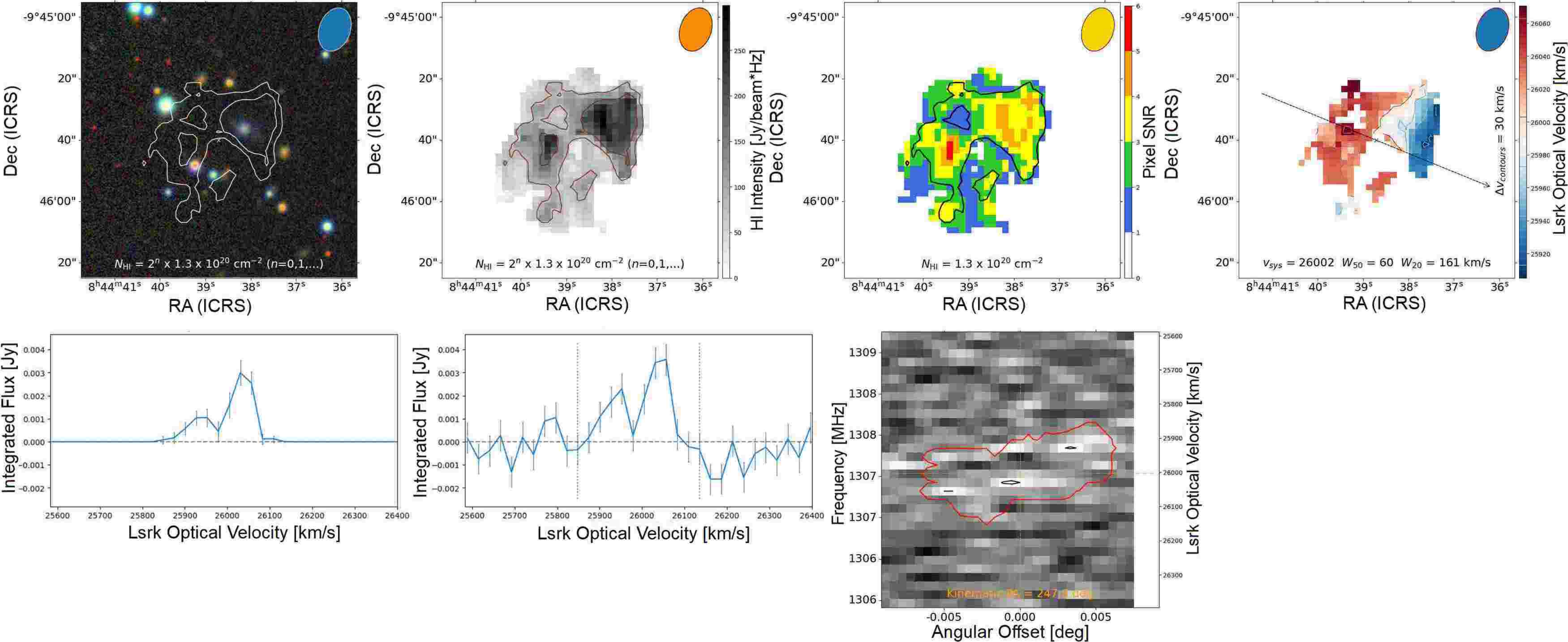}
\includegraphics[width=1.16\textwidth]{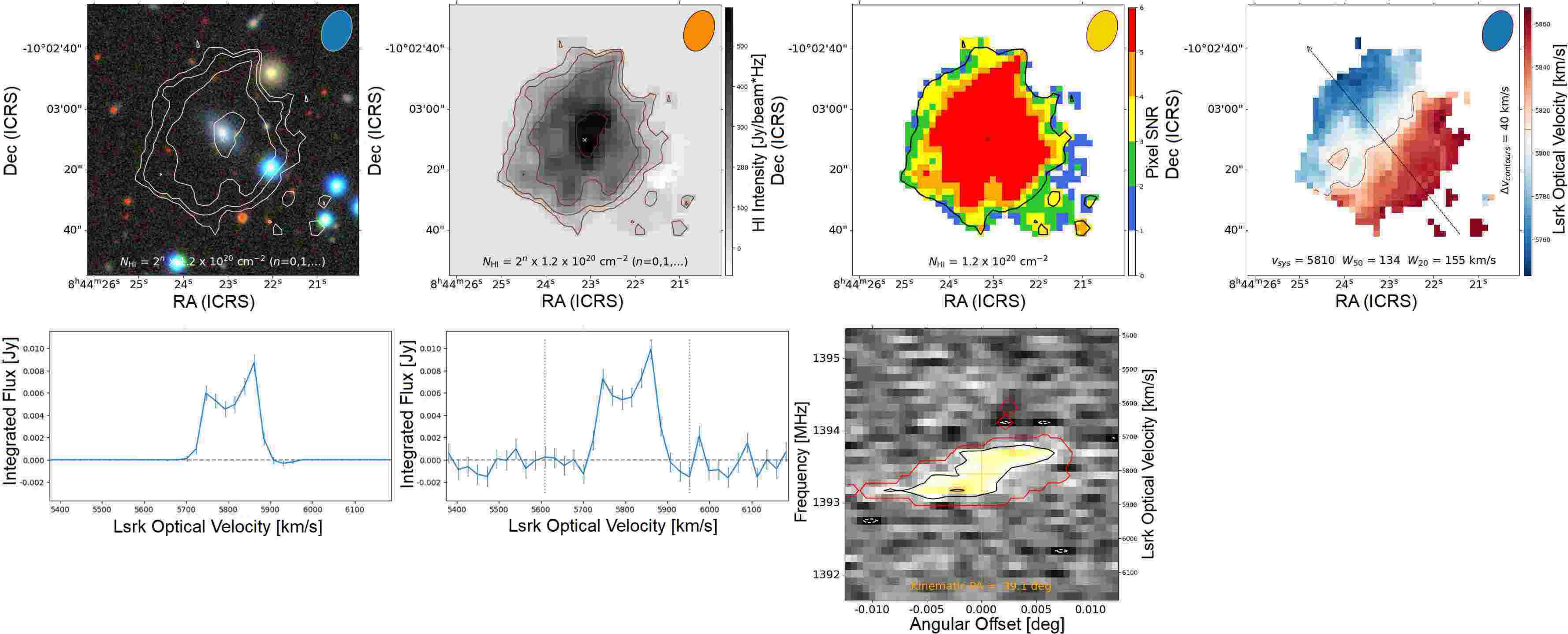}
\caption{Continued. SIP outputs for galaxy IDs 47 and 48.}
\end{figure}
\end{landscape}

\begin{landscape}
\begin{figure}
\ContinuedFloat
\includegraphics[width=1.16\textwidth]{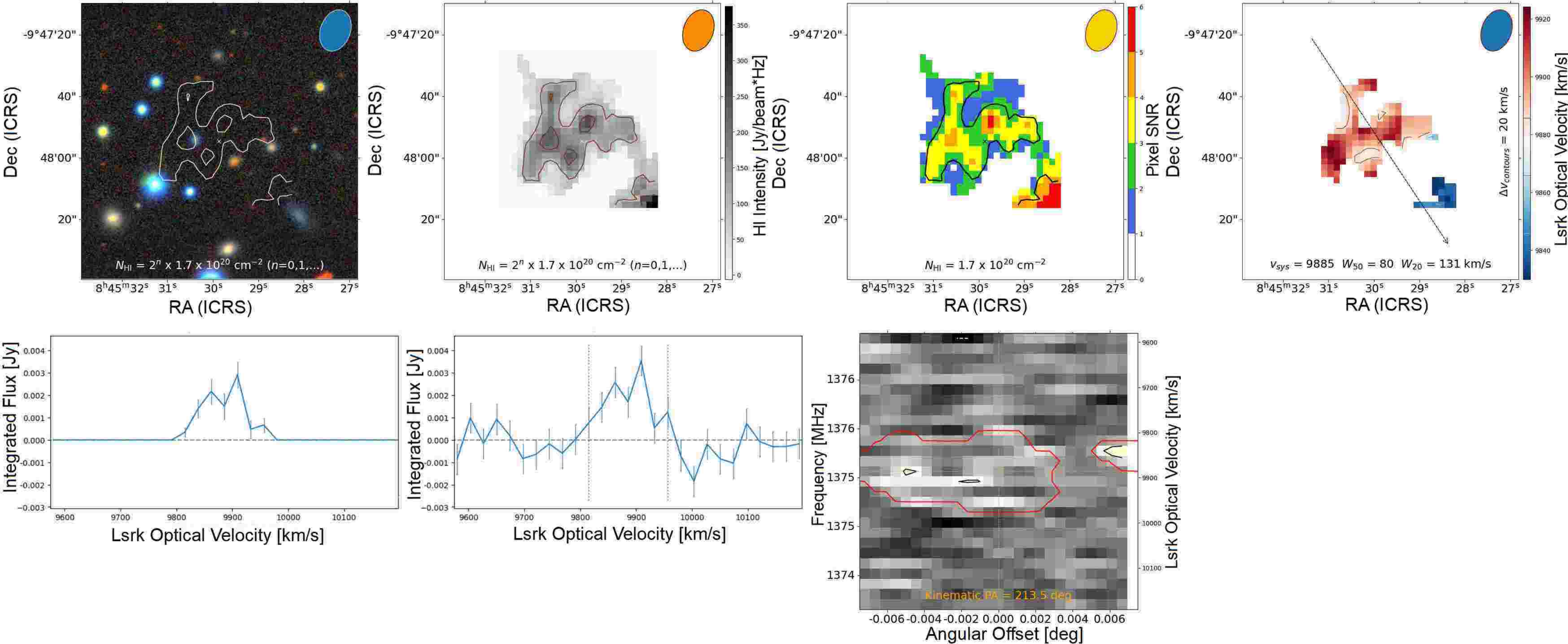}
\includegraphics[width=1.16\textwidth]{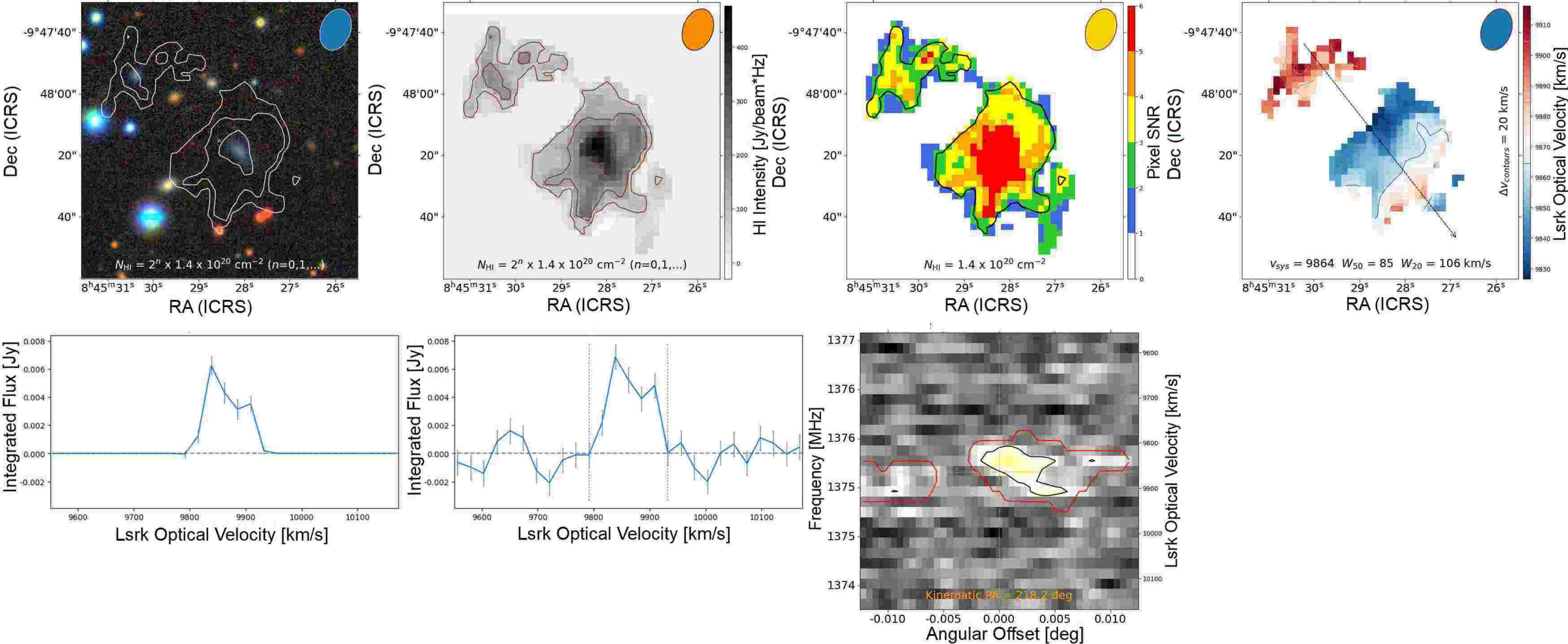}
\caption{Continued. SIP outputs for galaxy IDs 49, and for both ID 43 and 49 in the bottom set.}
\end{figure}
\end{landscape}

\begin{landscape}
\begin{figure}
\ContinuedFloat
\includegraphics[width=1.16\textwidth]{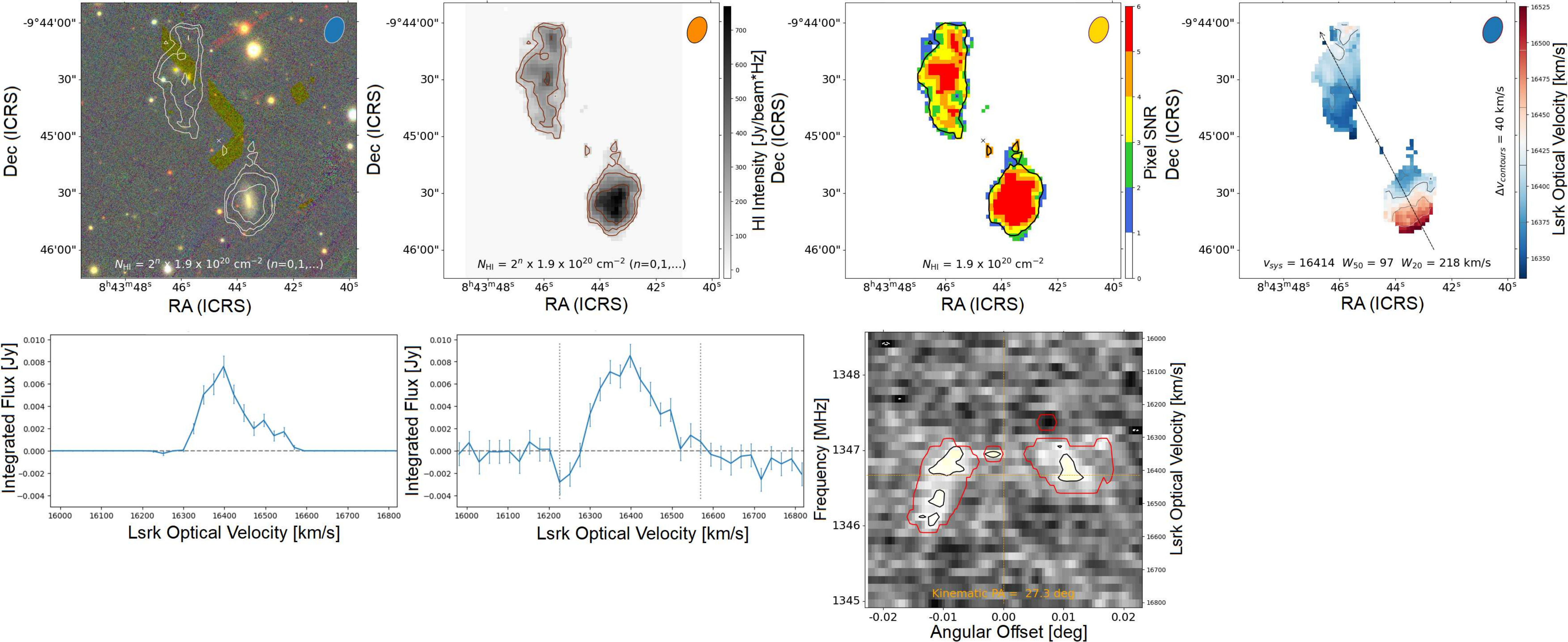}
\caption{Continued. SIP outputs for both galaxy IDs 16 and 17, which are near each other.}
\end{figure}
\end{landscape}

\begin{figure*}
\centering
\includegraphics[width=0.9\linewidth]{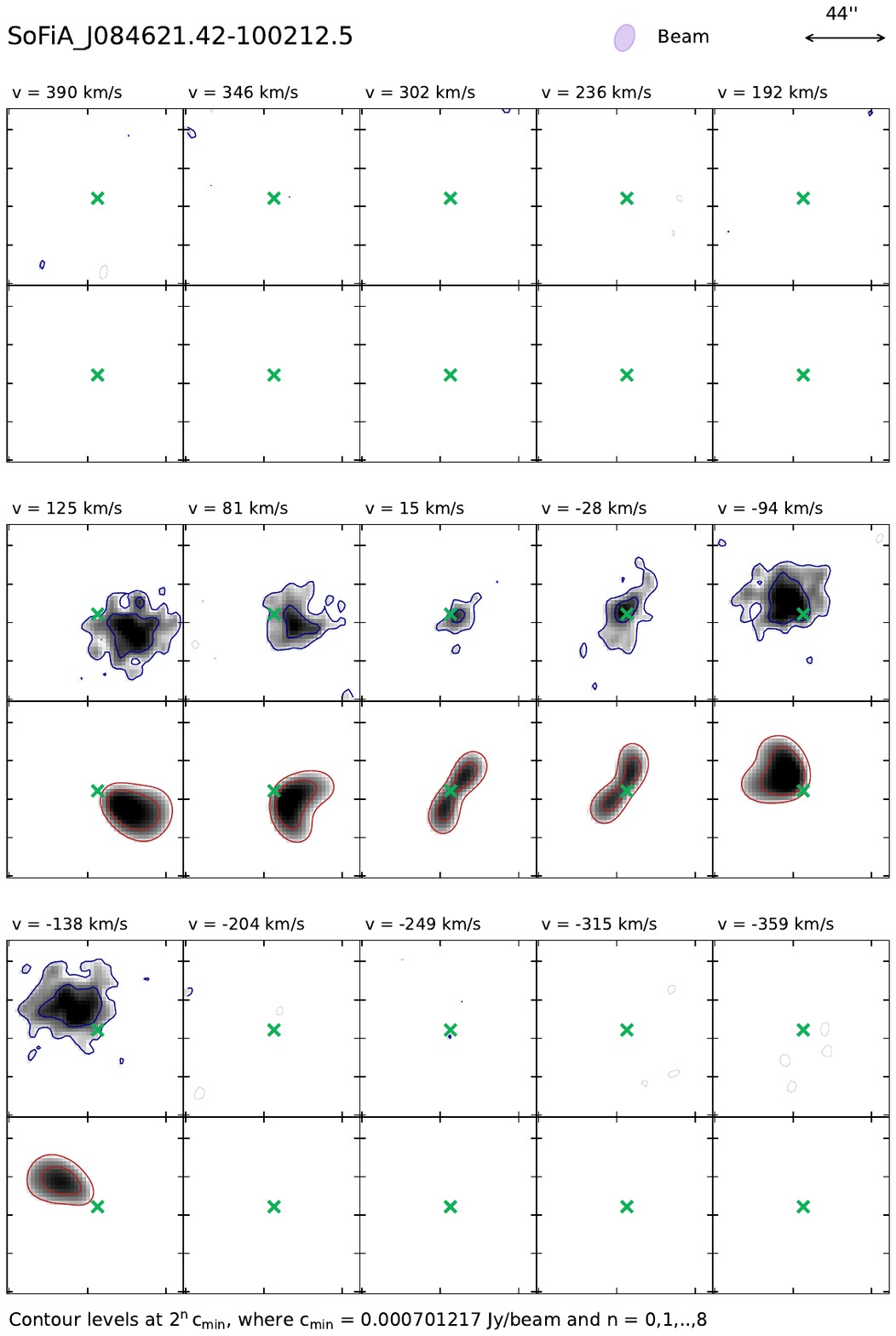}
\caption{Output from BBarolo for the channel-by-channel model fit to the \HI\ spectral line cube for ID18.}
\label{fig:app_bbarolochan}
\end{figure*}

\begin{figure*}
\centering
\includegraphics[width=0.9\linewidth]{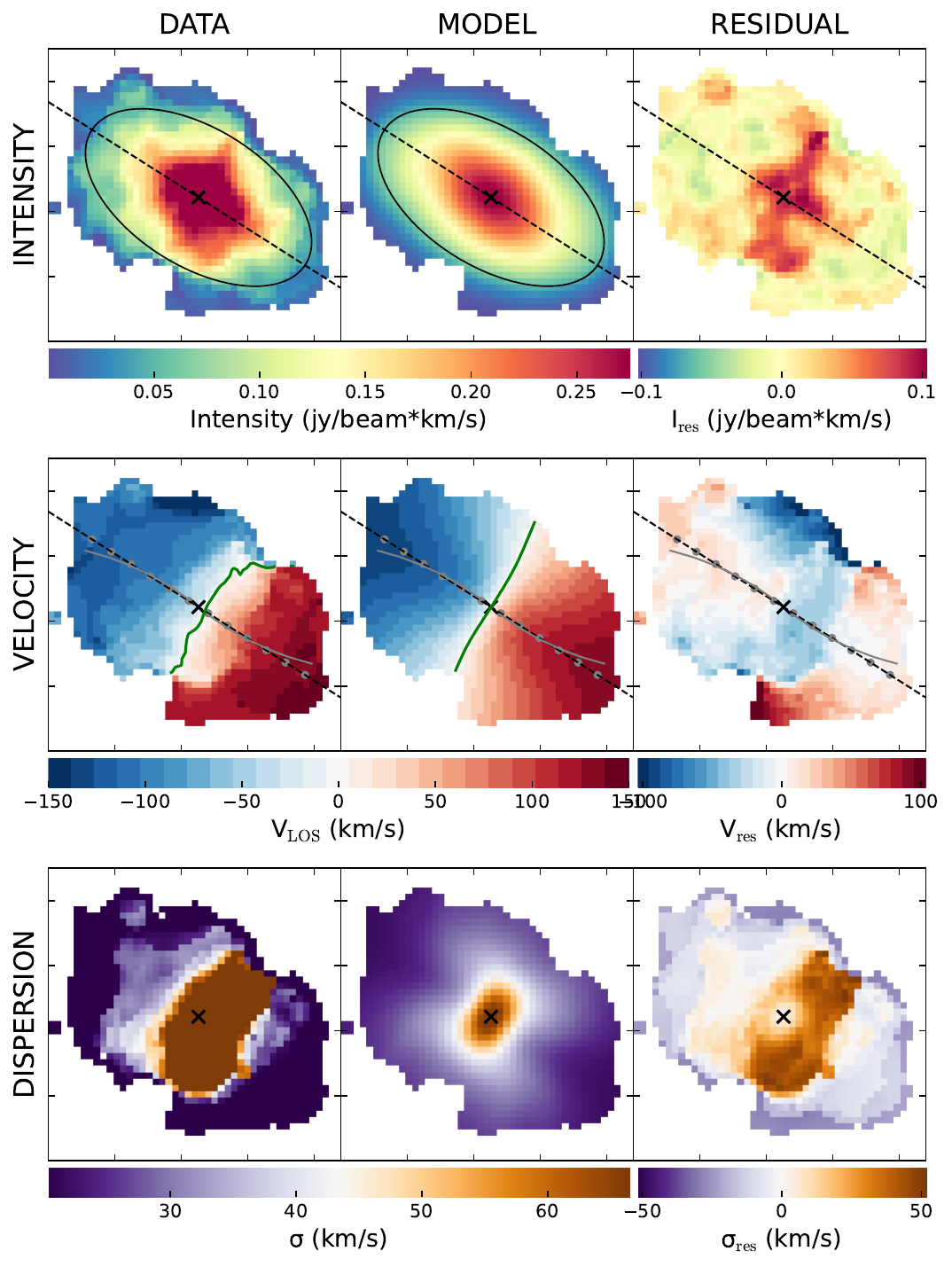}
\caption{Output from BBarolo for the moment maps for the model fit to the \HI\ spectral line cube for ID18.}
\label{fig:app_bbarolomommaps}
\end{figure*}



\begin{figure*}
\centering
\includegraphics[width=0.99\linewidth]{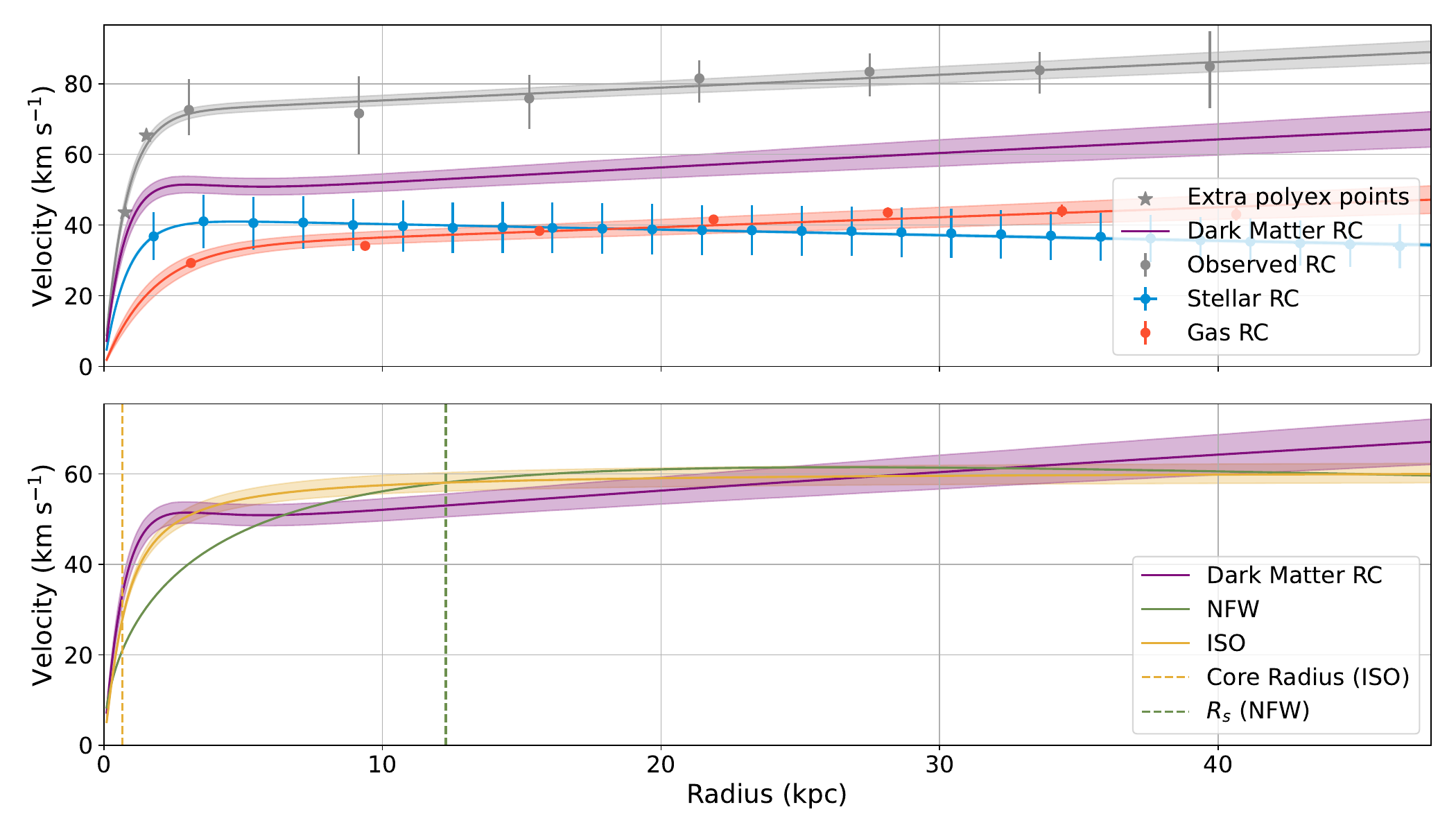}
\caption{As in Fig.~10 for ID8 (top two panels).}
\label{fig:app_rot_massmodel}
\end{figure*}

\begin{figure*}
\centering
\ContinuedFloat
\includegraphics[width=0.99\linewidth]{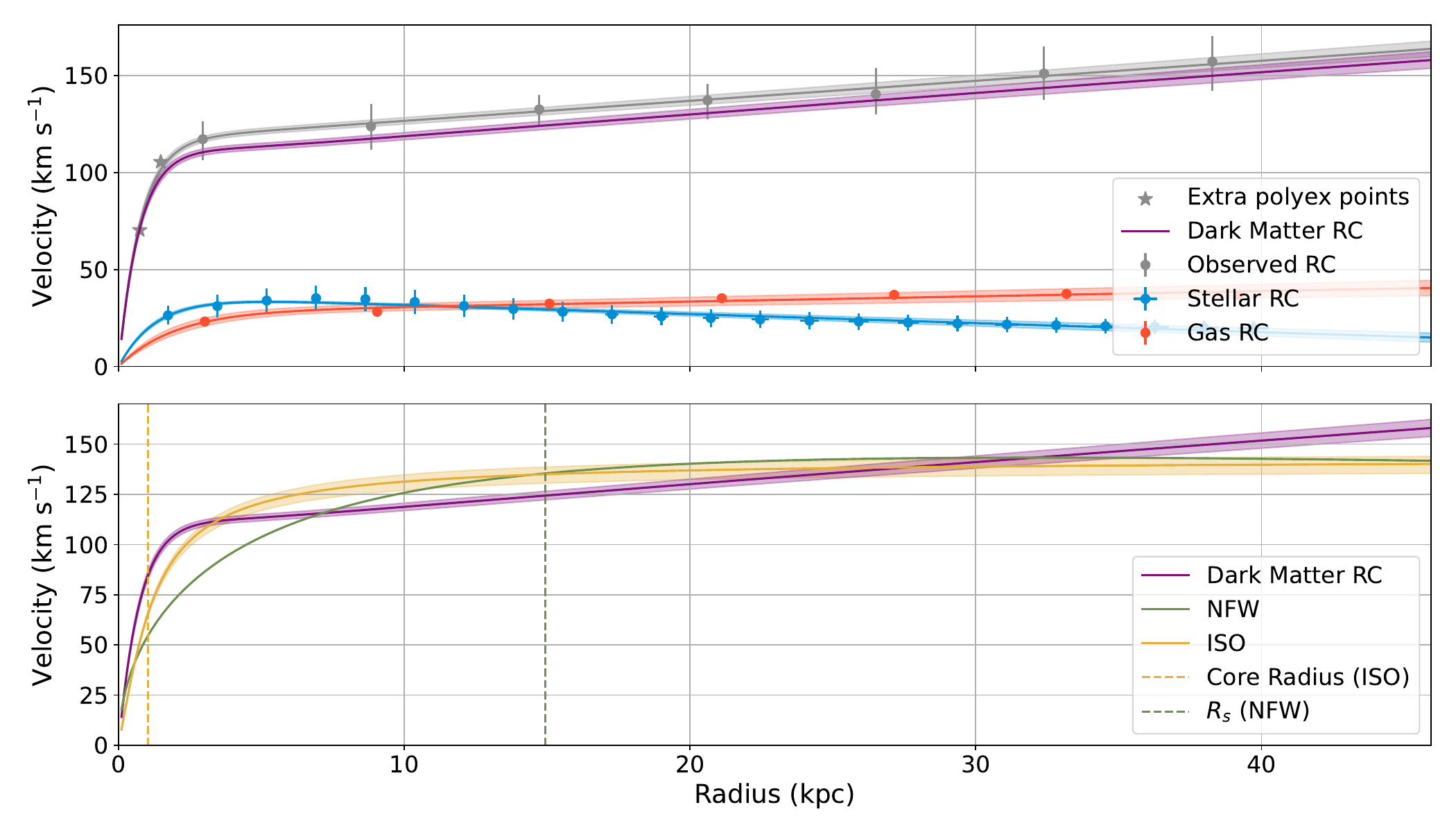}
\caption{As in Fig.~10 for ID10.}
\end{figure*}

\begin{figure*}
\centering
\ContinuedFloat
\includegraphics[width=0.99\linewidth]{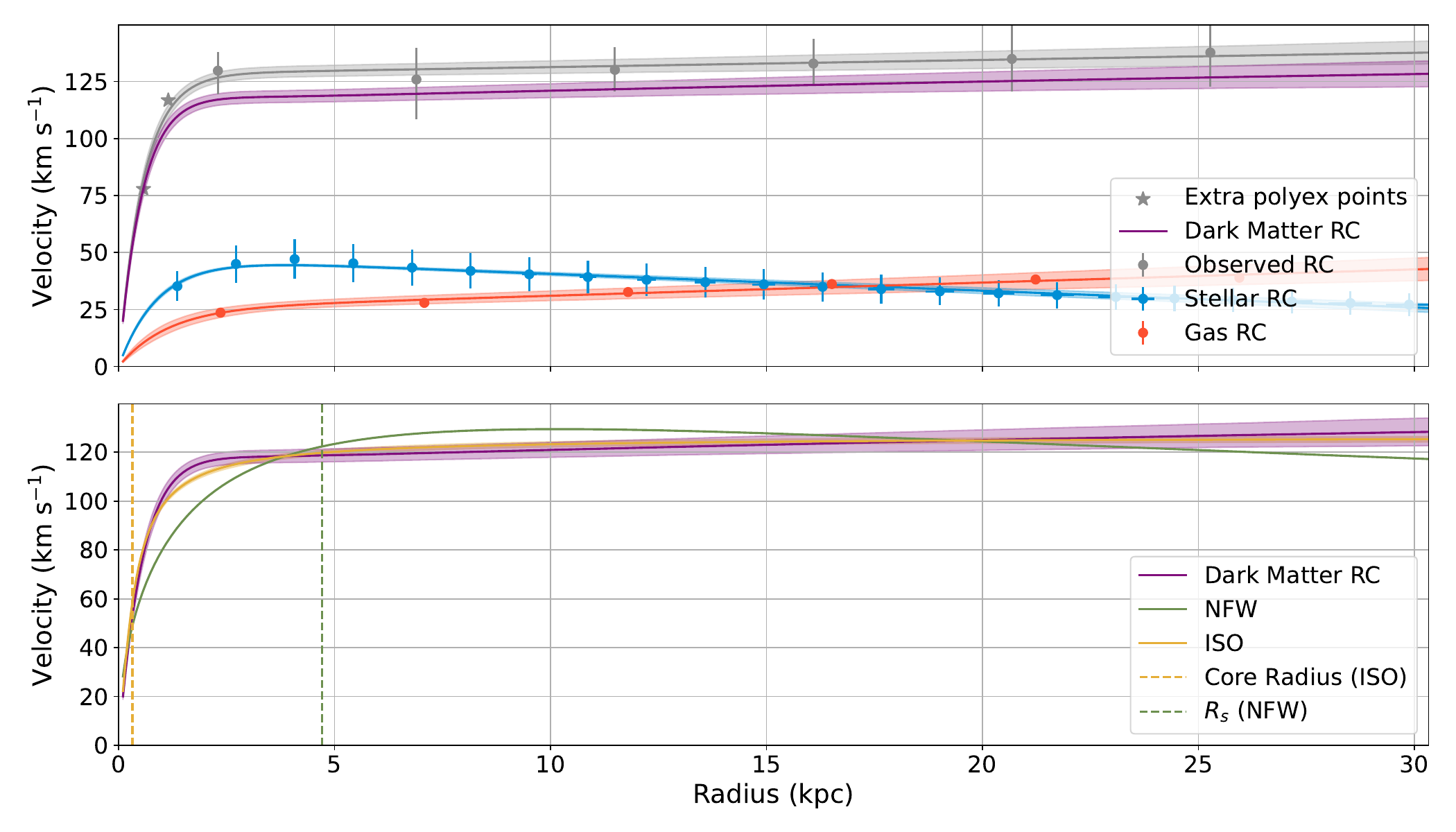}
\caption{As in Fig.~10 for ID19.}
\end{figure*}

\begin{figure*}
\centering
\ContinuedFloat
\includegraphics[width=0.99\linewidth]{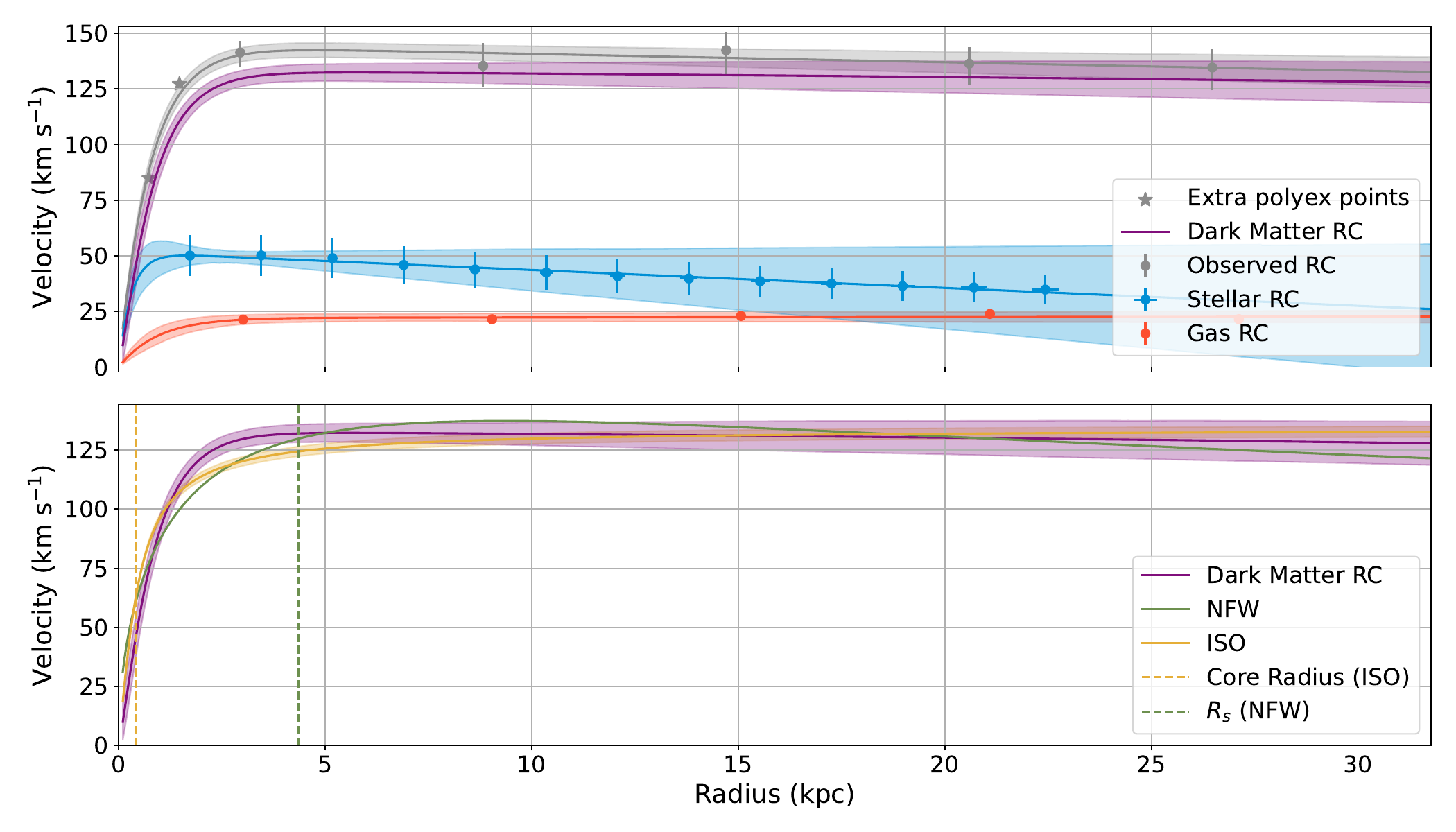}
\caption{As in Fig.~10 for ID24.}
\end{figure*}

\begin{figure*}
\centering
\ContinuedFloat
\includegraphics[width=0.99\linewidth]{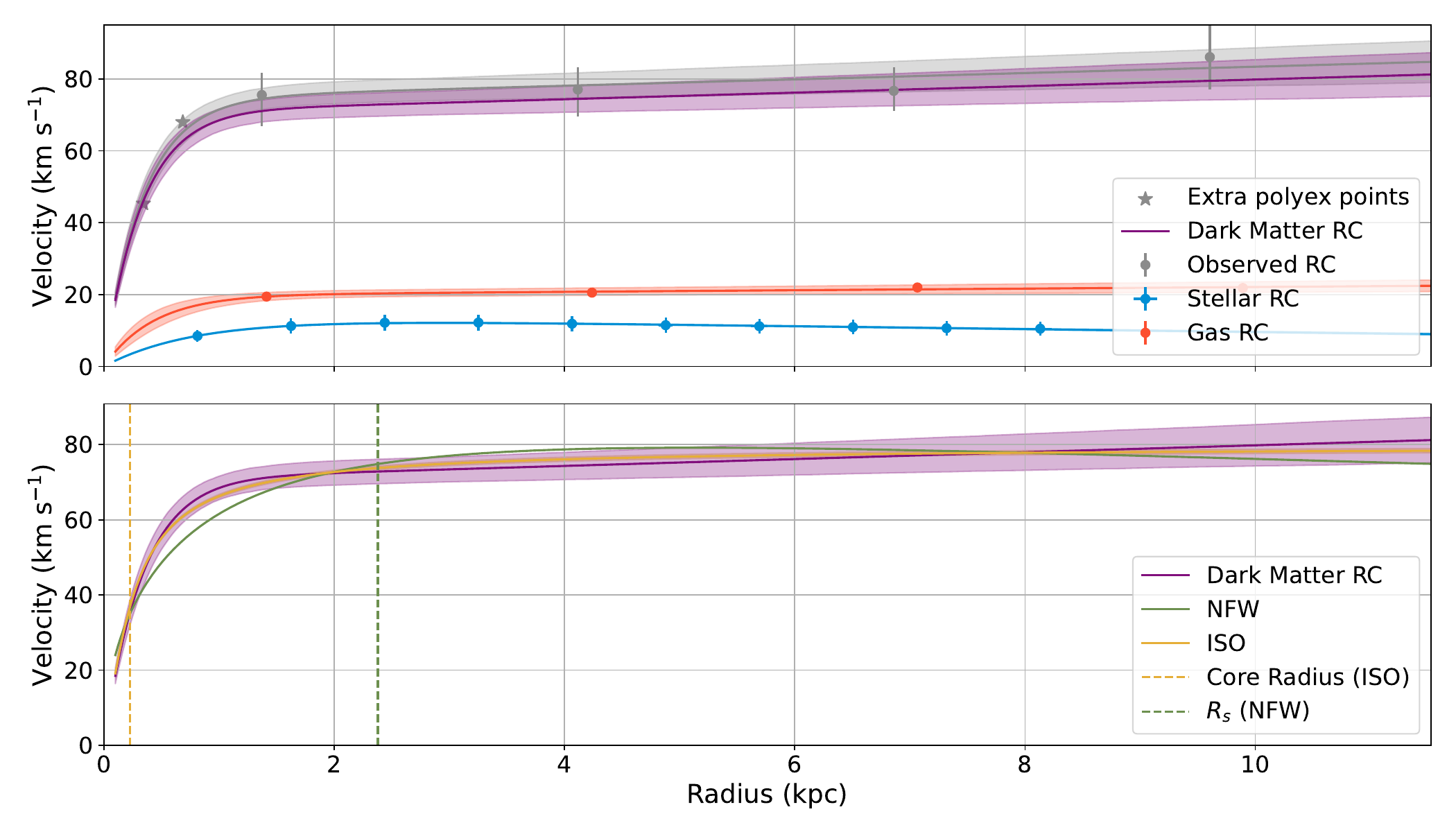}
\includegraphics[width=0.99\linewidth]{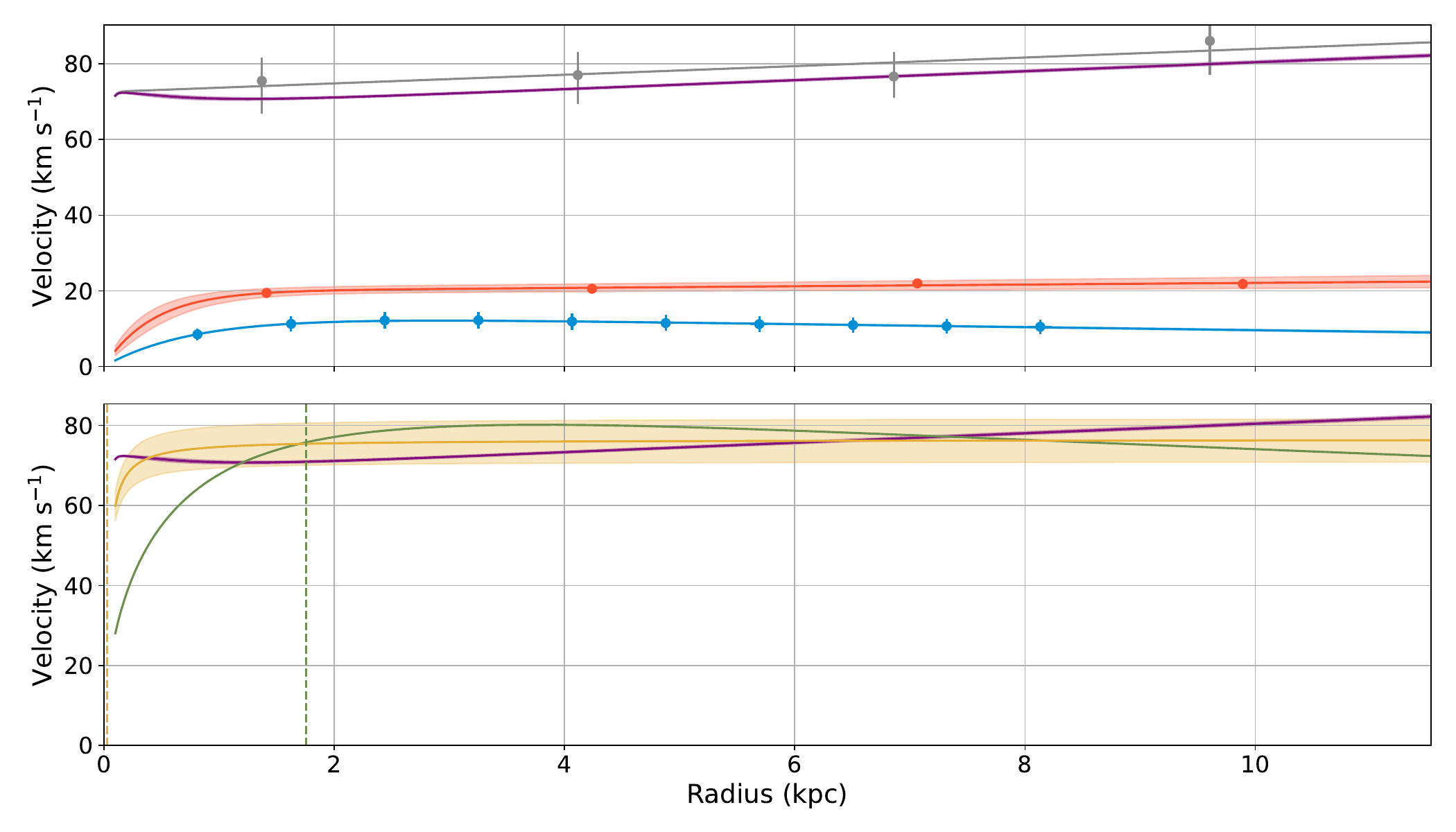}
\caption{As in Fig.~10 for ID48, {\myedit with the fitting results included without the two additional points (bottom two panels). Here we see a more realistic Polyex fit and corresponding ISO fit when including the two additional points.}}
\end{figure*}


\bsp	
\label{lastpage}
\end{document}